\documentclass[]{aa}
\usepackage{graphicx}
\usepackage{txfonts}
\usepackage{url}
\usepackage{multirow}
\usepackage{subcaption}
\usepackage{academicons}
\usepackage{makecell}
\usepackage{silence}
\usepackage{diagbox}
\usepackage{natbib,twoopt}
\usepackage[breaklinks=true]{hyperref}
\bibpunct{(}{)}{;}{a}{}{,}
\makeatletter
  \newcommandtwoopt{\citeads}[3][][]{\href{http://adsabs.harvard.edu/abs/#3}
    {\def\hyper@linkstart##1##2{}
     \let\hyper@linkend\@empty\citealp[#1][#2]{#3}}}
  \newcommandtwoopt{\citepads}[3][][]{\href{http://adsabs.harvard.edu/abs/#3}
    {\def\hyper@linkstart##1##2{}
     \let\hyper@linkend\@empty\citep[#1][#2]{#3}}}
  \newcommandtwoopt{\citetads}[3][][]{\href{http://adsabs.harvard.edu/abs/#3}
    {\def\hyper@linkstart##1##2{}
     \let\hyper@linkend\@empty\citet[#1][#2]{#3}}}
  \newcommandtwoopt{\citeyearads}[3][][]
    {\href{http://adsabs.harvard.edu/abs/#3}
    {\def\hyper@linkstart##1##2{}
     \let\hyper@linkend\@empty\citeyear[#1][#2]{#3}}}
\makeatother
\usepackage{orcidlink}
\newcommand{\PC}{Prox Cen}

\begin{document}

   \title{Time-resolved X-ray spectra of Proxima Centauri as seen by XMM-Newton}
   
   \author{
          A. Damonte
          \inst{1,2,3} \orcidlink{0009-0005-0345-6518}
          \and
          I. Pillitteri
          \inst{3} \orcidlink{0000-0003-4948-6550}
          \and
          A. Maggio
          \inst{3} \orcidlink{0000-0001-5154-6108}
          \and
          A. García Muñoz
          \inst{1} \orcidlink{0000-0003-1756-4825}
          \and
          G. Micela
          \inst{3} \orcidlink{0000-0002-9900-4751}
    }
          
    \institute{
                Universit\'e Paris Cit\'e, Universit\'e Paris-Saclay, CEA, CNRS, AIM, 91191, Gif-sur-Yvette, France
                \and
                Dip. di Fisica e Chimica, Università degli Studi di Palermo, Piazza del Parlamento 1, 90134 Palermo, Italy
                \and
                INAF – Osservatorio Astronomico di Palermo, Piazza del Parlamento 1, I-90134 Palermo, Italy
    }
             
    \date{Received 9 September 2025 / Accepted 16 December 2025}

   \abstract{
    Stellar soft X-ray ([1, 100] {\AA}) and Extreme Ultraviolet (also EUV, [100, 920] {\AA}; jointly, XUV) radiation affects the evolution and chemistry of exoplanet atmospheres. It is however uncertain to what extent the radiation's short-term variability contributes to these effects. The answer might indeed depend on the atmospheric composition in ways that remain largely unexplored. We are interested in what this variability might imply for planets around M dwarf stars, and focus on Proxima Centauri ({\PC}) for three reasons, namely: it is an active M dwarf with high levels of variability; it hosts a likely terrestrial exoplanet within its habitable zone (HZ) that will be a prime target for future direct imaging; and its proximity has led to extensive observations, yielding some of the best available X-ray data.}
    {We set out to produce time-resolved XUV spectra of {\PC} that will serve as input to atmospheric models, and to characterize the star’s intrinsic variability and  uncertainties in the inferred spectra.}
    {We analyzed the entire dataset of archival XMM-Newton observations for {\PC}. To derive the time-resolved X-ray spectra, we implemented a new pile-up correction, a new adaptive time-binning algorithm, and a time-dependent plasma model selection. The estimated EUV spectrum is based on a published template, that we scale with proposed relationships between X-ray and EUV fluxes.}
    {We produced spectra of {\PC} from 1 to 920 {\AA} over $\sim$260 ks of observations with unprecedented time resolution. The instantaneous X-ray flux of {\PC} varies between about 20 times and one-fifth of the average value over the available baseline, with significant differences between wavelengths. We further quantify how variability affects the estimated average flux when a limited number of snapshots (each typically of 30 ks exposure) are available, as is common in X-ray surveys. Future investigations of {\PC}'s planet atmospheres should fold in the time variability and uncertainties described here.}
    {}

    
    \maketitle

    \section{Introduction}\label{section:introduction}

    The evolution of planetary atmospheres is affected by the stellar environment. In particular, soft X-rays and extreme ultraviolet radiation may lead to strong heating of the upper atmosphere, causing hydrodynamic escape (or photoevaporation) and erosion \citep{yelle_aeronomy_2004, garcia_munoz_physical_2007, colombo_stellar_2025}. XUV-induced photoionization also affects the chemical balance of the atmosphere, occasionally resulting in detectable amounts of, for example, excited states of the H and He atoms \citep{yan_extended_2018, czesla_h_2022}. Stellar emission at these wavelengths varies by several orders of magnitude over timescales from minutes to Gyrs \citep{favata_stellar_2003, gudel_x-ray_2004, kowalski_stellar_2024}. The atmospheric response to XUV radiation is complex, and it remains unclear whether different temporal distributions of irradiation, even with the same total deposited energy, affect atmospheric evolution differently. It also remains unclear to what extent atmospheric composition dictates this response. Characterizing the XUV radiation over time and as a function of wavelength (or equivalently, energy) is critical for atmospheric modeling \citep{binder_x-ray_2024, amaral_impact_2025, chen_effects_2025, garcia_munoz_modeling_2025, rockcliffe_far-ultraviolet_2025}.

    Flares, responsible for most of the short-term (seconds to days) variability of a star's emission, are events in which the plasma in the corona, transition region, and chromosphere is impulsively heated by magnetic reconnection. They are stochastic phenomena, and it is generally agreed that their occurrence frequency follows a power-law distribution with energy, 
    \({dN}/{dE} \propto E^{-\alpha}\) \citep{kowalski_stellar_2024}. During flares, ionizing radiation increases by up to several orders of magnitude and the spectral shape of the emission hardens, with the XUV component being particularly enhanced. This hardening is especially relevant when considering high-metallicity planetary atmospheres, as the photoionization cross sections are typically large for the heavy elements in the X-ray band. It has been shown that in some systems, the flaring component enhances the atmospheric evaporation rate and shortens atmospheric lifetimes \citep{lee_effects_2018, amaral_contribution_2022, amaral_impact_2025}. Moreover, single high-energy flares drive rapid changes in the chemistry, which may alter the short-term climate  \citep{segura_effect_2010, konings_impact_2022, ridgway_3d_2023, chen_effects_2025}. Lastly, flares are frequently associated with coronal mass ejections (CMEs), events involving the release of large amounts of plasma from the star \citep{yashiro_statistical_2009, webb_coronal_2012, reames_solar_2021}, which can also impact atmospheric escape rates \citep{cohen_dynamics_2011, khodachenko_stellar_2014, hazra_magnetic_2025, vidotto_star-planet_2025}. This highlights how important it is to know the flaring properties of the host star when studying the evolution of a planet.

    Recently, the exoplanet community has been particularly interested in (low-mass, 0.08 < M$_{\star}$/M\textsubscript{$\odot$} < 0.6) M dwarf stars. They are very abundant in our galaxy and their small sizes enhance the planet-to-star size contrast. Because of their cooler effective temperatures, the habitable zone (HZ) \citep{kasting_habitable_1993} lies considerably closer than for more massive stars. This makes them excellent targets for the detection and characterization of small HZ exoplanets \citep{dressing_occurrence_2013, dressing_occurrence_2015}. HZ planets around M dwarfs are subject to XUV fluxes comparable to, or greater than, those received by HZ planets around more massive main-sequence stars. Furthermore, while the coronal activity of the latter usually decreases considerably after the first few hundreds of Myrs, low-mass stars usually remain active much longer \citep{pizzolato_stellar_2003, preibisch_evolution_2005}. Higher coronal activity results in both higher quiescent fluxes and an increased probability of large flares. For these reasons, the total amount of XUV energy absorbed by HZ planets around M dwarfs over their lifetimes is often considerably higher than for planets orbiting more massive stars \citep{johnstone_active_2021}.

    This work is motivated by two poorly explored questions, namely: to what extent can short-term stellar variability influence atmospheric retention over Gyr timescales? How significantly can it affect the strength of certain atmospheric observables? Before addressing them through in-detail atmospheric modeling, we must first characterize the stellar XUV emission with temporal resolution matching flare timescales (minutes to days). That is one of the goals of the present work, in which we also quantify how the observing cadence and different choices in data reduction and plasma modelling affect the derived spectra.

    We focus on {\PC}, the closest star to the Sun at a distance of 1.30 pc \citep{gaia_collaboration_dr3_vizier_2020}. {\PC} is an M4.5–M7 dwarf \citep{maldonado_hades_2020, bessell_late_1991, gaidos_trumpeting_2014} and one of the best-studied low-mass stars. It has a mass of 0.12 M$_\odot$, a radius of 0.15 R$_\odot$, and a rotation period of $\sim$87 days \citep{boyajian_stellar_2012, kiraga_age-rotation-activity_2007}. It hosts two small, non-transiting planets, {\PC} b and d, with minimum masses of $\sim$1.1 and $\sim$0.3 $M_\oplus$ and orbital periods of 11.2 and 5.1 days, respectively \citep{anglada-escude_terrestrial_2016, kipping_no_2017, mascareno_diving_2025}. Planet b orbits within the host star's HZ, making the system a key target for future direct imaging missions \citep{currie_direct_2023, blind_ristretto_2024}.
    Despite its estimated age of $\sim$4.9 Gyr \citep{bazot_uncertain_2016}, {\PC} remains magnetically active. An activity cycle is suspected but remains unconfirmed, particularly in the X-rays. \citet{wargelin_x-ray_2024} and \citet{mascareno_diving_2025} report possible cycles with periods of $\sim$8 and $\sim$18 years, respectively, although the modulation is mainly observed in the optical and infrared. A detailed X-ray analysis by \citet{ayres_landscape_2025} did not confirm any cycle, as often occurs in active stars near the saturated regime, characterized by high $L_{\mathrm{bol}}/L_X$ ratios. {\PC} flares frequently over most of the electromagnetic spectrum \citep{pavlenko_temporal_2019, howard_mouse_2022, gudel_flares_2004}. Due to its proximity, interstellar absorption of X-rays is minimal.

    These characteristics have led to extensive observations, making {\PC}'s coronal X-ray emission one of the best-characterized among M dwarfs, with broad multi-instrument coverage across X-ray and other wavelengths \citep{haisch_einstein_1980, haisch_solar-like_1995, gudel_flares_2004, fuhrmeister_multi-wavelength_2011, ribas_full_2017, wargelin_x-ray_2024}. Despite this, no time-series spectra of the star are currently available for atmospheric modeling. In this work, we produce such spectra in preparation for follow-up work that will investigate how various putative atmospheres for the {\PC} planets respond to them. To that end, we use archival data from the XMM-Newton space telescope \citep{schartel_xmm-newton_2024}.

    The work is organized as follows. In Section \ref{section:observations_and_data_reduction}, we present the observations of {\PC} that we analyzed, and the applied data reduction procedure. We give specific attention to the pile-up correction (Section \ref{section:pile-up_correction}), the time binning procedure (Section \ref{section:time_binning}), and the plasma modeling (Section \ref{section:plasma_modeling}). In Section \ref{section:regression_and_uncertainties}, we present the spectral regression method and approach used to estimate uncertainties, while Section \ref{section:EUV_scaling} applies literature-based scaling laws to derive EUV fluxes from the X-ray measurements. Section \ref{section:discussion} discusses the variability across different energy bands, compares the obtained spectra with previous literature, and evaluates the impact of limited observational coverage on the inferred fluxes. Finally, Section \ref{section:conclusions} summarizes the main conclusions.
    
    \section{Observations and data reduction}\label{section:observations_and_data_reduction}

    \subsection{Observations}\label{section:observations}
    
        We downloaded the {\PC} data from the XMM-Newton \citep{schartel_xmm-newton_2024} Science Archive\footnote{\url{nxsa.esac.esa.int/nxsa-web/}}. The observations span the period 2000–2019, with a total exposure time of about 250 ks. Table \ref{table:observations} contains specific information on them. We used data from all three European Photon Imaging Camera (EPIC) instruments, namely MOS1, MOS2, and pn \citep{struder_european_2001, turner_european_2001} and from the Optical Monitor (OM) \citep{mason_xmm-newton_2001}, which was equipped with the $U$ filter in all observations.
    
        \setlength{\tabcolsep}{3pt}
        \begin{table}
        \caption{XMM-Newton observations used in this work.}
            \centering
            \begin{tabular}{c c c c c}
                \hline
                \hline
                    OBSID        & Date       & Exposure (ks) & PI\\
                \hline
                \\
                    
                0049350101    & 2001-08-12 & 67        & M. Güdel \\\\
                
                0551120301    & 2009-03-10 & 29        & \multirow{3}{*}{C. Liefke}\\
                0551120201    & 2009-03-12 & 31        & \\
                0551120401    & 2009-03-14 & 29        & \\\\
                
                0801880201    & 2017-07-27 & 22        & \multirow{4}{*}{B.J. Wargelin}\\
                0801880301    & 2017-08-16 & 31        & \\
                0801880401    & 2017-09-07 & 27        & \\
                0801880501    & 2018-03-11 & 25        & \\\\
                    
                              &            & tot: 261  & \\\\
                \hline
            \end{tabular}\
            \tablefoot{Observation 0049350101 is analyzed in e.g. \cite{gudel_x-ray_2002, gudel_flares_2004} and \cite{reale_modeling_2004}. Observations 0551120301, 0551120201, and 0551120401 are analyzed in e.g. \cite{fuhrmeister_multi-wavelength_2011}. Observations 0801880201, 0801880301, 0801880401, and 0801880501 are analyzed in e.g. \cite{wargelin_x-ray_2024}.}
            \label{table:observations}
        \end{table}
        \setlength{\tabcolsep}{6pt}

    \subsection{Standard XMM-Newton pipeline}\label{section:standard_pipeline}
    
        We used the Current Calibration Files from release n°406\footnote{\url{cosmos.esa.int/web/xmm-newton/ccf-release-notes}} \citep{ballet_xmm-ccf-rel-0406_2024} and ran the standard pipeline of the XMM-Newton Science Analysis System (SAS\footnote{\url{cosmos.esa.int/web/xmm-newton/xsa}}, \cite{esa_xmm-newton_2023}, v. xmmsas\_20230412\_1735-21.0.0). X-ray raw data consist of a list of events, each corresponding to a detected photon with arrival time, sensor position, energy (in instrument channels), and pattern. The "pattern" is an integer identifying the geometry of the charge distribution generated by the photon on the sensor. These detections are hereafter referred to as "counts" or "events".
        
        Following the SAS user guide\footnote{\url{xmm-tools.cosmos.esa.int/external/xmm_user_support/documentation/sas_usg/USG/}} \citep{esa_xmm-newton_2023}, we processed MOS, pn, and OM data with the {\it emchain}, {\it epchain}, and {\it omfchain} tasks, respectively. For MOS events, we also ran {\it emanom} to exclude anomalous CCDs (Charge-Coupled Devices) \citep{esa_xmm-newton_soc_users_2023}. For each X-ray source, we identified its centroid and extracted the counts within a circular region. A 30 arcsec radius was used for pile-up correction (see Appendix \ref{appendix:pile-up}), while the final extraction radius was optimized for signal-to-noise via the {\it eregionanalyze} task.
        
        As in \cite{gudel_flares_2004} (from now on G+04), we selected background regions for MOS1 and MOS2 from outer CCDs to avoid source-contaminated areas in the main CCD. An annulus covering active, non-anomalous CCDs was used. For pn, the background was extracted from the same CCD as the source. Prior to this, we excluded all source regions listed by the pipeline processing system (PPS)\footnote{\url{cosmos.esa.int/web/xmm-newton/pipeline/}}. False positives near the target were removed manually.
        
        We checked for flaring soft proton background \citep[SPB,][]{kuntz_epic-mos_2008} in each observation. This was done by comparing out-of-field count rates with user guide thresholds \citep{esa_xmm-newton_soc_users_2023} and by examining the output of the {\it espfilt} task. SPB levels were low or negligible and significantly weaker than the source; therefore, no good-time-interval (GTI) correction was applied. Given the use of a wide background region, a standard background subtraction was deemed sufficient.
    
    \subsection{Pile-up correction}\label{section:pile-up_correction}
    
        When the photon flux is high, the probability of multiple photons hitting the camera during the same frame exposure, and close enough on the detector to be indistinguishable, becomes non-negligible. This instrumental effect, known as pile-up, causes the onboard software to interpret two or more photons as a unique event, leading to flux loss. The resulting event has an energy equal to the sum of the true photon energies (causing spectral distortion) and, often, a higher pattern value, as more pixels than usual are typically involved. This latter effect, called pattern migration \citep{ballet_pile-up_1999, jethwa_when_2015}, contributes to both the flux loss and the spectral distortion. If uncorrected, pile-up causes the retrieved spectrum to appear both fainter and harder than the true stellar emission.
        
        The available observations, except for observation 00493501, use a frame mode with a reduced active sensor area, enabling shorter readout times that reduce pile-up occurrence. This is enough to make pile-up negligible during quiescent periods, when the count rate stays well below the conservative thresholds of 3 s$^{-1}$ and 1.5 s$^{-1}$ for pn and MOS, respectively \citep{jethwa_when_2015}, but not during bright flares, when the count rate reaches up to 70 s$^{-1}$ and 22 s$^{-1}$ for pn and MOS respectively. We have devised a procedure to minimize the effects of pile-up that we briefly summarize in what follows and that is thoroughly described in Appendix \ref{appendix:pile-up}. As a takeaway message, we find that during the brightest flares, the flux loss (the relative number of counts missed when pile-up is uncorrected) can reach up to 22{\%} for pn and 13{\%} for MOS1/MOS2 as shown in figure \ref{fig:lc_pu_comparison}.
                
        Since pile-up depends on the instrument mode, a consistent correction is required when combining instruments or analyzing datasets taken with different modes. In addition, pile-up is time-dependent, as it scales with the instantaneous flux, and its implications are potentially amplified in the data reduction process. These considerations pose various challenges in time-resolved investigations like ours. 
        
        There is no single prescription to address pile-up. It is common to exclude the affected time intervals or remove the affected instrument from the entire analysis. After several tests, we found that the standard approaches (see Appendix \ref{appendix:pile-up}) do not provide a suitable correction for our pn+MOS1+MOS2 time-resolved analysis. For them, either the approach is not feasible across all instruments, or it reduces the number of counts too much, introducing uncertainties larger than the pile-up effect itself.
        
        Our newly developed pile-up correction is based on ideas from \cite{ballet_pile-up_1999} and \cite{molendi_assessing_2003}, which seek to enhance the single-pattern filtering strategy. The standard correction excludes all photon events that interact with more than one pixel on the detector. This strongly lowers the spectral distortion, as the probability of piled single-pattern events is typically negligible. A consistent amount of flux loss, however, remains even if the single-pattern filtering is accounted for in the response matrix, because of the pattern migration effect. Indeed, many events that were originally singles may have merged into doubles or higher patterns, which are now excluded. Accurately estimating how many of these excluded events should have been singles would require high signal-to-noise data and precise pattern-energy relations from the calibration. We simplified the correction by assuming that all flux loss is caused by singles merging into greater patterns (in principle there are also double+single and double+double). This corresponds to the worst-case scenario of flux loss, which we consider sufficient for our goals. 
                
        To estimate the number of missing events, we used the empirical relation between count rate and count loss from \cite{jethwa_when_2015}. The same number of events was generated using the energy distribution of the singles as a reference and these were added to the raw event list. Again, the full correction procedure, including the estimation of flux loss, generation of missing events, and results evaluation, is described in Appendix \ref{appendix:pile-up}.

    \subsection{Time binning}\label{section:time_binning}
        
        To obtain reliable X-ray fluxes model regression is required \citep{arnaud_xspec_1996}. Instrumental counts are grouped into time intervals and an instrumental spectrum, i.e. the spectrum formed by the counts per energy channel, is obtained for each one of them. A plasma model is then selected and convolved with the response matrix. Finally, the model is fitted to the spectrum to retrieve the physical quantities. When the source varies in flux and spectral shape, the time-binning strategy strongly affects the results. This is due not only to higher uncertainties at finer time resolution (there will be fewer counts per bin at a given count rate), but also to phase effects, i.e., where interval start times are placed. These effects are critical during flares, which can be smoothed and distorted by contamination from quiescent and cooling phases surrounding the peak.
        
        We here propose an innovative time-binning algorithm to optimize the balance between time-resolution and S/N (Signal-to-Noise ratio) while minimizing the impact of phase effects. The core of our strategy is a sliding window approach, where consecutive intervals overlap. Both the duration of the interval and the shift/step size are variable functions of the count rate/counts. This dynamic adjustment ensures that S/N and time resolution remain within predefined thresholds. The algorithm operates by iteratively selecting the interval length and then determining the start of the subsequent interval. To select the interval length $\Delta t$, we follow the set of rules described in equation \ref{eq:time-binning}. We start from the very first event of the event list. The interval length is selected as the maximum value between two: $\Delta t ( C_{min} )$, which is the interval length obtained by binning the first $C_{min}$ counts after the start of the interval, and $\Delta t_2$, which is obtained from the second equation of the system. $\Delta t_2$ is selected as the minimum between two values: $\delta_{max}$, a maximum time resolution, and $\Delta t(C_{max})$, that is the length defined by the $C_{max}$ counts after the start of the interval. We further comment on the reasons behind this algorithm below. To select the next interval start, i.e. the shift of the sliding window, we determine how many times the same counts are allowed to appear in different intervals. This is translated into a counts-percentage shift parameter, $\varsigma$, which defines the percentage of counts of the previously selected interval to be excluded from the event list. We will then restart from point (1) using the new event list.
        
        \begin{equation}\label{eq:time-binning}
        \begin{cases}
            \Delta t = \text{max}(\Delta t(C_{min}), \Delta t_2)\\
            \Delta t_2 = \text{min}(\delta_{max}, \Delta t(C_{max})) 
        \end{cases}
        \end{equation}
    
        We now comment on the reason for each parameter and the adopted values:
        \begin{itemize}
            \item  $C_{min}$ : ensures a minimum S/N; it limits time resolution during low-flux periods. It was fixed to $10^3$ counts.
            \item  $\delta_{max}$ : prevents excessively fine (unnecessary for our goals) time resolution at high count rates. If the interval for $C_{min}$ counts is shorter than $\delta_{max}$, counts are added until reaching $\delta_{max}$ to improve S/N without sacrificing the desired resolution. It was fixed to 180 s.
            \item  $C_{max}$ : limits the maximum counts per interval. If a spectrum has too many events, the dominant source of error changes from the low statistics to the instrumental+model systematic that renders comparisons less correct. It was set to $10^4$ counts.
            \item $\varsigma$ (step 2): a percentage defining how much of the previous interval's counts are excluded before starting the next interval. It was fixed at 33{\%} to balance overlap and independence of adjacent intervals and their total number.
        \end{itemize}
     
        We applied the algorithm to background-subtracted, pile-up-corrected data within the [0.18, 12] keV energy range, combining counts from all three instruments. Consequently, time resolution also depends on the number of active instruments at each time.
    
        A few consequences of our procedure are as follows. First, the time resolution is not uniform and depends on the S/N. Second, the superposition of consecutive intervals reduces the smoothing that naturally results from time binning. Third, consecutive intervals are not independent, as they share part of their counts. This does not affect the total fluxes, but it may need to be considered when performing certain time analyses.
        
        From eight observations, totaling 260 ks, the algorithm produced about 3200 intervals with an average duration of 300 s, ranging from 150 to 2000 s. Because of varying time resolution and interval overlap, we linearly interpolated the results using {\it scipy.interpolate} to generate a uniformly binned time series with 30-second resolution.

    \subsection{Plasma modeling}\label{section:plasma_modeling}
    
        The coronal plasma is usually considered an optically thin collisionally-ionized plasma (thermal plasma). \cite{gudel_flares_2004} confirmed that this assumption also holds for {\PC}. The emission is highly dependent on temperature, density and elemental abundances, properties that also vary with time and are non-uniformly distributed in space for most stars. Spectroscopic measurements necessarily average information both over time and over the visible surface of the star since for stars other than the Sun no spatial resolution is available. Flares further complicate the description of this plasma, being unpredictable in time, limited in space (over the star's surface), and spanning temperatures from hundreds of thousands to tens of millions of degrees.
        
        The standard approach involves looking at the thermal structure of the plasma. The emission measure (EM) quantifies how much plasma contributes to the emission at a given temperature. Measuring how this quantity varies as a function of temperature yields the differential emission measure (DEM). It is possible to derive it from high-resolution spectroscopy for the brightest sources. This is done looking at single emission lines of different elements (sensitive to different temperatures) and comparing their intensities.
        
        G+04 performed this analysis for {\PC} by dividing the major flaring events of observation 0049350101 into different time intervals. \cite{fuhrmeister_high_2022} (from now on F+22) did the same but using CXO (Chandra X-ray Observatory) and HST (Hubble Space Telescope) data. They divide the data into flaring and non-flaring intervals and analyze them separately. \cite{sanz-forcada_connection_2025} (from now on SF+25) used all the XMM-Newton available observations, and STIS (Space Telescope Imaging Spectrograph), FUSE (Far Ultraviolet Spectroscopic Explorer) and EUVE (Extreme Ultraviolet Explorer) observations. They did not distinguish between flaring and non-flaring intervals.
        
        When performing time-resolved analyses though, it is more common to use simpler models and low- to medium-resolution data, as we do in this paper. This is because the S/N is usually too low for a time-resolved high-resolution analysis. The usual strategy \citep{ciaravella_loop_1997, gudel_x-ray_2004, pandey_study_2008, osten_mouse_2010, pillitteri_x-ray_2022} is to approximate the temperature distribution with a few effective temperatures, either free or fixed on a grid. The standard model used to describe coronal plasmas at a fixed temperature is the Astrophysical Plasma Emission Code (APEC) \citep{smith_collisional_2001}. Depending on the S/N and on the goal, we can use the sum of one or more APEC components. The parameters of each component are the temperature, the elemental abundances and the norm (directly related to the EM). The abundances are usually given relative to a fixed reference table, usually the Solar one. In this work we use the abundance tables from \cite{wilms_absorption_2000}. In the APEC model the elemental ratios are fixed to that table and only the overall metallicity can vary. A vAPEC model also exists where each elemental abundance can be adjusted separately.
    
        Our model is described by the following expression: $\text{Model} = \text{TBabs}*(\text{APEC}_1+...+\text{APEC}_N)$. The sum of multiple APEC components describes the multi-temperature plasma.
        
        TBabs is a convolutional model that accounts for interstellar medium absorption (ISM). The hydrogen column value, i.e. the parameter of the TBabs model describing the ISM absorption, was fixed to $10^{18}\,\mathrm{cm}^2$ as done in \citep{wargelin_optical_2017, wargelin_x-ray_2024} for each of our models. This value is small enough to have a nearly negligible effect on the energy range we are considering, as expected for the closest star to us.
        
        When modeling low-resolution spectra the information in the data is often insufficient to retrieve detailed information about the structure of the plasma, in particular for a time-resolved analysis. This also depends on the available prior information about the star and its plasma. For temperature and normalization parameters, two approaches can be distinguished: 1) if the thermal structure is not known, a common approach is to represent the plasma with a few effective temperatures, usually between one and four. The temperatures are left free so that the dominant components are set by the data. The required number of components is the minimum that provides an acceptable fit, accounting for the S/N, luminosity, distance, absorption, and background. 2) Using either a previously derived DEM(T) or the temperatures obtained in point (1), a set of temperatures can be fixed while leaving only the normalizations free. This reduces the number of free parameters and allows more complex models to be fitted in a stable and tractable way.
    
        The plasma metallicity is the last parameter. In principle, it varies during flares, as detected in previous studies \citep{favata_spectroscopic_1999, gudel_xmm-newton_2001}, but this issue is complex \citep{gudel_x-ray_2009} and beyond the scope of this work. Moreover, since our intervals contain few counts and have low spectral resolution, we did not attempt to constrain elemental abundances, either as a function of time or globally. We excluded high-resolution RGS data because their time resolution cannot be matched to ours, and abundance estimates from longer integrations were already reported by G+04, SF+25, and F+22. RGS analyses also cannot constrain the H abundance and require adopting a reference ratio. The Fe/H ratio is usually taken as reference, since Fe lines dominate the emission, ensuring that the main impact of abundances on the flux remains fixed even when other lines are constrained by the RGS measurements (assuming that the relative abundances found for the other elements are not excessively large, otherwise their contribution could be compared to the Fe one). To test whether a more detailed abundance calibration could improve the fits, we modeled the flare in observation 0049350101 using six vAPEC components selected following the RGS time-resolved analysis of the same flare in G+04. As shown in Figure \ref{fig:model_selection}, this model performs worse than the simpler one in terms of information criteria (AIC). We therefore deemed further tests with vAPEC components and fine-tuned abundances unnecessary.
        
        In the APEC models the metallicity was fixed to 0.5, as done in G+04 and SF+25. Although other works (e.g., F+22) assumed solar metallicity, a subsolar value is closer to the average obtained when the abundances are allowed to vary in our retrieval. Indeed, when the metallicity is left free, its retrieved value ranges from 0.2 to 1.4 (depending on the number of temperature components), leading to large variations in the derived EMs while the fluxes are less affected. To avoid possible nonphysical trends in the EMs, we fixed the metallicity to a value drawn from the literature, since allowing it to vary does not significantly affect the fluxes. We tested a subset of observations to evaluate the impact of metallicity as another possible source of uncertainty. Varying the metallicity between 0.2 and 1 changes the total flux by at most $\sim$20{\%}, but the median difference remains below 2{\%}, and therefore within the flux uncertainties.
        
        Additionally, the model choice may affect the final fluxes, and, even more, the spectral shape, as shown in Appendix \ref{appendix:further_comments_on_model_selection}. We use different models to test different hypotheses and the sensitivity of the retrieved synthetic spectra to the model choice:
       
        \begin{enumerate}
            \item 2 APEC with free temperatures: the structure of this model is described in detail in Table \ref{table:modelparameters1}. No prior information is assumed except for the abundances, fixed to 0.5. It corresponds to description (1) above. We decided to use two temperature components. Our goal with this model is to get an idea of the temperature range of the coronal plasma and compare the results with the DEMs described in the literature. Specifically, we want to test which temperatures are reached for tuning the subsequent models.
            
            \item 3 APEC: we fix the temperatures (3 of them in total) and only look for the norm parameters. The temperature values we use are 0.2 keV, 0.7 keV, 1.2 keV. These values were based on the results from the 2 APEC model and the DEM$(T)$ in G+04 and SF+25. 
    
            \item 5 APEC: in this case we add two hotter temperatures to the 3 APEC case that cannot reproduce the high energy emission in the simpler model, in particular during flares. The values we use are 0.2 keV, 0.7 keV, 1.2 keV, 1.5 keV and 2.7 keV. 
            
            \item 6 vAPEC: in this case, starting from the G+04 analysis of the flare in observation 0049350101, we assume that the 6 DEMs derived for different time intervals (that they identify as quiescence, rise, flare peak, decay, secondary flare peak, following decay) may be used as a base to describe the general behavior of Proxima. We approximate each DEM with a vAPEC component with effective temperature fixed to the DEM's highest value. In this model, when available from \citet{gudel_flares_2004}, we fix abundances to the reported ones, otherwise, we adopt solar abundances. 
        \end{enumerate}
        \begin{table}
                \caption{Model 1 parameter relationships.}
                \centering
                \begin{tabular}{c c c c c c c}
                    \hline
                    \hline
                        Par & State & Val & Link & Delta & Min & Max \\ 
                    \hline
                        T1 & Free & 0.2 & -- & 1 & 0.1 & 5      \\
                        Z1 & Fixed & 0.5 & -- & -- & -- & --       \\
                        norm1 & Free & 1e-3 & -- & 1 & 1e-7 & 2   \\
                        T2 & Linked & -- & T1+$\Delta T$ & -- & -- & --   \\
                        Z2 & Fixed & 0.5 & -- & -- & -- & --        \\
                        norm2 & Free & 1e-3 & -- & 1 & 1e-7 & 2    \\
                        N$_H$  & Fixed & $10^{18}$ & -- & -- & -- & -- \\
                        $\Delta T$ & Free & 0.5 & -- & 1 & 0.1 & 5   \\
                    \hline
                \end{tabular}
                \tablefoot{Redshift is fixed to 0. The free parameter $\Delta T$ is defined as a custom additive model with norm fixed to 0. T$_i$ are the temperatures in keV, Z$_i$ are the relative abundances, and norm$_i$ the norm values. N$_H$ stands for the hydrogen column density, in units of atoms/cm$^2$.}
                \label{table:modelparameters1}
            \end{table}

    \subsection{Regression and uncertainties}\label{section:regression_and_uncertainties}
    
        We simultaneously fitted the MOS1, MOS2, and pn data. We used the SAS tasks {\it evselect}, {\it backscale}, {\it arfgen}, {\it rmfgen} to extract the source spectrum and its relative background, together with their response matrices, for each time interval identified with the binning algorithm. The task parameters were left unchanged from their default values except for the point spread function (PSF) energy that was set to 1 keV as {\PC}'s spectrum is usually centered around this value. We used the energy range within which all the instruments are calibrated i.e. [0.18, 12.0] keV.
        
        To perform the spectral regression we used XSPEC \citep{arnaud_xspec_1996}. We used C-stat\footnote{Strictly speaking, XSPEC uses the B-stat when background is subtracted.}\; statistic, given the few counts per interval. As usually prescribed for the C-stat implementation in XSPEC, the energy bins were grouped to contain at least one count each. To maximize the likelihood and estimate the uncertainties of the parameters we used the XSPEC built-in Monte Carlo Markov Chain (MCMC) method. To get the starting values for the chains we first fitted the data with a Levenberg–Marquardt likelihood maximization algorithm. Comparison of the final results with those obtained using the initial Levenberg–Marquardt algorithm shows that, for all our models, the best fit does not change significantly. Using the MCMC speeds up and simplifies the error computation.
        
        We generated synthetic unabsorbed spectra using the best-fit parameters and a resolution of 0.01 {\AA}. To consider uniform intervals and use a standard range for X-rays, the synthetic spectra are generated in the range [0.12, 12.0] keV corresponding to [1, 100] {\AA}.

        Uncertainties at the 1$\sigma$ level were estimated for the model parameters using the XSPEC {\it error} command, which automatically uses the loaded MCMC chains to compute them. For the flux errors, using the standard {\it cflux} convolutional model would have needed an additional fit for each energy bin so we did not use this approach. We used the information contained in the chains instead. We selected the 68 percentile of vectors with the best statistic from the chains. From them, we selected 100 representative vectors and for each of them we computed the total flux. We took the maximum and minimum fluxes obtained in this way as upper and lower errors. 
    
        The 100 vectors were selected as follows: 98 of these vectors of parameters were randomly extracted. To these we added the vectors with the smallest and the biggest likelihood from the selection of vectors that are within 1-sigma. In this way we account for different shapes in the energy space as the {\it cflux} method does, but without the need of performing MCMC and error computation again. To estimate the errors for each energy bin of the synthetic spectrum we consider the maximum and minimum values occurring in each bin among the 100 spectra. This overestimates the errors because the energy bins are not independent but strongly related to the spectral shape, but we consider it reasonable as a conservative approach that also provides error estimates as a function of energy.
        
        We highlight that 1-sigma refers to the criterion used to select the chain vectors. It is related to the errors of the fluxes/parameters, but the latter are not necessarily Gaussian and may be asymmetric, so when considering them we need to track the upper and the lower errors separately. From now on, when using the terms lower/upper error, we will always refer to these errors or to errors propagated from them.
    
\begin{figure}
            \centering
            \includegraphics[width=0.99\linewidth]{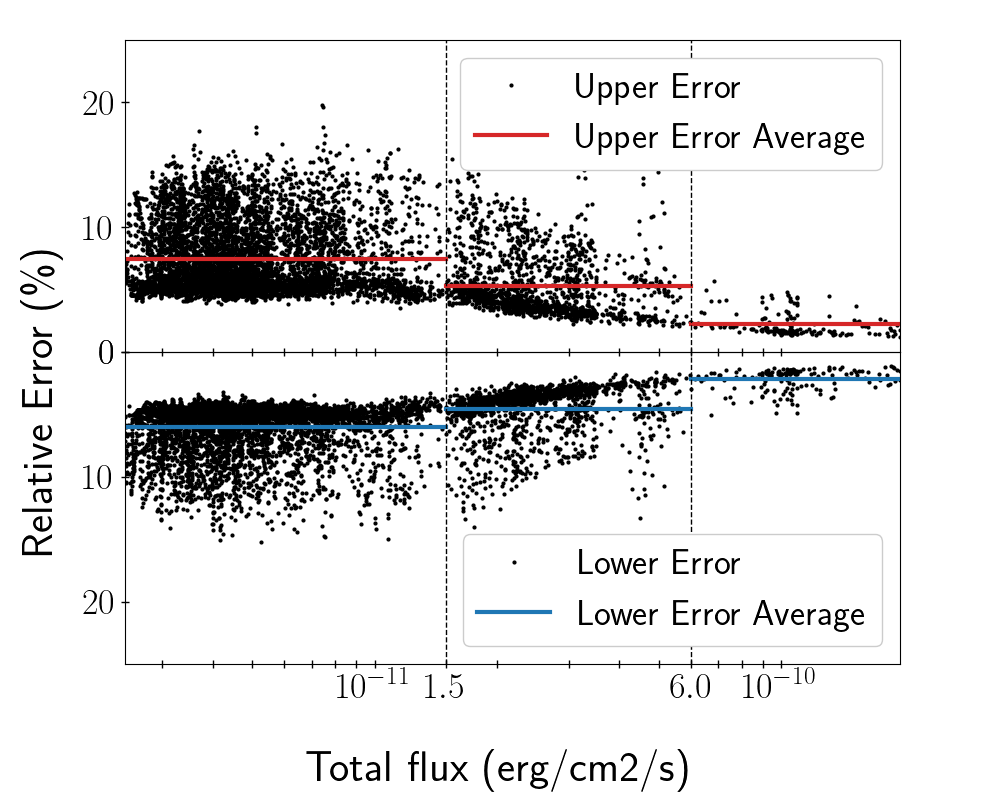}
            \caption{Upper and lower relative errors on the total integrated X-ray flux as a function of total flux for the 3-temperature model. Vertical dashed lines delimit the by-eye–selected ranges. Red and blue segments represent the average errors in the corresponding ranges.}
            \label{fig:total_flux_errors}
        \end{figure}

        Errors on the total flux for the different time intervals are shown in Figure \ref{fig:total_flux_errors} as a function of the total flux. We identified three ranges by eye: fluxes smaller than $1.5 \times 10^{-11}$ erg s$^{-1}$ cm$^{-2}$, fluxes greater than $6 \times 10^{-11}$ erg s$^{-1}$ cm$^{-2}$ and intermediate fluxes. For each interval, we computed the average upper and lower uncertainties, represented as horizontal segments in the Figure. The spread is moderate, with uncertainties consistently below 20{\%}. Nearly all flux values above the upper threshold originate from the peak of the flare in observation 0049350101.
    
        Based on this and on the results of the pile-up correction presented in Appendix \ref{appendix:pile-up}, we apply the correction only when the estimated flux loss exceeds the uncertainty on the total flux, specifically for fluxes above $1.5 \times 10^{-11}$ erg s$^{-1}$ cm$^{-2}$, corresponding to the lower of the two thresholds defined above. This way we retain a better S/N ratio during the low-counts time intervals, where pile-up is less significant. The total flux reference to perform this selection is obtained as the average between the fluxes obtained with the 3 APEC models for the two different datasets. We use the 3-temperature model because we used it to define the thresholds.

    \subsection{Time-dependent model selection}\label{section:time-dependent_model_selection}
    
        We next select the best-fitting model as a function of time by using the Akaike information criterion. We compute it for the 4 different models described in Section \ref{section:plasma_modeling}, each one with a different number of temperatures (2, 3, 5 and 6) and settings (free temperatures, temperature grid, APEC or vAPEC models). For each time interval we select between the dataset without pile-up correction (NOCORR; see \ref{appendix:pile-up}) and the one obtained with the new correction (NEWCORR). The selection is based on the total flux threshold defined in Section \ref{section:regression_and_uncertainties}. 
        
        Assuming that these are the only possible models, we can use the Bayesian approach, to quantify the relative probability of each model being the one that generates the data. For this purpose, we compute the Akaike weights $\omega_i$, defined as:
        \begin{equation}\label{eq:akaike-weights}
            \omega_i = \frac{\exp{(-\frac{1}{2}(\text{AIC}_i - \text{AIC}_{\text{min}})})}{\sum_{m=1}^{M}\exp{(-\frac{1}{2}(\text{AIC}_m - \text{AIC}_{\text{min}})})}
        \end{equation}
        
\begin{figure}
            \centering
            \includegraphics[width=0.99\linewidth]{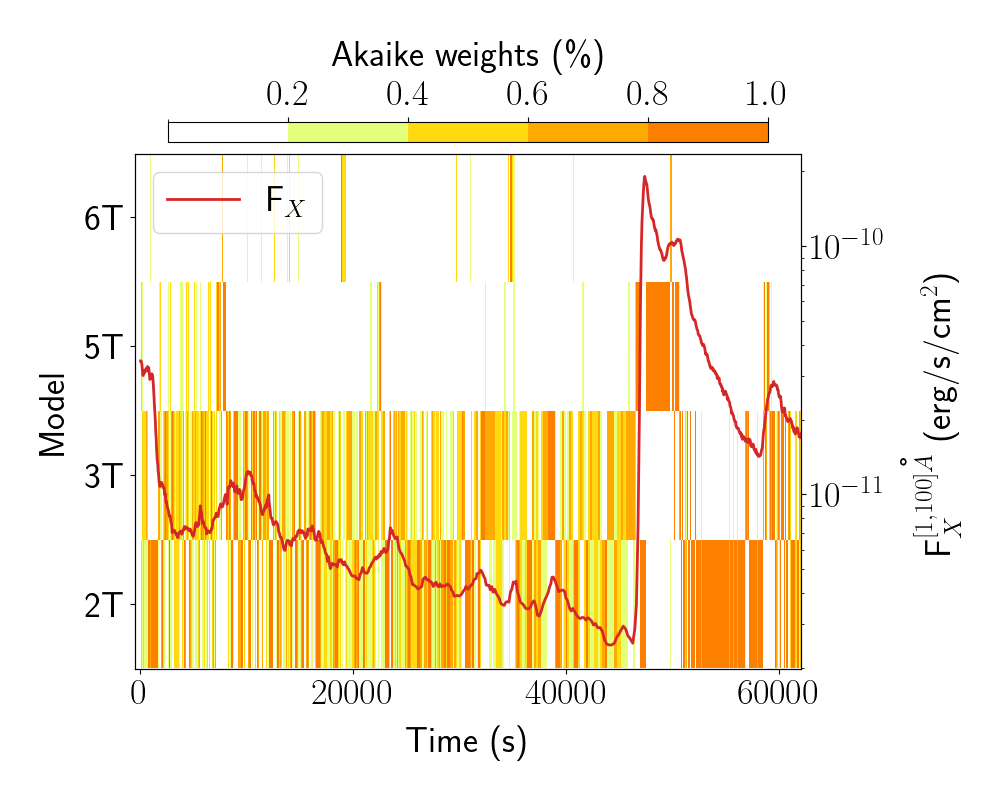}
            \caption{Comparison of $\omega_i$ for the different intervals of observation 0049350101. The left axis represents the model, the right axis shows the wavelength-integrated flux light curve ([1, 100] {\AA}). The values of the Akaike weights for each model as a function of time are color coded as indicated by the colorbar.}
            \label{fig:model_selection}
        \end{figure}

        In this equation, AIC$_i$ is the Akaike information criterion as defined in \cite{burnham_model_1998}. $i$ is the index of the model, and the sum goes over the four models ($M=4$). $\text{AIC}_{\text{min}}$ represents the minimum value obtained among the four models. In Figure \ref{fig:model_selection} we show the results of this exercise for observation 0049350101. For the four models we chose, we represent the weights as a function of time. For each interval we have a set of weights that add up to unity. We divide the space of the weights into five intervals, which we represent with a colorbar. The ratio of the Akaike weights can be interpreted as the ratio between the probabilities that the corresponding models are the true ones. Accordingly, in our analysis we choose the model with the highest $\omega_i$ value at each time step. We also plot the wavelength-integrated flux light curve ([1, 100] {\AA}), previously used to select the pile-up correction. We note that it is common for the best-fitting model weight to be smaller than 80\%.
    
        We highlight some common behaviors. As one may expect, the number of temperatures of the "best" model overall increases with the total flux. This is because a more complex model may be favored when more counts are available, and because higher fluxes are usually due to flares, which correspond to higher-temperature plasma requiring additional components. The best-fitting model tends to change rapidly in time, even when no evident variation is present in the light curve, and disregards the fact that adjacent intervals are not independent (at least in groups of three, as the shifting parameter is set to 33{\%}) because of the binning procedure. This highlights that the data are rarely sufficient to strongly discriminate between the models. The 6-temperature model is almost always discarded, even during the very same flare for which it was calibrated (Figure \ref{fig:model_selection}). This is probably due to its complexity relative to the data quality, and for this reason we exclude it from the analysis hereafter. The 2-temperature model is the preferred one during the peaks of the two major events. We attribute this to the fact that at these higher temperatures, two components with free temperature parameters perform better than many fixed-temperature components, since the non-flaring emission is negligible and the temperature of the flaring plasma is changing rapidly and is hotter than the fixed components.
        
        Hereafter, we exclude the 6-temperature model. We repeat the selection but considering only the other three models recomputing the weights with just the three of them. From the selection we obtain a reference time series of spectra. Since the time intervals are not uniform, we perform a linear interpolation on the data to obtain evenly sampled time series with 30 s spacing. Further comments on model selection and how this affect the spectral shape, specifically in the end of the (X-ray) wavelength range, are presented in Appendix \ref{appendix:further_comments_on_model_selection}.

    \subsection{EUV scaling}\label{section:EUV_scaling}
    
\begin{figure}[htbp]
            \centering
            \begin{subfigure}{0.48\textwidth}
                \includegraphics[width=0.99\linewidth]{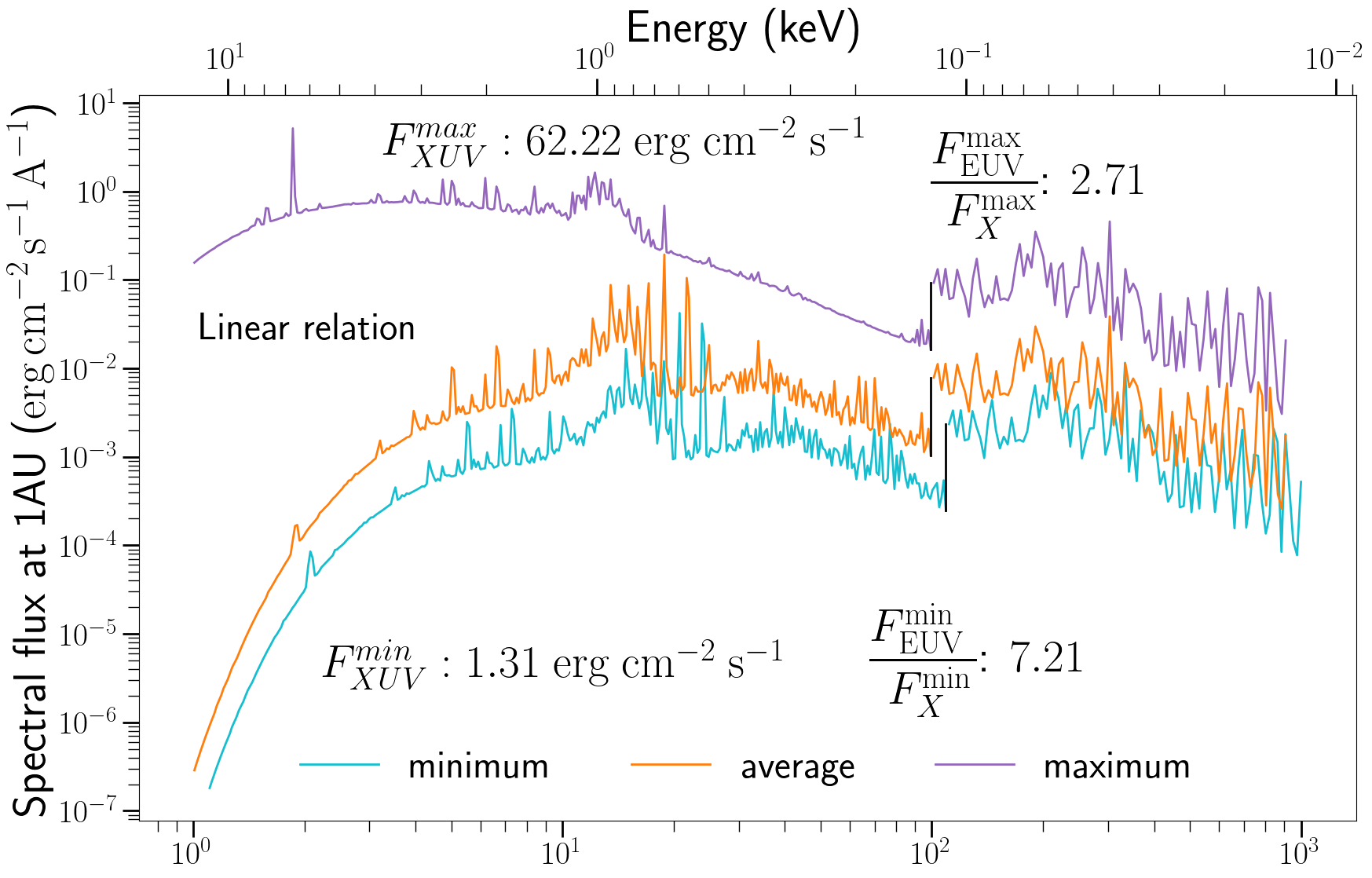}
            \end{subfigure}
            \hfill
            \begin{subfigure}{0.48\textwidth}
                \includegraphics[width=0.99\linewidth]{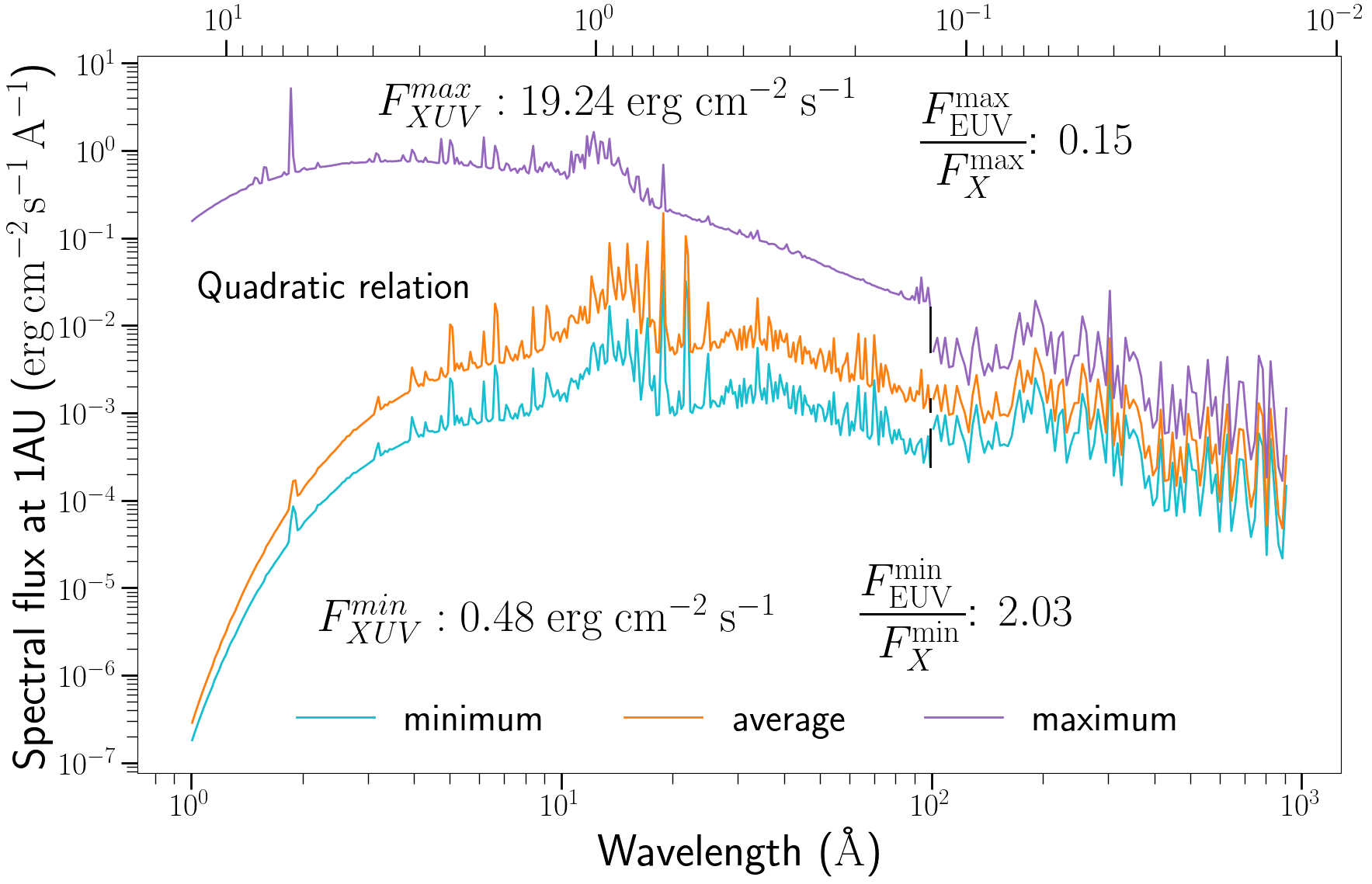}
            \end{subfigure}
            \caption{XUV spectra obtained using the linear (top panel) and quadratic (lower panel) relations from SF+25, scaled from the minimum, average and maximum X-ray fluxes. The XUV-integrated fluxes are reported on the plot together with the EUV/X-ray ratio for maximum and minimum spectra.}
            \label{fig:EUV_scaling}
        \end{figure}

        Thus far we have focused on the X-ray energy range covered by the XMM-Newton data. The EUV range is as important in determining the planetary atmospheric evolution. Unfortunately, measurements in the EUV range are difficult due to severe absorption by the ISM and almost always unavailable. To obtain the EUV flux there are two possibilities. Both use information from the adjacent ranges of X-rays and FUV. The best approach is to model the stellar outer layers (chromosphere, transition region, and corona) and their plasma, deriving their DEM structure with high-resolution spectroscopy and plasma simulations. This method is beyond the scope of this work and difficult to apply in a time-resolved way. A more pragmatic way involves the use of scaling laws, which is the one used in this work. In this case many assumptions are made and results may change considerably from law to law, so it is important to take these results with a grain of salt. In the literature, various scaling laws are available to derive the EUV luminosity from the X-ray luminosity \citep[e.g.,][]{linsky_intrinsic_2014,johnstone_active_2021, maggio_xuv_2024, sanz-forcada_connection_2025}. Depending on the adopted strategy and on the amount of data at our disposal, the reconstructed EUV flux may vary greatly. This adds significant uncertainty to the estimated XUV flux and, in turn, to the atmospheric model outputs. To give a quantitative idea of the uncertainties we can only compare the results from different scaling laws at different levels of emission. These differences dominate over the other sources of error. We comment on this in the text below, while discussing Figure \ref{fig:EUV_scaling}. Unfortunately, none of the available scaling laws takes into account flares, as they are calibrated with the average luminosity.  The use of these scaling laws in time-resolved analyses adds uncertainty, but at the moment it remains the most practical approach. This is the approach we take here. 
    
        To gain insight into the uncertainties introduced by the scaling laws, we experiment with two of them and a fixed EUV spectral shape. Our reference work is SF+25, from which we take both the scaling laws and the average EUV spectrum of {\PC}. The first scaling law (equations 2 and 3 of SF+25) is linear, adapted from \cite{sanz-forcada_estimation_2011}: the logarithm of the EUV luminosity is expressed as a linear function of the logarithm of the X-ray luminosity. The second scaling law (equations 4 and 5 of SF+25) expresses the logarithm of the EUV luminosity, normalized to the bolometric luminosity, as a second-order polynomial function of the logarithm of the similarly normalized X-ray luminosity. This new relation is introduced in SF+25, where it is argued that it is more accurate. These relations are presented in SF+25 for two spectral ranges. We use the ones calibrated over the [100, 920] {\AA} range (respectively equations 2 and 4 of SF+25).
        
        We proceed as follows. Firstly, we obtain an EUV-integrated flux time series by feeding the X-ray fluxes into the scaling laws. Then we use these values to scale the average fixed-shape EUV spectrum of {\PC}. For each time interval, the resulting EUV spectrum is stitched to the X-ray spectrum to obtain a time series of XUV spectra.
        
        This procedure is not ideal for different reasons. The first reason is that the relation is based on the average EUV flux of the population. By using it to infer the time-varying EUV flux of an individual star, we implicitly assume that the variability of a single star follows the relation derived from average luminosities of different stars. Secondly, we are using this relation to scale the fluxes measured during short time intervals. These intervals might be dominated by flares or dimmings. This type of emission is not represented by the average, on which the scaling law is calibrated. Moreover, we use a fixed EUV spectral shape for all the time intervals, while at X-ray and presumably also at EUV wavelengths, the spectral shape changes considerably with time, specifically during flares. For these reasons, this procedure often leads to a discontinuity in the spectrum where the X-ray and EUV parts are stitched together. We argue that the discontinuity found for the average flux can be used as a criterion to compare different reconstructions. We stress the need for EUV spectra that describe the star in different flaring states. There is also a need for X-ray vs. EUV relations during these conditions that can be used as simple prescriptions for future reconstructions.
        
        In Figure \ref{fig:EUV_scaling} we show the reconstructed XUV spectra obtained for conditions of average, maximum and minimum X-ray flux, using both the linear (top panel) and quadratic (lower panel) scaling laws. We also report the XUV-integrated flux at 1 AU and the ratio between the EUV-integrated and the X-ray-integrated fluxes for maximum and minimum X-ray conditions. The two scaling laws yield ratios that differ by significant factors.
    
        The quadratic relation is presumably the most accurate as it minimizes the discontinuities in the reconstructed spectrum at 100 {\AA}, according to our criterion. For these reasons, we will focus on the corresponding reconstructed spectra. For the brightest flares the X-ray fluxes exceed the reliability threshold defined in SF+25. We are aware of the limitations of this scaling law but, lacking a viable alternative, we accept them and we decide to push the relation outside its recommended limits.

        To quantify the uncertainty in the EUV flux (not reported in the final spectra), we compare values derived from two different relations in the three regimes: minimum, average, and maximum. For the average flux, the ratio of the total EUV fluxes produced by the two methods is 5.4: the linear relation yields 3.85 erg cm$^{-2}$ s$^{-1}$ (at 1 AU), while the quadratic relation yields 0.71 erg cm$^{-2}$ s$^{-1}$. The corresponding ratios are 3.6 for the minimum flux and 18 for the maximum. The discrepancies for the minimum and average fluxes lie within the expectations for the relations reported in SF+25. The large discrepancy for the maximum flux arises because the flare-specific relation was applied outside the recommended parameter range in SF+25.
        
        \subsection{Data products}
        
        The EUV and X-ray spectra are stitched together to obtain the time series of XUV spectra over the 1-920 {\AA} range. Each X-ray spectrum is associated with a set of plasma parameters retrieved from the regression. Lightcurves are obtained by integrating the spectral flux time series over different energy ranges. In the data products, we also include the lightcurves measured with the U filter, which cover longer wavelengths. Lightcurves and spectra are presented and discussed in the next section.   
        
    \section{Results and discussion}\label{section:discussion}
    
    Next, we discuss the variability in our data. We are particularly interested in the energy incident on {\PC} b, both instantaneously and as a time average. The EUV range is included in the analysis where relevant.
    
    \subsection{Short-term variability and flare diversity}\label{section:short-term_variability}
    
\begin{figure}[htbp]
                \centering
                \begin{subfigure}{0.48\textwidth}
                    \includegraphics[width=0.99\linewidth]{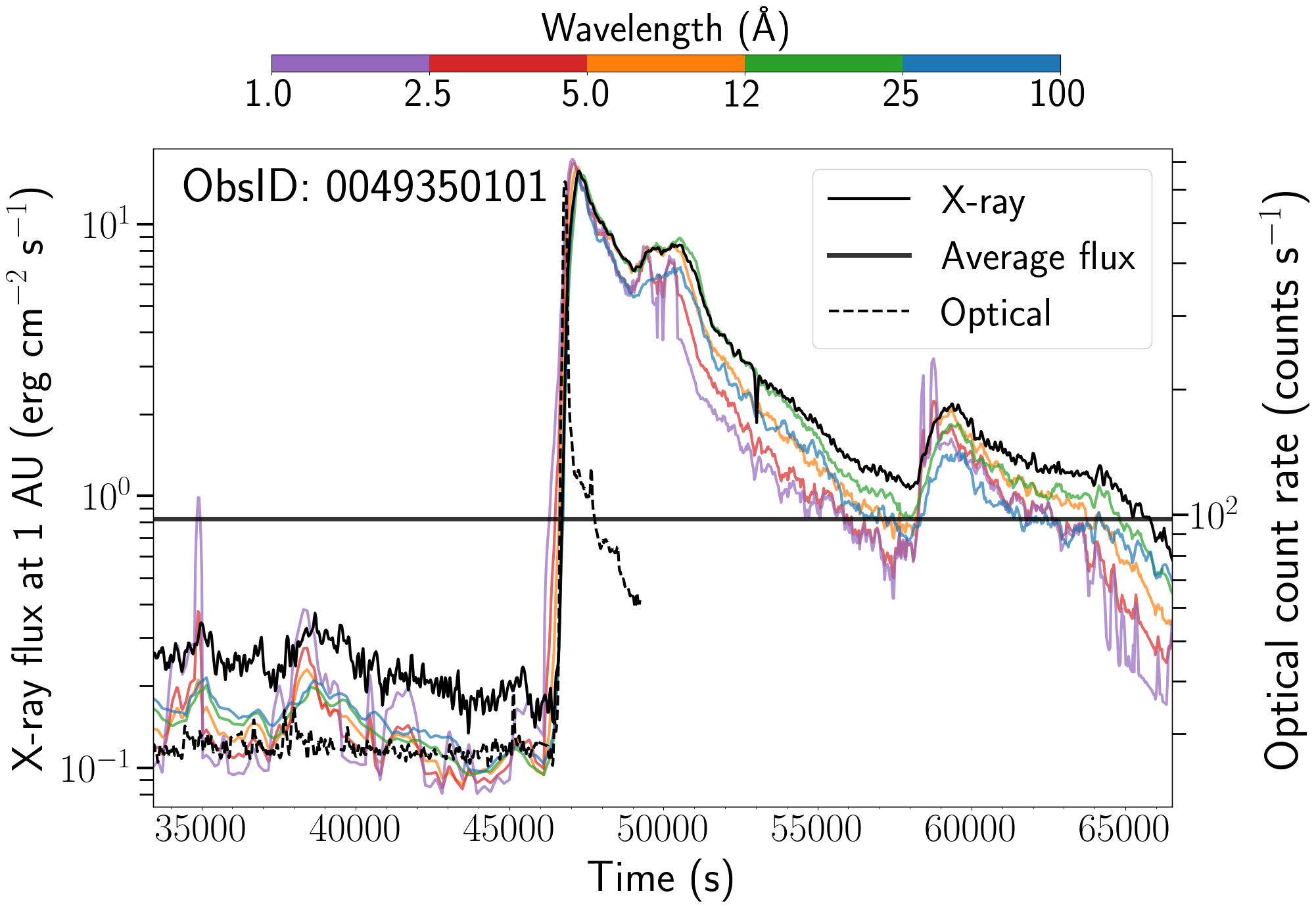}
                \end{subfigure}
                \hfill
                \begin{subfigure}{0.48\textwidth}
                    \includegraphics[width=0.99\linewidth]{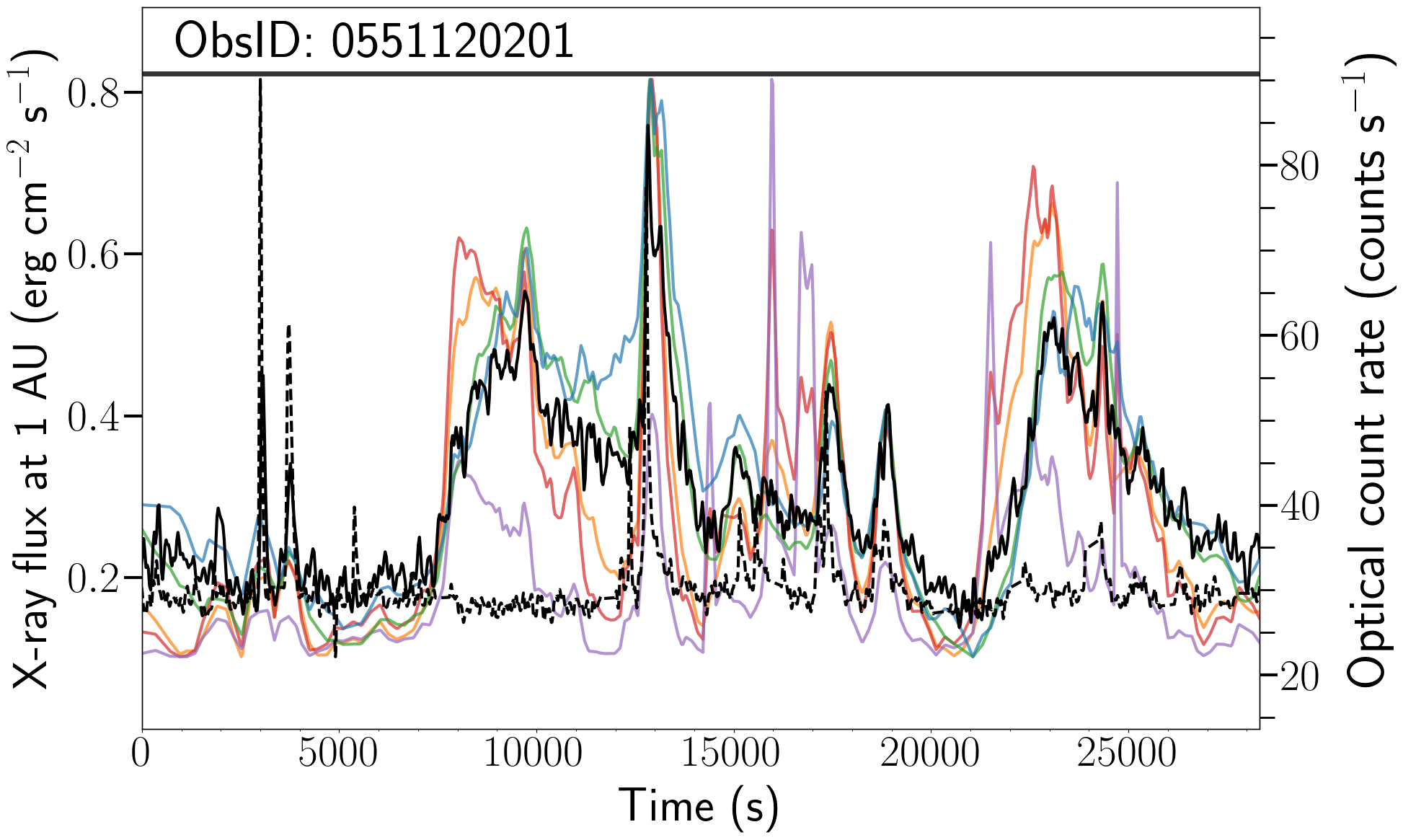}
                \end{subfigure}
                \hfill
                \begin{subfigure}{0.48\textwidth}
                    \includegraphics[width=0.99\linewidth]{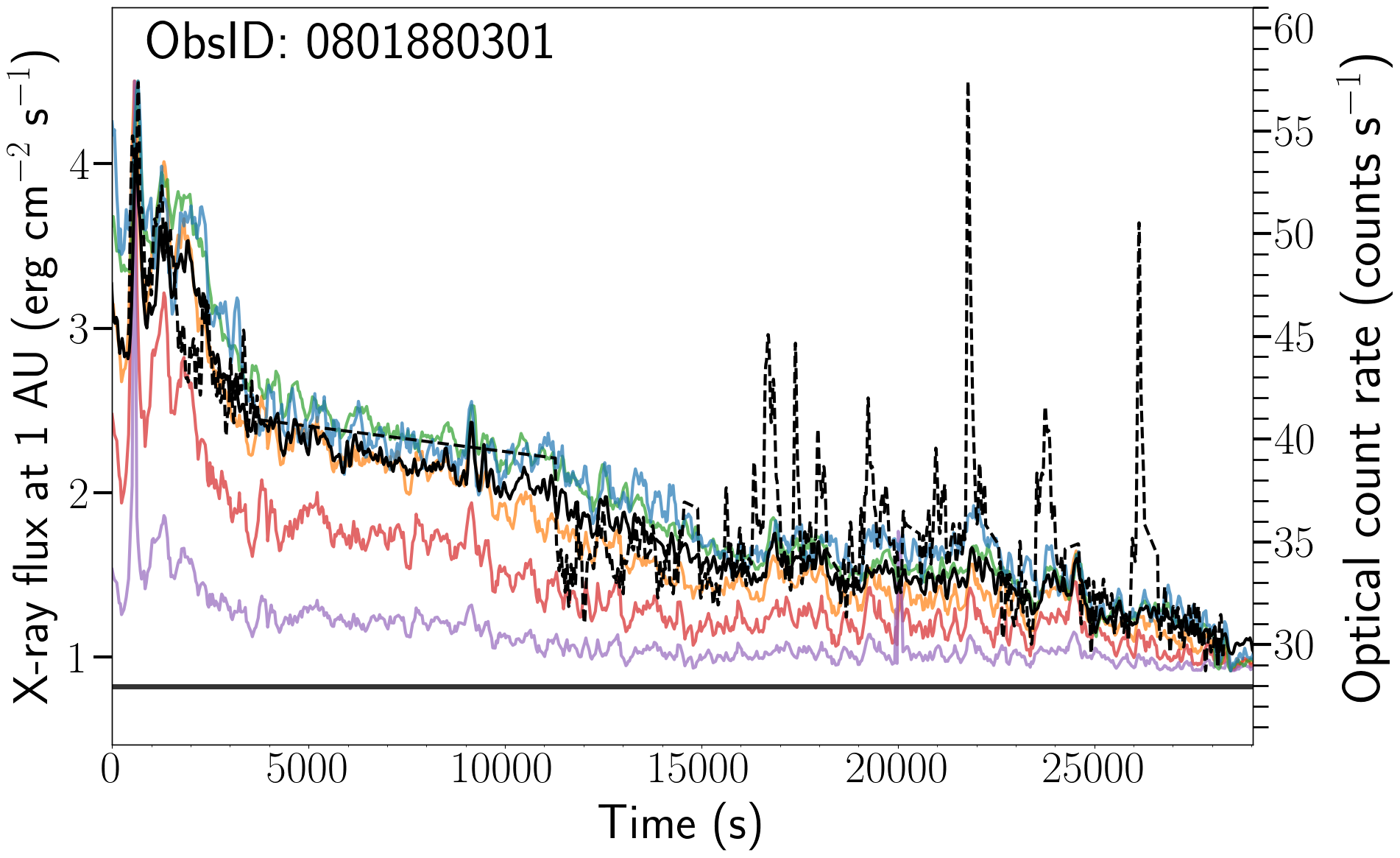}
                \end{subfigure}
                \caption{X-ray flux (1-100 {\AA}; count s$^{-1}$; black-solid line) is plotted for selected snapshot observations, together with the X-ray average over all observations. The optical count rate is plotted as a black dashed line. Colored lines (see the color bar at the top panel) indicate lightcurves in different energy bands, obtained by integrating the spectral fluxes. No axis labels or ticks are displayed for these, as they are plotted only for qualitative comparison. Limits are set to (0.8 x min, 1.2 x max) separately for each band.}
                \label{fig:LCs}
            \end{figure}

        In Figure \ref{fig:LCs} we select three snapshots in time of the lightcurves. The top panel (second half of observation 0049350101) shows the strongest flare recorded in our dataset, justifying the use of a logarithmic scale. The middle panel (observation 0551120201) depicts one of the three observations in which the stellar emission always stays below average. The lower panel (observation 0801880301) shows the only observation during which the flux is always above the average. 
            
        {\PC} is a frequently flaring star, and its flare strengths vary significantly over time. Many of them, usually the most energetic ones, are detected at both optical and X-ray wavelengths, with the optical peak preceding the X-ray peak, as described by the Neupert effect \citep{neupert_comparison_1968}. Heating, and subsequent cooling, are also evident for many intermediate events, with peaks appearing first in the hard and later in the soft bands. This behavior is visible in the flare at 12,500 s in the middle panel of Figure \ref{fig:LCs}. This variability shows itself also in the more quiescent periods, as noted by F+22, who report that the quiescent count rates vary by about a factor of 2 within a few days.
            
        Many peaks appear for some energies but not for others, which is plausibly explained by heating acting different in different layers of the star. For example, the main event in the top panel of Figure \ref{fig:LCs}, the two events at $\sim$2,500 s of the middle panel and the event at $\sim$12,500 s in the same observation, probably involve a wide region from the chromosphere to the corona as the peak is clearly seen at all wavelengths (including the optical). The loss of resolution in the peaks of the two events at $\sim$2,500 s due to time-binning is noticeable.
            
        Other events, instead, are visible only at certain wavelengths. For example, in the middle panel of Figure \ref{fig:LCs}, at $\sim$5,000 s, a small optical flare does not seem to have an evident X-ray counterpart. At the start of the observation in the top panel and at $\sim$15,000 s in the middle one, instead, there are peaks at high-energy X-ray wavelengths (purple and red) that are undetected at both softer X-ray (green, blue) and optical wavelengths. This might be interpreted either as the energy release due to magnetic reconnection happening only in the corona, or because the foot-points of loop generating the event are behind the limb, concealed from us. In a different type of event, as seen at $\sim$7,500 s in the middle panel, the high-energy X-ray flux rises with seemingly no optical counterpart. The flux increases by a factor of 3, which is comparable to medium-intensity flares. Its timescale is less than 5,000 s, and therefore the event is relatively short.
        
        Observation 0801880301 (lower panel of Figure \ref{fig:LCs}) constantly remains above the average flux. Observations like this one could considerably bias our estimate of the star’s average behavior, specifically for more distant targets, for which the S/N does not allow time-resolved or spectral analysis. We also highlight how, towards the end of the observation, high variability is measured in the optical range, but not in the X-ray one. Similar patterns were already reported by F+22 looking at X-ray and FUV data.
        
        The major event in the top panel increases the total flux relative to pre-flare values by two orders of magnitude. This rise is even more pronounced in the shortest wavelengths ranges. At the shortest available wavelength (purple curve), the intensity increases by four orders of magnitude.
    
        Variability over timescales between five minutes and ten hours is seen in all observations. For these observations, the emission intensities vary by up to a factor of 30.

    \subsection{Flux distribution}\label{section:flux_distribution}
    
\begin{figure}
            \centering
            \includegraphics[width=\linewidth]{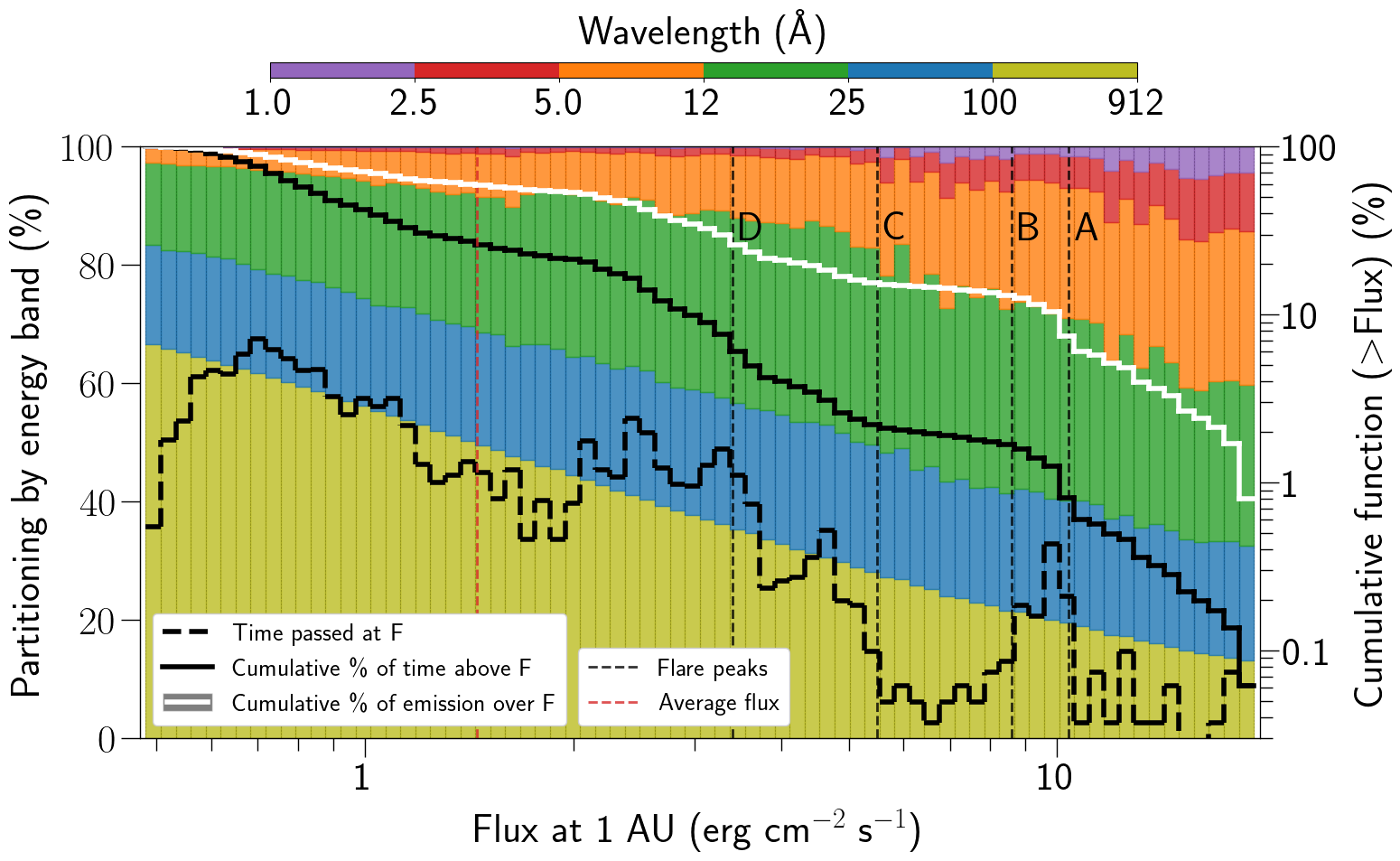}
            \caption{The x-axis shows the flux of the star at 1 AU (integrated from 1 to 920 {\AA}). The flux is logarithmically binned. The right-hand-side y-axis refers to the three bold black and white lines. The three lines share this axis because all quantities are expressed as percentages. The black-dashed line represents the percentage of time {\PC} is emitting at the corresponding x-axis flux bin. The black continuous line is the cumulative curve of the black dashed line. It is showing the percentage of time the star is emitting above the corresponding x-axis flux value. The white continuous line corresponds to the percentage of the total emitted energy of {\PC} above the corresponding flux value. The vertical black-dashed lines indicate important relative maxima in the observations, excluding the absolute maximum, which would have been plotted at the extreme right. From right to left, in order from the most to the least intense, they refer to: the secondary peak of the main flare of observation 0049350101 (A), the peak of the main flare of observation 0551120401 (B), the maximum of observation 0801880301 (C), and the peak at the start of observation 0049350101 (D). The red-dashed vertical line is the average flux over all the observations. The left-hand-side y-axis describes the partitioning by wavelength range of each flux bin. They are color-coded by the top color bar. This partitioning is obtained by averaging all spectra in the corresponding flux bin.}
            \label{fig:hist}
        \end{figure}

        Next, we study how {\PC}'s emission is distributed over different flux levels. Figure \ref{fig:hist} shows three curves, all based on the total flux of {\PC} (at 1 AU, integrated over 1–920 {\AA}) on the x-axis. The total flux is binned logarithmically. The curves represent: (black-dashed) the percentage of time the star emits at the flux level of the corresponding x-axis bin; (black-solid) the percentage of time the emission is above that flux; (white-solid) the percentage of the total emitted energy above that flux. All the percentages are relative to the whole set of observations. The three curves share the right-hand-side y-axis.
        
        For each flux bin, we plot (left-hand-side y-axis) stacked bars of different colors to show how the emitted energy is partitioned across six selected wavelength intervals. This partitioning is obtained by averaging all spectra falling in the corresponding flux bin. This conveys the spectral hardness of the emission. The different colors always sum up to 100 and the wavelength intervals are described by the top colorbar. 
            
        Vertical dashed black lines indicate important relative maxima throughout the observations. The absolute maximum (over all the observations) reached during the main flare of observation 0049350101 is not included. From right to left, they refer to the secondary peak of the main flare of observation 0049350101 (A), the main flare of observation 0551120401 (B), the maximum of observation 0801880301 (C), and the peak at the start of observation 0049350101 (D). These events appear as clear changes in the shape of the cumulative distributions. This also highlights that single major events can significantly bias the time-averaged emission.
            
        For different flux levels (x-axis values), the contribution of the EUV wavelengths ranges from nearly 70{\%} to less than 20{\%}. Correspondingly, the contribution of the 1-5 {\AA} wavelengths is typically negligible. Exceptionally, when the stellar emission is strong, the very energetic wavelengths can contribute up to 10{\%} to the total XUV flux. The fact that even rare events contribute significantly to the total emission makes a strong case for including flares in atmospheric models, ideally with the correct spectral shape.

\begin{figure}
            \centering
            \includegraphics[width=0.9\linewidth]{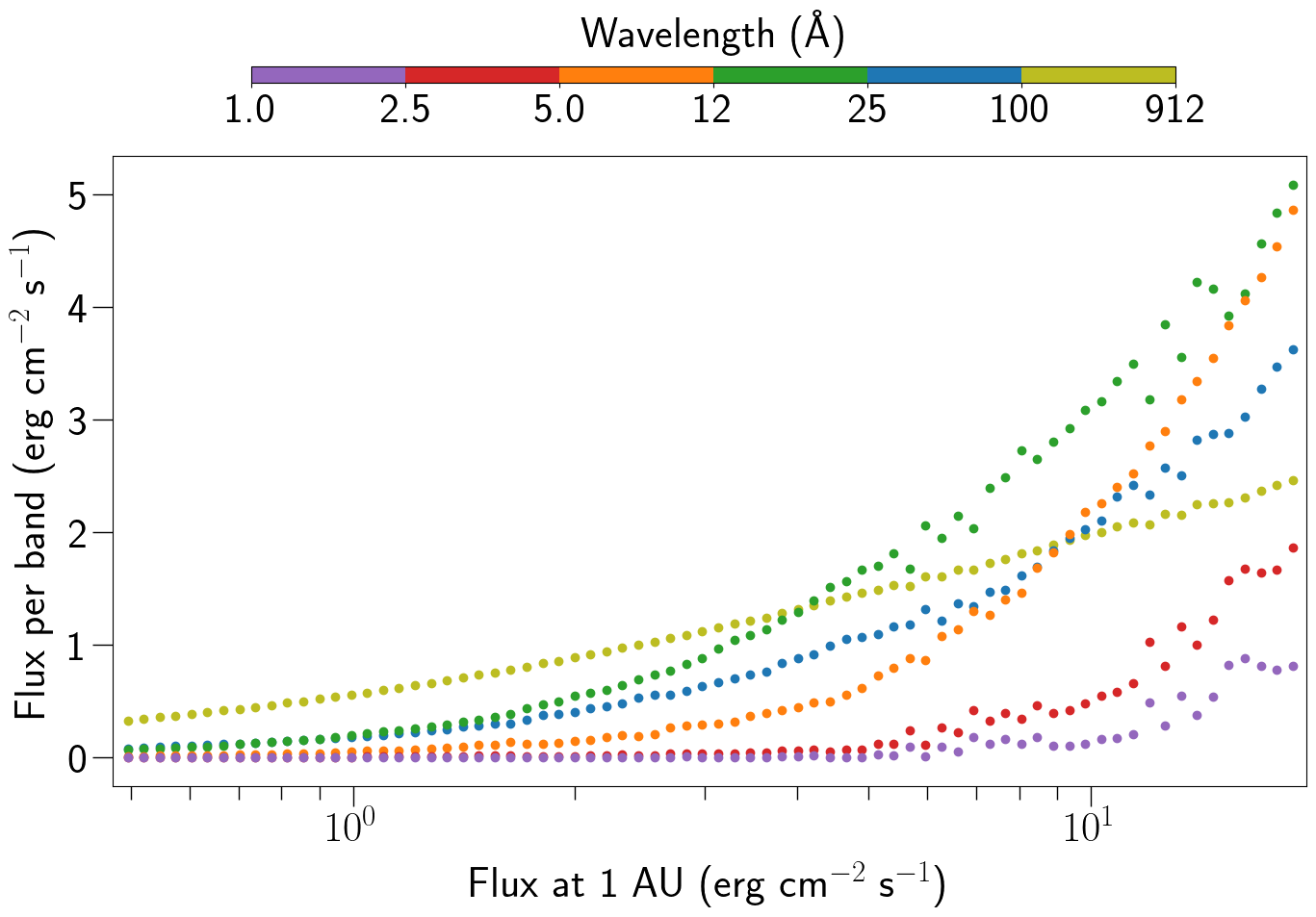}
            \caption{The x-axis and the color bar for the different wavelength intervals are the same as in Fig. \ref{fig:hist}. The y-axis shows, for each flux bin, the average flux over all intervals falling within that bin for each wavelength interval. As a reminder, the EUV fluxes are less and less accurate as the flux rises (during flares specifically).}
            \label{fig:histnew}
        \end{figure}

        Figure \ref{fig:histnew} shows the average flux for each wavelength interval within each flux bin. This figure is analogous to Figure \ref{fig:hist}, but uses absolute values in terms of flux at 1 AU. Wavelengths shorter than 5 {\AA} contribute to the total flux only during the highest emission levels.
        
        Again, we note that the EUV flux strongly depends on the adopted scaling relation, particularly for values far from the average. Therefore, these results should be regarded as one possible realization among several, but still, highlight how different could be the X-ray to EUV ratio as a function of time.
    
    \subsection{Reduction uncertainties and literature comparison}\label{section:reduction_uncertainties_and_literature_comparison}
    
        We assess the effects that different reduction procedures and observational baselines have on the reconstructed spectra, and the associated uncertainties. We also compare our fluxes with those from other analyses reported in the literature. These spectra are always averaged over multiple intervals, which leads to significant smoothing of the peaks and to large differences in their relative maxima.
            
        Firstly, we summarize the literature on the XUV spectrum of {\PC}. We downloaded the published spectra from the corresponding references and, for an easier comparison, we scaled them to 1 AU and binned them onto a common energy grid:
        
        \cite{ribas_full_2017}. One average spectrum based on ROSAT and XMM-Newton data. For the latter dataset, a correction of 0.83 is applied to the measured fluxes to compensate for the presumed high activity of the star at the time of observation \citep{wargelin_optical_2017}. Spectral range: [7, 100] {\AA}.
        
        \cite{loyd_muscles_2018}. Two versions of an average spectrum based on XMM-Newton data, each with a slightly different spectral range and resolution. The first one (APEC) has the raw resolution of the XSPEC model. The second one (Multi) is smoothed and extended over a broader spectral range. A 2-temperature model is used for both. Spectral range: [5, 100] {\AA} for APEC, [2.5, 100] {\AA} for Multi.
        
        SF+25. An average spectrum,  reconstructed from emission lines in high-resolution XMM-Newton data. Spectral range: {[}1, 100{]} {\AA}.
        
        \cite{binder_x-ray_2024}. Five reference spectra, representative of conditions described by the labels: quiescent, elevated, descending, rising, flare. The fitting model includes either 2 or 3 temperature components. Spectral range: {[}1, 100{]} {\AA}.
    
        For comparison, we consider 5 spectra based on our work in the {[}1, 100{]} {\AA} spectral range. We label them as:
        
        \begin{itemize}
            \item Minimum: spectrum with the minimum integrated flux in the time series.
            \item Average: average spectrum over all times.
            \item Median: corresponding to the 50th percentile in integrated flux.
            \item 80th percentile: corresponding to the 80th percentile in integrated flux.
            \item Maximum: spectrum with the maximum integrated flux in the time series.
        \end{itemize}
    
\begin{figure}[htbp]
            \centering
            \includegraphics[width=0.8\linewidth]{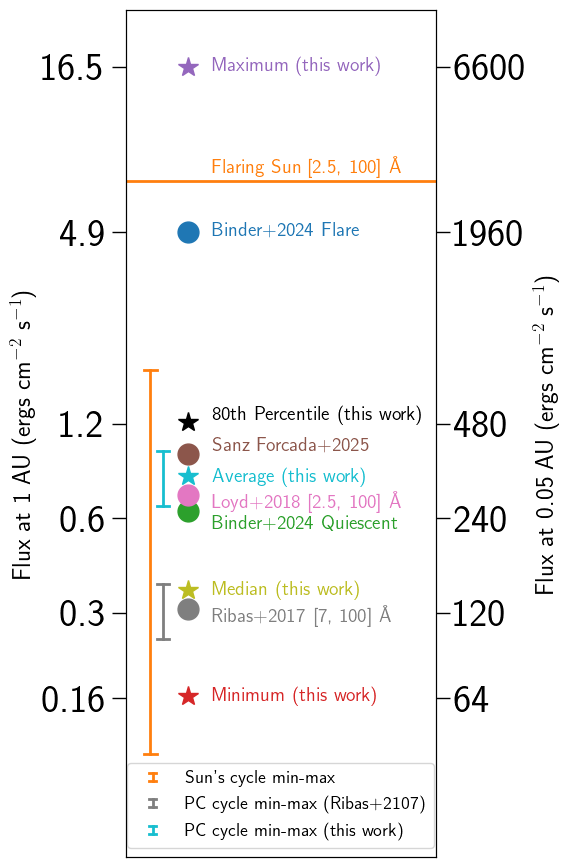}
            \caption{Integrated fluxes are shown at 1 AU (right-side axis) and at 0.05 AU (left-side axis, corresponding to the orbital distance of {\PC} b). When not included in the figure, the integration range is [1, 100] {\AA}. Star-shaped markers indicate results from this work, while round markers indicate results from other works. For the Sun (integrated over [4.1, 100] {\AA}), the variability range over its activity cycle in quiescence and a flaring reference are taken from \cite{peres_sun_2000}. Assuming a cycle maximum-to-minimum ratio of 1.5 for {\PC} \citep{wargelin_x-ray_2024}, we show the star’s quiescent variability due to its activity cycle. To provide a reference for the amplitude of the activity cycle on the plot, we use two example spectra, the \cite{ribas_full_2017} spectrum (gray) and our average spectrum (azure), as mid-cycle quiescent flux references.}
            \label{fig:TotalFluxMarkers}
        \end{figure}

\begin{figure*}[htbp]
            \centering
            \includegraphics[width=0.9\linewidth]{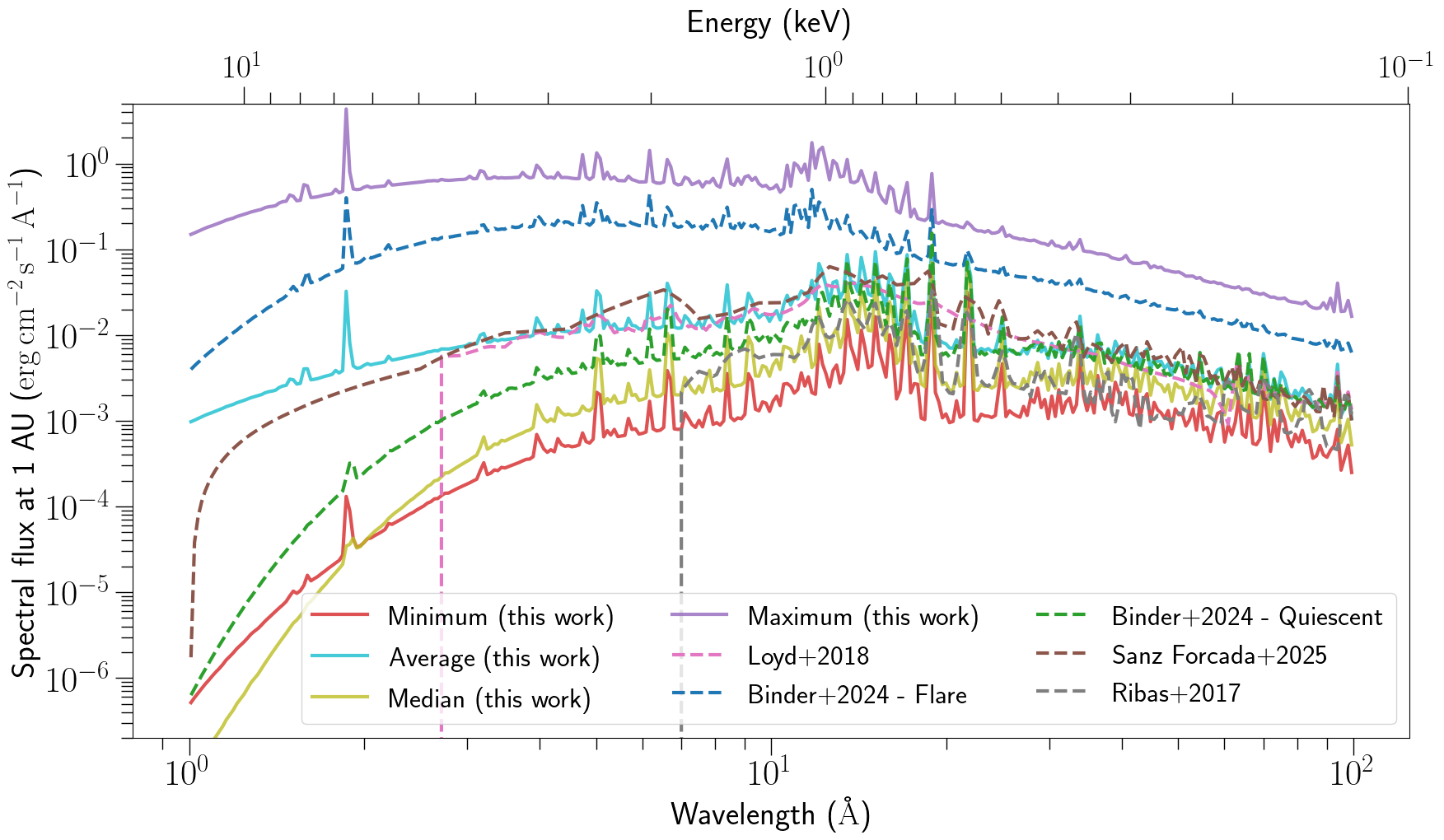}
            \caption{Spectra fluxes at 1 AU. Results from this work (minimum, average and maximum total fluxes in the [1, 100] {\AA} range) are compared with those from \cite{binder_x-ray_2024} (quiescent and flaring, [1, 100] {\AA} range), \cite{ribas_full_2017} ([7, 100] {\AA} range), SF+25, and \cite{loyd_muscles_2018} ([1, 100] {\AA} range).}
            \label{fig:FluxComparison}
        \end{figure*}

        In Figure \ref{fig:TotalFluxMarkers}, we plot the integrated fluxes over the available wavelength ranges. In Figure \ref{fig:FluxComparison}, we plot a selection of our spectra and others reported in the literature. Table \ref{table:ratiostable} contains the ratios between the references selected above and the average integrated flux obtained with this work. In the first row, unlike in the other rows, we show the flux values (integrated over the corresponding wavelength range) for the time-averaged spectrum.

        These values are computed both in the full spectral range and in different wavelength ranges. The third row of the table, unlike the others, reports the time-averaged flux at 1 AU from {\PC} calculated in this work. Here are two examples illustrating how to read the table. In the [5, 10] {\AA} band, the flaring flux reported in \cite{binder_x-ray_2024} is $12^{+1}_{-1}$ times stronger than the average flux of this work. In the [10, 20] {\AA} band, the flux from \cite{ribas_full_2017} is $2.7^{+0.3}_{-0.2}$ times smaller than our average. Errors include only uncertainties from this work. Dividing two values in the same column allows comparison between different references; for example, in the [1, 100] {\AA} range, the maximum flux is $5 \times 20 = 100$ times smaller than the minimum flux.
        
        The [1, 100] {\AA} column of Table \ref{table:ratiostable} can be used to quantify the relations between the values in Figure \ref{fig:TotalFluxMarkers}. In Figure \ref{fig:TotalFluxMarkers}, we also show, for reference, the Sun's activity cycle variability range and a flaring Sun value (in orange) from \cite{peres_sun_2000}, covering [4.1, 100] {\AA}.
    
        \newlength{\mywidth}
        \setlength{\mywidth}{\widthof{Loyd et al. MUSCLES APECaa}}
    
        \renewcommand{\arraystretch}{1.6}
        \begin{table*}
        \caption{Ratios between the reference spectra described in the text and the time-averaged flux of this work.}
        \begin{tabular}{c|c|cccccc}
        
        \hline
        \diagbox[width=\mywidth,height=1cm]{Spectrum}{E. range}                       & [1, 100] {\AA}       & [1, 2.5] {\AA} & [2.5, 5] {\AA} & [5, 10] {\AA} & [10, 20] {\AA} & [20, 50] {\AA} & [50,100] {\AA}\\
        
        \hline
        \makecell{\\[-7pt](this work) Average\\[3pt](erg cm$^{-2}$ s$^{-1}$) at 1 AU\\[-8pt]\hspace{2pt}}
                                                        & $0.81^{+0.10}_{-0.08}$    
        
                                                        & $0.006^{+0.001}_{-0.001}$ 
                                                        & $0.024^{+0.004}_{-0.003}$ 
                                                        & $0.083^{+0.009}_{-0.007}$ 
                                                        & $0.34^{+0.03}_{-0.03}$ 
                                                        & $0.23^{+0.02}_{-0.02}$ 
                                                        & $0.13^{+0.01}_{-0.01}$ \\
                                                        
        \hline
        DA+26 Minimum                             & $1/{5.0^{+0.7}_{-0.5}}$
        
                                                        & $1/{126^{+115}_{-86}}$ 
                                                        & $1/{27^{+13}_{-8}}$ 
                                                        & $1/{13^{+4}_{-2}}$ 
                                                        & $1/{6.3^{+1.0}_{-0.6}}$ 
                                                        & $1/{3.6^{+0.4}_{-0.4}}$ 
                                                        & $1/{3.4^{+0.5}_{-0.3}}$ \\

        DA+26 Median                              & $1/{2.3^{+0.4}_{-0.3}}$   
        
                                                        & $1/{121^{+908}_{-30}}$ 
                                                        & $1/{12^{15}_{-2}}$ 
                                                        & $1/{5.1^{+1.7}_{-0.6}}$ 
                                                        & $1/{2.6^{+0.4}_{-0.4}}$ 
                                                        & $1/{1.8^{+0.3}_{-0.2}}$ 
                                                        & $1/{1.7^{+0.3}_{-0.2}}$ \\[2pt]
                                                        
        DA+26 80th percentile                     & ${1.5^{+0.2}_{-0.2}}$    
        
                                                        & $1/{18^{+36}_{-8}}$ 
                                                        & $1/{2.4^{+1}_{-0.4}}$ 
                                                        & $1.0^{+0.2}_{-0.1}$ 
                                                        & ${1.5^{+0.2}_{-0.2}}$ 
                                                        & ${1.6^{+0.2}_{-0.2}}$ 
                                                        & ${1.6^{+0.2}_{-0.2}}$ \\
                                                        
        DA+26 Maximum                             & ${20.1^{+2.1}_{-1.8}}$   
        
                                                        & ${123^{+26}_{-32}}$ 
                                                        & ${71^{+12}_{-11}}$ 
                                                        & ${39^{+4}_{-4}}$ 
                                                        & ${17^{+2}_{-2}}$ 
                                                        & ${14^{+1}_{-1}}$ 
                                                        & ${11^{+1}_{-1}}$ \\

        \hline
                                                        
        BI+24 Quiescent        & $1/{1.3^{+0.1}_{-0.1}}$ 
        
                                                        & $1/{20^{+4}_{-5}}$ 
                                                        & $1/{3.7^{+0.6}_{-0.5}}$ 
                                                        & $1/{2.2^{+0.2}_{-0.2}}$ 
                                                        & $1/{1.3^{+0.1}_{-0.1}}$ 
                                                        & $1/{1.1^{+0.1}_{-0.1}}$ 
                                                        & $1.0^{+0.1}_{-0.1}$ \\
                                                        
        BI+24 Elevated               & ${1.1^{+0.1}_{-0.1}}$ 
        
                                                        & $1/{16^{+3.3}_{-3.7}}$ 
                                                        & $1/{2.6^{+0.4}_{-0.3}}$ 
                                                        & $1/{1.4^{+0.2}_{-0.1}}$ 
                                                        & ${1.1^{+0.1}_{-0.1}}$ 
                                                        & ${1.2^{+0.1}_{-0.1}}$ 
                                                        & ${1.3^{+0.1}_{-0.1}}$ \\
                                                        
        BI+24 Rising                 & ${1.3^{+0.1}_{-0.1}}$ 
        
                                                        & $1/{4.9^{+1.0}_{-1.1}}$ 
                                                        & $1/{1.2^{+0.2}_{-0.2}}$ 
                                                        & ${1.1^{+0.1}_{-0.1}}$ 
                                                        & ${1.3^{+0.1}_{-0.1}}$ 
                                                        & ${1.4^{+0.1}_{-0.1}}$ 
                                                        & ${1.4^{+0.2}_{-0.1}}$ \\
                                                        
        BI+24 Descending             & ${1.5^{+0.2}_{-0.1}}$ 
        
                                                        & $1/{15^{+3}_{-3}}$ 
                                                        & $1/{1.8^{+0.3}_{-0.25}}$ 
                                                        & ${1.2^{+0.1}_{-0.1}}$ 
                                                        & ${1.8^{+0.2}_{-0.2}}$ 
                                                        & ${1.4^{+0.1}_{-0.1}}$ 
                                                        & ${1.5^{+0.2}_{-0.1}}$ \\
                                                        
        BI+24 Flare                  & ${6.0^{+0.6}_{-0.5}}$ 
        
                                                        & ${15^{+3}_{-4}}$ 
                                                        & ${18^{+3}_{-3}}$ 
                                                        & ${12^{+1}_{-1}}$ 
                                                        & ${4.8^{+0.4}_{-0.4}}$ 
                                                        & ${4.9^{+0.4}_{-0.4}}$ 
                                                        & ${4.2^{+0.4}_{-0.3}}$ \\
                                                        
        SF+25 Average     & ${1.2^{+0.1}_{-0.1}}$ 
        
                                                        & $1/{2.1^{+0.4}_{-0.5}}$ 
                                                        & ${1.0^{+0.2}_{-0.1}}$ 
                                                        & ${1.3^{+0.1}_{-0.1}}$ 
                                                        & ${1.2^{+0.1}_{-0.1}}$ 
                                                        & ${1.3^{+0.1}_{-0.1}}$ 
                                                        & $1.0^{+0.1}_{-0.1}$ \\
                                                        
        RI+17 Average                  & $1/{2.6^{+0.3}_{-0.2}}$ 
        
                                                        & --
                                                        & --
                                                        & $1/{5.6^{+0.6}_{-0.5}}\;(\star)$
                                                        & $1/{2.7^{+0.3}_{-0.2}}$ 
                                                        & $1/{2.4^{+0.3}_{-0.2}}$ 
                                                        & $1/{1.8^{+0.2}_{-0.1}}$ \\
                                                        
        LO+18 MUSCLES APEC     & $1/{1.2^{+0.1}_{-0.1}}$
        
                                                        & --
                                                        & --
                                                        & $1/{1.1^{+0.2}_{-0.1}}$ 
                                                        & $1/{1.1^{+0.1}_{-0.1}}$ 
                                                        & $1/{1.1^{+0.1}_{-0.1}}$ 
                                                        & $1/{1.1^{+0.1}_{-0.1}}$ \\
                                                        
        LO+18 MUSCLES Multi          & $1/{1.3^{+0.1}_{-0.1}}$    
        
                                                        & --
                                                        & $1/{1.1^{+0.2}_{-0.2}}$ 
                                                        & $1/{1.1^{+0.2}_{-0.1}}$ 
                                                        & $1/{1.2^{+0.1}_{-0.1}}$ 
                                                        & $1/{1.1^{+0.1}_{-0.1}}$ 
                                                        & $1/{5.6^{+0.6}_{-0.4}}$ \\
        \hline
        \end{tabular}
        \tablebib{
        DA+26: this work;
        BI+24: \citet{binder_x-ray_2024};
        SF+25: \citet{sanz-forcada_connection_2025};
        RI+17: \citet{ribas_full_2017};
        LO+18: \citet{loyd_muscles_2018};
        }
        \tablefoot{The first row shows the total flux values in the different wavelength ranges for the average. Absolute 1-$\sigma$ errors are propagated considering only the uncertainties of the references from this work. For example, the value $1/{1.1^{+0.1}{-0.2}}$ indicates that in that band the referenced spectrum is ${1.1^{+0.1}{-0.2}}$ times fainter than our time-averaged spectrum (first row, same column), whereas the value ${18^{+3}{-3}}$ indicates that the referenced spectrum is ${18^{+3}{-3}}$ times brighter than our time-averaged spectrum (first row, same column).\\ {\small $(\star)$ The reference spectrum for this value is integrated over [7, 100] {\AA} instead of [5, 100] {\AA}.}}
        \label{table:ratiostable}
        \end{table*}
        \renewcommand{\arraystretch}{1}

        The average fluxes presented in Figure \ref{fig:FluxComparison} (gray, pink, brown, and azure colors for different sources) are generally in good agreement. Integrated over [1, 100] {\AA}\footnote{If the integration range differs, it is reported in the corresponding table or figure.} (see Table \ref{table:ratiostable} and Figure \ref{fig:TotalFluxMarkers}), the time-averaged values range from 2.6 times lower to 1.2 times higher than the time-averaged value obtained in this work. This implies that the lowest and highest time-averaged values differ by a factor of $\sim 3$. These differences arise from a combination of factors, namely the observational data and instruments used, the calibration and reduction procedures, the plasma modeling and assumed ISM absorption, and the marginally different integration ranges adopted in \cite{ribas_full_2017} and \cite{loyd_muscles_2018}.

        With a time resolution of 150 s during major events, we are able to generate spectra that capture the emission maxima and minima with little or no smoothing from time binning. This, for example, explains differences between our work and that of \cite{binder_x-ray_2024}. For higher fluxes, the pile-up effect also contributes to increasing the difference in the results.

        Next, we assess to what extent time resolution affects the average spectrum. In principle, the average could differ depending on whether it is computed as the mean of the time-resolved spectra or from a single spectrum accumulated over all photon events. We test this by the following procedure. For each observation, we fit a 3-temperature model to the spectrum resulting from all the data from that observation. This spectrum is then compared to the average of the time-resolved spectra for the same observation. The difference in the integrated flux always stays below 10{\%} for all observations.
        
        \cite{wargelin_x-ray_2024} report an 8-year X-ray activity cycle for {\PC}, with a maximum-to-minimum amplitude of $\sim$1.5. In this respect, we note the following: 1) over the timescale of XMM-Newton observations ($\sim$8 days spanning 17 years), as well as additional observations from other telescopes, the flux differences reach a factor of $\sim3$, resulting solely from different datasets combination and data analysis procedures. We note that this factor is double the amplitude of the cycle; 2) On timescales of hours or less, the variability reaches two orders of magnitude, while in the [2.5, 5] {\AA} and [1, 2.5] {\AA} ranges, the variability reaches three and four orders of magnitude, respectively.
        
        \citeauthor{mascareno_diving_2025} report a $\sim$18-year activity cycle, which differs significantly from the 8-year cycle reported by \cite{wargelin_x-ray_2024}. If the longer period is correct, it turns out that {\PC} also flares significantly during the cycle minimum, as the XMM-Newton observations would not always coincide with the cycle maximum anymore.

        The spectra that we have reconstructed and presented in Figure \ref{fig:FluxComparison} lead us to question what we should define as quiescent conditions. In the higher energies the median (yellow-olive) spectrum is noticeably lower than the average (azure) spectrum, more sensitive to extreme values, indicating that flares have a substantial effect on the total emitted energy in this range. More importantly, the median obtained in this work is lower than the quiescence spectrum of \cite{binder_x-ray_2024} that was using exactly the same dataset. We wonder about the significance of defining a quiescent flux when the median value is smaller.
         
       It is interesting to assess the effects that may arise when observations are sparse, short, and widely separated in time. For most stars, there are at best a few observations, usually lasting no more than one day each, which may also be separated by years. If the star is variable over these timescales, the data may not represent the star’s average behavior. Indeed, over timescales of years, some G stars exhibit high-amplitude activity cycles that change their X-ray fluxes by more than an order of magnitude \citep{aschwanden_irradiance_1994, favata_x-ray_2008, robrade_coronal_2012}, and the same may hold for M stars. In addition, flares can change the X-ray flux by large factors over hour- or day-long timescales. We tested how this could affect {\PC} by performing a simple combinatorial exercise, comparing the average flux over all observations with that obtained from a subset.
        
        We have 8 observations, for a total exposure time of $\sim$250 ks. We split the first observation into two segments, obtaining a total of 9 observations with approximately equal durations. Using the average over all 9 observations as a reference, we consider cases in which only a subset (1–8) of them is available. For each case, we consider all unique combinations without repetition and compute the corresponding averages. Results for integrated XUV fluxes ([1,100] {\AA}) are shown in Table \ref{table:total_fluxes_combinatory}.
    
       Each row of Table \ref{table:total_fluxes_combinatory} contains a histogram showing how many combinations of observations produce a given ratio between the average flux over the full dataset and that obtained from the subset. The vertical line in each plot represents unity. Bins to the left of this line indicate that the subset average is less than the full-sample average, while bins to the right indicate an overestimate. The ratio of left-to-right combinations is shown as U/O (Underestimated/Overestimated), indicating whether under- or overestimation of the flux is more likely. Values near 1 (i.e., bins adjacent to the vertical line) are excluded from this analysis. The label “inf” indicates that there are no cases of flux overestimation. Obviously, increasing the number of observations reduces the deviation from the full-sample average. We are therefore primarily interested in the first rows, which correspond to cases with fewer observations. Regardless of the number of observations, the probability of underestimating the average always exceeds that of overestimating it. Using only 1 to 3 observations, the flux can be underestimated by up to a factor of 3 and overestimated by a factor of 2–3. This effect likely contributed to the factor-of-3 discrepancy observed between averages reported by different authors. Variability is amplified at higher energies (wavelengths shorter than 5 {\AA}). Even when using half of the observations, the flux can be underestimated by up to an order of magnitude in the [1, 2.5] {\AA} band and by up to a factor of 5 in the [2.5, 5] {\AA} band.
    
        \begin{table}
        \centering
        \resizebox{\columnwidth}{!}{
        \begin{tabular}{c|cc|c}
        {\small \# obs} &
        \multicolumn{2}{c|}{$\dfrac{\text{All-observations average flux}}{\text{Sub-set average flux}}$} &
        {\small \# obs} \\
        \hline
        1 &
        \raisebox{-0.5\height}{\includegraphics[width=3.4cm]{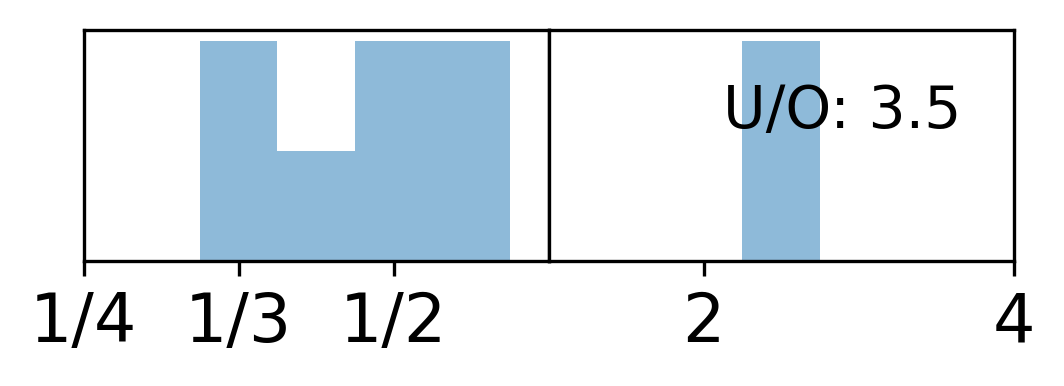}} &
        \raisebox{-0.5\height}{\includegraphics[width=3.4cm]{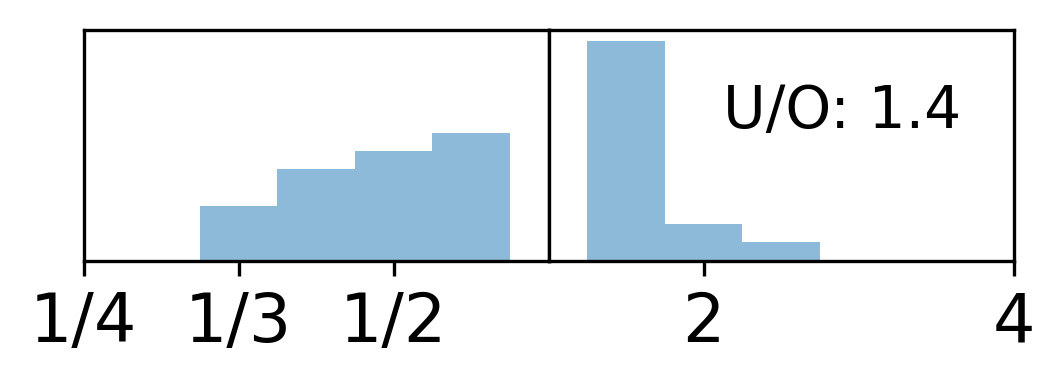}} &
        2 \\
        3 &
        \raisebox{-0.5\height}{\includegraphics[width=3.4cm]{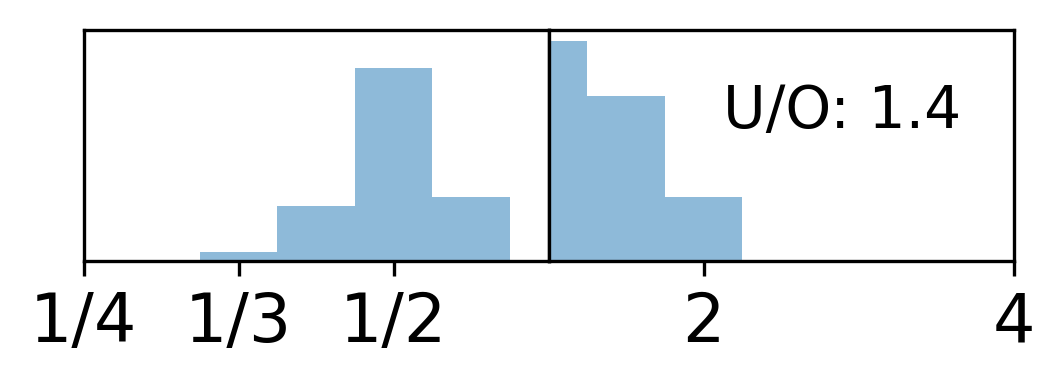}} &
        \raisebox{-0.5\height}{\includegraphics[width=3.4cm]{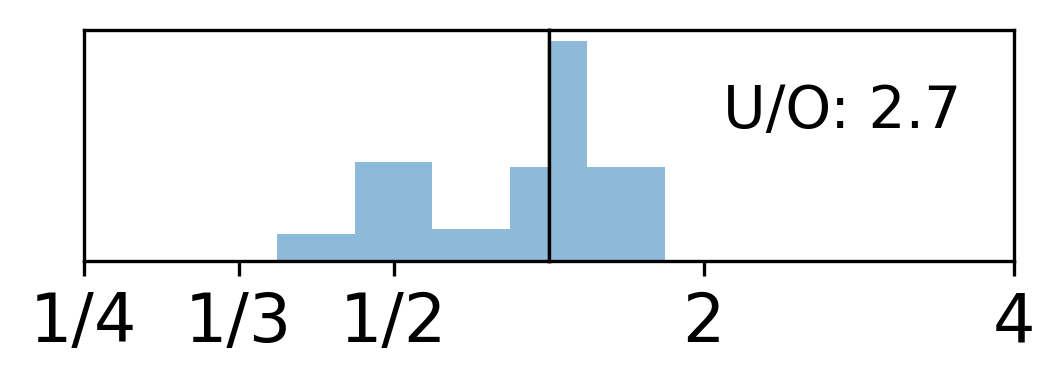}} &
        4 \\
        5 &
        \raisebox{-0.5\height}{\includegraphics[width=3.4cm]{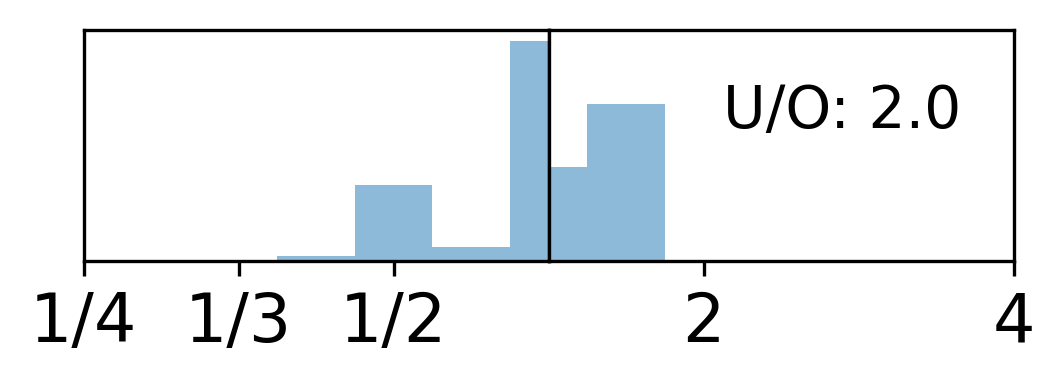}} &
        \raisebox{-0.5\height}{\includegraphics[width=3.4cm]{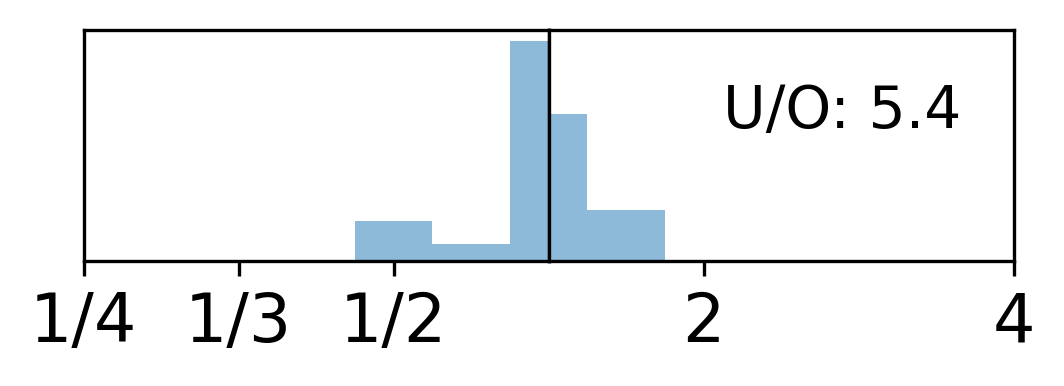}} &
        6 \\
        7 &
        \raisebox{-0.5\height}{\includegraphics[width=3.4cm]{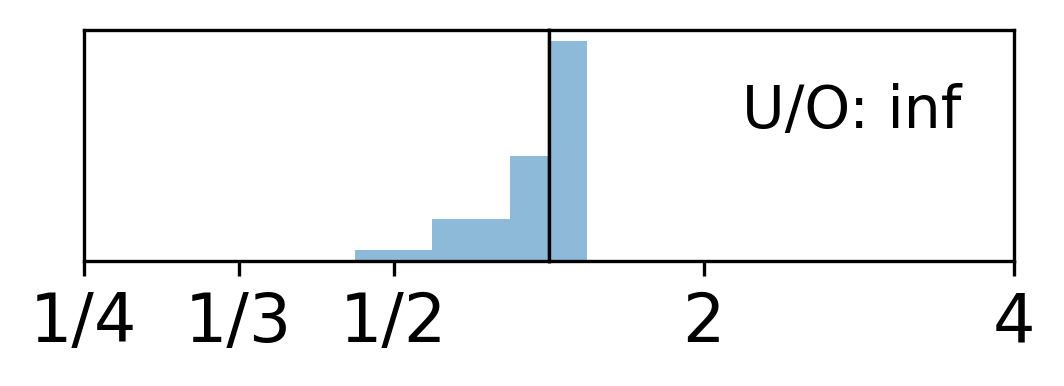}} &
        \raisebox{-0.5\height}{\includegraphics[width=3.4cm]{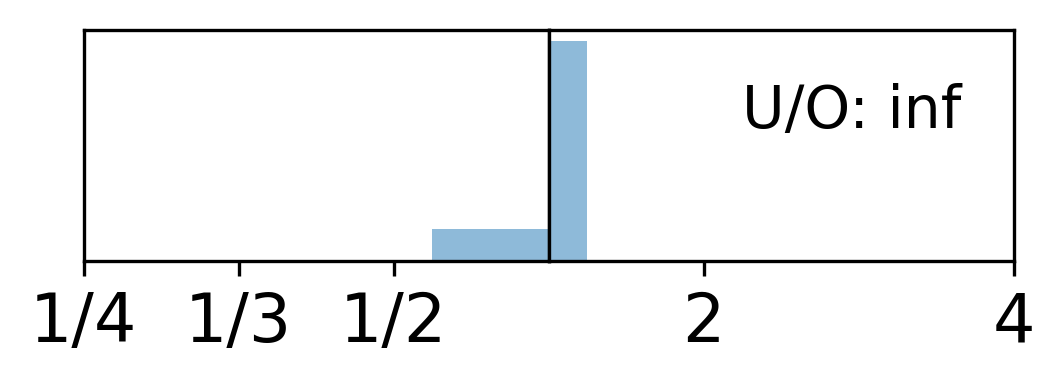}} &
        8 \\
        \end{tabular}
        }
        \caption{Ratio between the average flux over all eight observations (the first one is split into two) and that obtained by randomly sampling a given number of observations (considering all possible combinations without repetition). The vertical line at the center represents unity. Values to its left indicate underestimation, while values to its right indicate overestimation. The ratio of underestimating to overestimating cases is reported as U/O. Cases falling within the –1 to 1 bins are excluded from this ratio.}
        \label{table:total_fluxes_combinatory}
        \end{table}
        \renewcommand{\arraystretch}{1}
        \setlength{\tabcolsep}{4pt}

    \section{Conclusions}\label{section:conclusions}

    The main goal of this work was to provide high time-resolution X-ray fluxes from {\PC} that can be used to model the response of planetary atmospheres to short-term stellar variability. We analyzed all available XMM-Newton X-ray observations of {\PC} and produced time series of spectra covering the 1–100 {\AA} range, spanning nearly three days of data collected over 17 years. In addition to this, an average {\PC} EUV spectrum and the X-ray to EUV scaling law from SF+25 were used to extend the time series of spectra up to 920 {\AA}.

    Time resolution is important in the study of exoplanets for three reasons: (1) atmospheric responses are non-linear, so the timing of energy release can be as important as the total energy; (2) observations of exoplanet atmospheres are instantaneous snapshots, which may occur shortly after flaring events or at different phases of the activity cycle, making knowledge of the full flux variability crucial for accurate interpretation; (3) averaging the stellar spectrum before or after the model regression can lead to different results, although in our case this effect alters the integrated flux by less than 10{\%}, a value comparable with the uncertainties.

    Taking advantage of the high number of photon counts we get from {\PC} due to its proximity, we implemented a novel time-binning strategy, achieving an average time-resolution of 300 s. The high number of counts though, also causes non-negligible pile-up effect. To solve this, a new correction for pile-up was applied. We showed that for {\PC}, during flares, this effect can reduce the measured fluxes by up to 30{\%} if left uncorrected.
    
    Thermal plasma models with different numbers of components were tested, and the best-fitting model was selected for each time interval; in many cases, multiple models provided comparably good fits. We compared our fluxes with those reported in previous studies and discussed how methodological choices affect them. Finally, we examined the impact of selecting only a subset of the available observations on the inferred average flux and applied two scaling laws to reconstruct the 100–920 {\AA} flux from the 1–100 {\AA} range.

    {\PC}’s intrinsic variability is characterized across different wavelengths. In the 1–100 {\AA} band, the flux varies by about two orders of magnitude between minimum and maximum, with fluctuations of factors 2–10 occurring commonly on timescales of minutes to hours. Variability is even stronger in the [1–20] {\AA} range, where changes span 3–4 orders of magnitude. Continuous variability, largely attributable to flares over a broad intensity range, precludes the precise definition of a stellar quiescent state. Even the lowest observed flux levels may correspond to phases affected by coronal dimming following previous flares, and thus cannot be reliably used as a quiescent reference. We note that, if a cycle with an amplitude of approximately 1.5 were confirmed, its effects on planetary atmospheres would likely be negligible compared to those of flares and short-term variability.
    
    Over the full observational baseline, we studied the average, median, minimum, maximum, and 80th-percentile X-ray fluxes and compared them with flux estimates from previous works. We found that time-resolved analyses are essential to properly capture peaks and minima. Labeling intervals as flaring or quiescent and then averaging them, as commonly done to improve the S/N, may lead to differences of up to a few factors, which should be considered when studying the response of planetary atmospheres to transient events.
    
    The average flux is strongly influenced by stellar variability and significantly differs from the quiescent level, unlike the case of less active stars. Specifically, the average flux can vary by up to a factor of $\sim$3 when computed over different combinations of observations, implying that single or sparse observations of active targets may not be representative of the long-term average. For this reason, an additional uncertainty should be considered beyond the statistical ones, its amplitude depending on the activity level and total observing time.

    With the study of the specific case of {\PC} we identified different sources of uncertainties, some common to all the sources, some others specific of bright sources like {\PC}. Data reduction choices (e.g. the time binning strategy and the target time-resolution), observational setup (e.g., instruments used, their sensitivity, and observing conditions), and assumptions on the hydrogen column density, the metallicity (see e.g. the up to 20{\%} differences in the flux reported in Section \ref{section:plasma_modeling}), and plasma models are generally common sources in most of the analyses. For bright sources pile-up can also introduce further uncertainties.
    
    When looking at total fluxes, these uncertainties are likely negligible compared to the reported factor of $\sim$3 related to the limited observational baseline. However, they may dominate in specific situations, such as during high-flux flares or in high-energy wavelength intervals. In these cases, uncertainties increase due to modeling limitations: multiple components can overlap during high-flux intervals, or the S/N becomes very low at high energies, making the hot plasma components poorly constrained. In addition, when not corrected, pile-up can be significant during flares.

    Furthermore, using two different X-ray–to–EUV scaling relations from the same authors introduces another possible factor of $\sim$3 variation in the average flux. This further emphasizes the need to account for flux differences of several factors when modeling atmospheric responses.

    These results emphasize that, when modeling exoplanetary atmospheres, it is important to test a range of fluxes that may be considerably broader than implied by the "standard" uncertainties alone, even for a well-studied target such as {\PC}. Our analysis further highlights the need for long-term monitoring to accurately characterize a star's X-ray emission and, in turn, better constrain the response of its exoplanets.
    
    \section*{Data availability}
    Data products from this work are available at the CDS via anonymous ftp to cdsarc.u-strasbg.fr (130.79.128.5) or via \url{http://cdsweb.u-strasbg.fr/cgi-bin/qcat?J/A+A/}. Additional data will be provided upon reasonable request to \href{mailto:andrea.damonte@inaf.it}{andrea.damonte@inaf.it} or \href{mailto:andrea.damonte@cea.fr}{andrea.damonte@cea.fr}.

    \begin{acknowledgements}
          The work by AD has been funded through the VINCI 2023 Programme of the Université Franco-Italienne. This work was supported by the Programme Nationale de Planétologie (PNP) - Action Thématique Exosystèmes of CNRS/INSU, co-funded by CEA and CNES. IP acknowledges support from Bando per il Finanziamento della Ricerca Fondamentale 2024 dell’Istituto Nazionale di Astrofisica (INAF). AMa and IP acknowledge financial contribution from the INAF grant 2023 for data analysis. GM acknowledges the support of the ASI-INAF agreement 2021-5-HH.0 and its addendum. Based on observations obtained with XMM-Newton, an ESA science mission with instruments and contributions directly funded by ESA Member States and NASA.
    \end{acknowledgements}

    \bibliography{references}

@article{schellenberger_using_2025,
	title = {Using the {XMM}-{Newton} small window mode to investigate systematic uncertainties in the particle background of {X}-ray charge-coupled device detectors},
	volume = {11},
	issn = {2329-4124, 2329-4221},
	url = {https://www.spiedigitallibrary.org/journals/Journal-of-Astronomical-Telescopes-Instruments-and-Systems/volume-11/issue-1/018006/Using-the-XMM-Newton-small-window-mode-to-investigate-systematic/10.1117/1.JATIS.11.1.018006.full},
	doi = {10.1117/1.JATIS.11.1.018006},
	number = {1},
	urldate = {2025-12-18},
	journal = {JATIS},
	author = {Schellenberger, Gerrit and Kraft, Ralph and Nulsen, Paul and Miller, Eric D. and Bautz, Marshall W. and Grant, Catherine E. and Wilkins, Dan and Allen, Steven and Molendi, Silvano and Burrows, David N. and Falcone, Abraham D. and Fioretti, Valentina and Foster, Richard F. and Hall, David and Hubbard, Michael W. J. and Perinati, Emanuele and Poliszczuk, Artem and Rau, Arne and Sarkar, Arnab and Schneider, Benjamin},
	month = mar,
	year = {2025},
	pages = {018006},
}

@inproceedings{blind_ristretto_2024,
	title = {{RISTRETTO}: a {VLT} {XAO} design to reach {Proxima} {Cen} b in the visible},
	volume = {13097},
	shorttitle = {{RISTRETTO}},
	url = {https://www.spiedigitallibrary.org/conference-proceedings-of-spie/13097/130976U/RISTRETTO--a-VLT-XAO-design-to-reach-Proxima-Cen/10.1117/12.3019992.full},
	doi = {10.1117/12.3019992},
	urldate = {2025-12-18},
	booktitle = {Adaptive {Optics} {Systems} {IX}},
	publisher = {SPIE},
	author = {Blind, N. and Shinde, M. and Dinis, I. and Restori, N. and Chazelas, B. and Fusco, T. and Guyon, O. and Kühn, J. and Lovis, C. and Martinez, P. and Motte, M. and Sauvage, J.-F. and Spang, A.},
	month = aug,
	year = {2024},
	pages = {1622--1634},
}

@article{peres_sun_2000,
	title = {The {Sun} as an {X}-{Ray} {Star}. {II}. {Using} the {Yohkoh}/{Soft} {X}-{Ray} {Telescope}-derived {Solar} {Emission} {Measure} versus {Temperature} to {Interpret} {Stellar} {X}-{Ray} {Observations}},
	volume = {528},
	issn = {0004-637X},
	url = {https://iopscience.iop.org/article/10.1086/308136/meta},
	doi = {10.1086/308136},
	language = {en},
	number = {1},
	urldate = {2025-07-07},
	journal = {ApJ},
	author = {Peres, G. and Orlando, S. and Reale, F. and Rosner, R. and Hudson, H.},
	month = jan,
	year = {2000},
	pages = {537},
}

@article{sanz-forcada_estimation_2011,
	title = {Estimation of the {XUV} radiation onto close planets and their evaporation},
	volume = {532},
	issn = {0004-6361},
	url = {https://ui.adsabs.harvard.edu/abs/2011A&A...532A...6S},
	doi = {10.1051/0004-6361/201116594},
	urldate = {2025-03-12},
	journal = {A\&A},
	author = {Sanz-Forcada, J. and Micela, G. and Ribas, I. and Pollock, A. M. T. and Eiroa, C. and Velasco, A. and Solano, E. and García-Álvarez, D.},
	month = aug,
	year = {2011},
	pages = {A6},
}

@article{robrade_coronal_2012,
	title = {Coronal activity cycles in nearby {G} and {K} stars - {XMM}-{Newton} monitoring of 61 {Cygni} and α {Centauri}},
	volume = {543},
	copyright = {© ESO, 2012},
	issn = {0004-6361, 1432-0746},
	url = {https://www.aanda.org/articles/aa/abs/2012/07/aa19046-12/aa19046-12.html},
	doi = {10.1051/0004-6361/201219046},
	language = {en},
	urldate = {2025-09-03},
	journal = {A\&A},
	author = {Robrade, J. and Schmitt, J. H. M. M. and Favata, F.},
	month = jul,
	year = {2012},
	pages = {A84},
}

@article{yashiro_statistical_2009,
	title = {Statistical relationship between solar flares and coronal mass ejections},
	volume = {257},
	issn = {1743-9221},
	url = {https://ui.adsabs.harvard.edu/abs/2009IAUS..257..233Y/abstract},
	doi = {10.1017/S1743921309029342},
	language = {en},
	urldate = {2025-06-11},
	journal = {IAU Symp.},
	author = {Yashiro, Seiji and Gopalswamy, Nat},
	month = mar,
	year = {2009},
	pages = {233--243},
}

@article{wilms_absorption_2000,
	title = {On the {Absorption} of {X}-{Rays} in the {Interstellar} {Medium}},
	volume = {542},
	issn = {0004-637X},
	url = {https://ui.adsabs.harvard.edu/abs/2000ApJ...542..914W},
	doi = {10.1086/317016},
	urldate = {2025-02-18},
	journal = {ApJ},
	author = {Wilms, J. and Allen, A. and McCray, R.},
	month = oct,
	year = {2000},
	pages = {914--924},
}

@article{webb_coronal_2012,
	title = {Coronal {Mass} {Ejections}: {Observations}},
	volume = {9},
	shorttitle = {Coronal {Mass} {Ejections}},
	url = {https://ui.adsabs.harvard.edu/abs/2012LRSP....9....3W/abstract},
	doi = {10.12942/lrsp-2012-3},
	language = {en},
	number = {1},
	urldate = {2025-06-11},
	journal = {Living rev. sol. phys.},
	author = {Webb, David F. and Howard, Timothy A.},
	month = dec,
	year = {2012},
	pages = {3},
}

@article{wargelin_optical_2017,
	title = {Optical, {UV}, and {X}-ray evidence for a 7-yr stellar cycle in {Proxima} {Centauri}},
	volume = {464},
	issn = {0035-8711, 1365-2966},
	url = {https://academic.oup.com/mnras/article-lookup/doi/10.1093/mnras/stw2570},
	doi = {10.1093/mnras/stw2570},
	language = {en},
	number = {3},
	urldate = {2024-08-07},
	journal = {MNRAS},
	author = {Wargelin, B. J. and Saar, S. H. and Pojmański, G. and Drake, J. J. and Kashyap, V. L.},
	month = jan,
	year = {2017},
	pages = {3281--3296},
}

@article{wargelin_x-ray_2024,
	title = {X-{Ray}, {UV}, and {Optical} {Observations} of {Proxima} {Centauri}’s {Stellar} {Cycle}},
	volume = {977},
	issn = {0004-637X},
	url = {https://dx.doi.org/10.3847/1538-4357/ad8faa},
	doi = {10.3847/1538-4357/ad8faa},
	language = {en},
	number = {2},
	urldate = {2025-01-09},
	journal = {ApJ},
	author = {Wargelin, Bradford J. and Saar, Steven H. and Irving, Zackery A. and Slavin, Jonathan D. and Ratzlaff, Peter and Nascimento, José-Dias do},
	month = dec,
	year = {2024},
	pages = {144},
}

@article{turner_european_2001,
	title = {The {European} {Photon} {Imaging} {Camera} on {XMM}-{Newton}: {The} {MOS} cameras},
	volume = {365},
	issn = {0004-6361},
	shorttitle = {The {European} {Photon} {Imaging} {Camera} on {XMM}-{Newton}},
	url = {https://www.aanda.org/component/article?access=bibcode&bibcode=&bibcode=2001A%2526A...365L..27TFUL},
	doi = {10.1051/0004-6361:20000087},
	urldate = {2025-01-09},
	journal = {A\&A},
	author = {Turner, M. J. L. and Abbey, A. and Arnaud, M. and Balasini, M. and Barbera, M. and Belsole, E. and Bennie, P. J. and Bernard, J. P. and Bignami, G. F. and Boer, M. and Briel, U. and Butler, I. and Cara, C. and Chabaud, C. and Cole, R. and Collura, A. and Conte, M. and Cros, A. and Denby, M. and Dhez, P. and Di Coco, G. and Dowson, J. and Ferrando, P. and Ghizzardi, S. and Gianotti, F. and Goodall, C. V. and Gretton, L. and Griffiths, R. G. and Hainaut, O. and Hochedez, J. F. and Holland, A. D. and Jourdain, E. and Kendziorra, E. and Lagostina, A. and Laine, R. and La Palombara, N. and Lortholary, M. and Lumb, D. and Marty, P. and Molendi, S. and Pigot, C. and Poindron, E. and Pounds, K. A. and Reeves, J. N. and Reppin, C. and Rothenflug, R. and Salvetat, P. and Sauvageot, J. L. and Schmitt, D. and Sembay, S. and Short, A. D. T. and Spragg, J. and Stephen, J. and Strüder, L. and Tiengo, A. and Trifoglio, M. and Trümper, J. and Vercellone, S. and Vigroux, L. and Villa, G. and Ward, M. J. and Whitehead, S. and Zonca, E.},
	month = jan,
	year = {2001},
	pages = {L27--L35},
}

@article{struder_european_2001,
	title = {The {European} {Photon} {Imaging} {Camera} on {XMM}-{Newton}: {The} pn-{CCD} camera},
	volume = {365},
	issn = {0004-6361},
	shorttitle = {The {European} {Photon} {Imaging} {Camera} on {XMM}-{Newton}},
	url = {https://www.aanda.org/component/article?access=bibcode&bibcode=&bibcode=2001A%2526A...365L..18SFUL},
	doi = {10.1051/0004-6361:20000066},
	urldate = {2025-01-09},
	journal = {A\&A},
	author = {Strüder, L. and Briel, U. and Dennerl, K. and Hartmann, R. and Kendziorra, E. and Meidinger, N. and Pfeffermann, E. and Reppin, C. and Aschenbach, B. and Bornemann, W. and Bräuninger, H. and Burkert, W. and Elender, M. and Freyberg, M. and Haberl, F. and Hartner, G. and Heuschmann, F. and Hippmann, H. and Kastelic, E. and Kemmer, S. and Kettenring, G. and Kink, W. and Krause, N. and Müller, S. and Oppitz, A. and Pietsch, W. and Popp, M. and Predehl, P. and Read, A. and Stephan, K. H. and Stötter, D. and Trümper, J. and Holl, P. and Kemmer, J. and Soltau, H. and Stötter, R. and Weber, U. and Weichert, U. and von Zanthier, C. and Carathanassis, D. and Lutz, G. and Richter, R. H. and Solc, P. and Böttcher, H. and Kuster, M. and Staubert, R. and Abbey, A. and Holland, A. and Turner, M. and Balasini, M. and Bignami, G. F. and La Palombara, N. and Villa, G. and Buttler, W. and Gianini, F. and Lainé, R. and Lumb, D. and Dhez, P.},
	month = jan,
	year = {2001},
	pages = {L18--L26},
}

@article{smith_collisional_2001,
	title = {Collisional {Plasma} {Models} with {APEC}/{APED}: {Emission}-{Line} {Diagnostics} of {Hydrogen}-like and {Helium}-like {Ions}},
	volume = {556},
	issn = {0004-637X},
	shorttitle = {Collisional {Plasma} {Models} with {APEC}/{APED}},
	url = {https://ui.adsabs.harvard.edu/abs/2001ApJ...556L..91S},
	doi = {10.1086/322992},
	urldate = {2025-03-17},
	journal = {ApJ},
	author = {Smith, Randall K. and Brickhouse, Nancy S. and Liedahl, Duane A. and Raymond, John C.},
	month = aug,
	year = {2001},
	pages = {L91--L95},
}

@incollection{schartel_xmm-newton_2024,
	title = {{XMM}-{Newton}},
	isbn = {978-981-19-6960-7},
	url = {https://doi.org/10.1007/978-981-19-6960-7_41},
	language = {en},
	urldate = {2025-02-12},
	booktitle = {Handbook of {X}-ray and {Gamma}-ray {Astrophysics}},
	publisher = {Springer Nature},
	author = {Schartel, Norbert and González-Riestra, Rosario and Kretschmar, Peter and Kirsch, Marcus and Rodríguez-Pascual, Pedro and Rosen, Simon and Santos-Lleó, Maria and Smith, Michael and Stuhlinger, Martin and Verdugo-Rodrigo, Eva},
	editor = {Bambi, Cosimo and Santangelo, Andrea},
	year = {2024},
	pages = {1501--1538},
}

@article{sanz-forcada_connection_2025,
	title = {Connection between planetary {He} {I} \$λ\$10830 Å absorption and extreme-ultraviolet emission of planet-host stars},
	issn = {0004-6361, 1432-0746},
	url = {http://arxiv.org/abs/2501.03716},
	doi = {10.1051/0004-6361/202451680},
	urldate = {2025-01-08},
	journal = {A\&A},
	author = {Sanz-Forcada, J. and López-Puertas, M. and Lampón, M. and Czesla, S. and Nortmann, L. and Caballero, J. A. and Osorio, M. R. Zapatero and Amado, P. J. and Murgas, F. and Orell-Miquel, J. and Pallé, E. and Quirrenbach, A. and Reiners, A. and Ribas, I. and Sánchez-López, A. and Solano, E.},
	month = jan,
	year = {2025},
}

@article{rockcliffe_far-ultraviolet_2025,
	title = {Far-ultraviolet flares and variability of the young {M} dwarf {AU} {Mic}: a non-detection of planet c in transit at {Lyman}-alpha},
	volume = {169},
	issn = {0004-6256, 1538-3881},
	shorttitle = {Far-ultraviolet flares and variability of the young {M} dwarf {AU} {Mic}},
	url = {http://arxiv.org/abs/2505.17197},
	doi = {10.3847/1538-3881/adccc7},
	number = {6},
	urldate = {2025-05-26},
	journal = {AJ},
	author = {Rockcliffe, Keighley E. and Newton, Elisabeth R. and Youngblood, Allison and Duvvuri, Girish M. and Gilbert, Emily A. and Plavchan, Peter and Gao, Peter and Müller, Hans-R. and Feinstein, Adina D. and Barclay, Thomas and Lopez, Eric D.},
	month = jun,
	year = {2025},
	pages = {321},
}

@article{ridgway_3d_2023,
	title = {{3D} modelling of the impact of stellar activity on tidally locked terrestrial exoplanets: atmospheric composition and habitability},
	volume = {518},
	issn = {0035-8711},
	shorttitle = {{3D} modelling of the impact of stellar activity on tidally locked terrestrial exoplanets},
	url = {https://ui.adsabs.harvard.edu/abs/2023MNRAS.518.2472R/abstract},
	doi = {10.1093/mnras/stac3105},
	language = {en},
	number = {2},
	urldate = {2025-08-21},
	journal = {MNRAS},
	author = {Ridgway, R. J. and Zamyatina, M. and Mayne, N. J. and Manners, J. and Lambert, F. H. and Braam, M. and Drummond, B. and Hébrard, E. and Palmer, P. I. and Kohary, K.},
	month = jan,
	year = {2023},
	pages = {2472--2496},
}

@article{ribas_full_2017,
	title = {The full spectral radiative properties of {Proxima} {Centauri}},
	volume = {603},
	issn = {0004-6361},
	url = {https://ui.adsabs.harvard.edu/abs/2017A%26A...603A..58R/abstract},
	doi = {10.1051/0004-6361/201730582},
	language = {en},
	number = {A58},
	urldate = {2025-05-15},
	journal = {A\&A},
	author = {Ribas, Ignasi and Gregg, Michael D. and Boyajian, Tabetha S. and Bolmont, Emeline},
	month = jul,
	year = {2017},
	pages = {12},
}

@book{reames_solar_2021,
	series = {Lect. {Notes} {Phys}.},
	title = {Solar {Energetic} {Particles}: {A} {Modern} {Primer} on {Understanding} {Sources}, {Acceleration} and {Propagation}},
	volume = {978},
	isbn = {978-3-030-66401-5 978-3-030-66402-2},
	url = {http://link.springer.com/10.1007/978-3-030-66402-2},
	language = {en},
	urldate = {2025-06-11},
	publisher = {Springer International Publishing},
	author = {Reames, Donald V.},
	year = {2021},
}

@article{preibisch_evolution_2005,
	title = {The {Evolution} of {X}-{Ray} {Emission} in {Young} {Stars}},
	volume = {160},
	issn = {0067-0049},
	url = {https://iopscience.iop.org/article/10.1086/432094/meta},
	doi = {10.1086/432094},
	language = {en},
	number = {2},
	urldate = {2025-04-29},
	journal = {ApJS},
	author = {Preibisch, Thomas and Feigelson, Eric D.},
	month = oct,
	year = {2005},
	pages = {390},
}

@article{pandey_study_2008,
	title = {A study of {X}-ray flares - {I}. {Active} late-type dwarfs},
	volume = {387},
	issn = {0035-8711},
	url = {https://ui.adsabs.harvard.edu/abs/2008MNRAS.387.1627P/abstract},
	doi = {10.1111/j.1365-2966.2008.13342.x},
	language = {en},
	number = {4},
	urldate = {2025-07-03},
	journal = {MNRAS},
	author = {Pandey, J. C. and Singh, K. P.},
	month = jul,
	year = {2008},
	pages = {1627--1648},
}

@article{osten_mouse_2010,
	title = {The {Mouse} {That} {Roared}: {A} {Superflare} from the {dMe} {Flare} {Star} {EV} {Lac} {Detected} by {Swift} and {Konus}-{Wind}},
	volume = {721},
	issn = {0004-637X},
	shorttitle = {The {Mouse} {That} {Roared}},
	url = {https://ui.adsabs.harvard.edu/abs/2010ApJ...721..785O/abstract},
	doi = {10.1088/0004-637X/721/1/785},
	language = {en},
	number = {1},
	urldate = {2025-07-03},
	journal = {ApJ},
	author = {Osten, Rachel A. and Godet, Olivier and Drake, Stephen and Tueller, Jack and Cummings, Jay and Krimm, Hans and Pye, John and Pal'shin, Valentin and Golenetskii, Sergei and Reale, Fabio and Oates, Samantha R. and Page, Mat J. and Melandri, Andrea},
	month = sep,
	year = {2010},
	pages = {785--801},
}

@article{khodachenko_stellar_2014,
	title = {Stellar {CME} activity and its possible influence on exoplanets' environments: {Importance} of magnetospheric protection},
	volume = {300},
	issn = {1743-9221},
	shorttitle = {Stellar {CME} activity and its possible influence on exoplanets' environments},
	url = {https://ui.adsabs.harvard.edu/abs/2014IAUS..300..335K/abstract},
	doi = {10.1017/S1743921313011174},
	language = {en},
	urldate = {2025-07-28},
	journal = {IAU Symp.},
	author = {Khodachenko, Maxim L. and Sasunov, Yury and Arkhypov, Oleksiy V. and Alexeev, Igor I. and Belenkaya, Elena S. and Lammer, Helmut and Kislyakova, Kristina G. and Odert, Petra and Leitzinger, Martin and Güdel, Manuel},
	month = jan,
	year = {2014},
	pages = {335--346},
}

@article{konings_impact_2022,
	title = {Impact of stellar flares on the chemical composition and transmission spectra of gaseous exoplanets orbiting {M} dwarfs},
	volume = {667},
	copyright = {© T. Konings et al. 2022},
	issn = {0004-6361, 1432-0746},
	url = {https://www.aanda.org/articles/aa/abs/2022/11/aa43436-22/aa43436-22.html},
	doi = {10.1051/0004-6361/202243436},
	language = {en},
	urldate = {2025-10-28},
	journal = {A\&A},
	author = {Konings, T. and Baeyens, R. and Decin, L.},
	month = nov,
	year = {2022},
	pages = {A15},
}

@article{neupert_comparison_1968,
	title = {Comparison of {Solar} {X}-{Ray} {Line} {Emission} with {Microwave} {Emission} during {Flares}},
	volume = {153},
	issn = {0004-637X},
	url = {https://ui.adsabs.harvard.edu/abs/1968ApJ...153L..59N/abstract},
	doi = {10.1086/180220},
	language = {en},
	urldate = {2025-07-31},
	journal = {ApJ},
	author = {Neupert, Werner M.},
	month = jul,
	year = {1968},
	pages = {L59},
}

@article{mason_xmm-newton_2001,
	title = {The {XMM}-{Newton} optical/{UV} monitor telescope},
	volume = {365},
	copyright = {© ESO, 2001},
	issn = {0004-6361, 1432-0746},
	url = {https://www.aanda.org/articles/aa/abs/2001/01/aaxmm33/aaxmm33.html},
	doi = {10.1051/0004-6361:20000044},
	language = {en},
	number = {1},
	urldate = {2025-01-09},
	journal = {A\&A},
	author = {Mason, K. O. and Breeveld, A. and Much, R. and Carter, M. and Cordova, F. A. and Cropper, M. S. and Fordham, J. and Huckle, H. and Ho, C. and Kawakami, H. and Kennea, J. and Kennedy, T. and Mittaz, J. and Pandel, D. and Priedhorsky, W. C. and Sasseen, T. and Shirey, R. and Smith, P. and Vreux, J.-M.},
	month = jan,
	year = {2001},
	pages = {L36--L44},
}

@article{pavlenko_temporal_2019,
	title = {Temporal changes of the flare activity of {Proxima} {Centauri}},
	volume = {626},
	issn = {0004-6361},
	url = {https://ui.adsabs.harvard.edu/abs/2019A&A...626A.111P},
	doi = {10.1051/0004-6361/201834258},
	urldate = {2024-08-07},
	journal = {A\&A},
	author = {Pavlenko, Ya. V. and Suárez Mascareño, A. and Zapatero Osorio, M. R. and Rebolo, R. and Lodieu, N. and Béjar, V. J. S. and González Hernández, J. I. and Mohorian, M.},
	month = jun,
	year = {2019},
	pages = {A111},
}

@article{mascareno_diving_2025,
	title = {Diving into the planetary system of {Proxima} with {NIRPS} -- {Breaking} the metre per second barrier in the infrared},
	volume = {700},
	issn = {0004-6361, 1432-0746},
	url = {http://arxiv.org/abs/2507.21751},
	doi = {10.1051/0004-6361/202553728},
	urldate = {2025-07-31},
	journal = {A\&A},
	author = {Mascareño, Alejandro Suárez and Artigau, Etienne and Mignon, Lucile and Delfosse, Xavier and Cook, Neil J. and Bouchy, François and Doyon, René and Hernández, Jonay I. González and Vandal, Thomas and Leão, Izan de Castro and Stefanov, Atanas K. and Faria, João and Cadieux, Charles and Lamontagne, Pierrot and Baron, Frédérique and Barros, Susana C. C. and Benneke, Björn and Bonfils, Xavier and Bryan, Marta and Martins, Bruno L. Canto and Cloutier, Ryan and Cowan, Nicolas B. and Freitas, Daniel Brito de and Medeiros, Jose Renan De and Delgado-Mena, Elisa and Figueira, Pedro and Dumusque, Xavier and Ehrenreich, David and Lafrenière, David and Lovis, Christophe and Malo, Lison and Melo, Claudio and Mordasini, Christoph and Pepe, Francesco and Rebolo, Rafael and Rowe, Jason and Santos, Nuno C. and Ségransan, Damien and Udry, Stéphane and Valencia, Diana and Wade, Gregg and Abreu, Manuel and Aguiar, José L. A. and Moulla, Khaled Al and Allain, Guillaume and Allart, Romain and Arial, Tomy and Auger, Hugues and Bazinet, Luc and Blind, Nicolas and Bohlender, David and Boisse, Isabelle and Boucher, Anne and Bourrier, Vincent and Bovay, Sébastien and Broeg, Christopher and Brousseau, Denis and Cabral, Alexandre and Carmona, Andres and Carteret, Yann and Challita, Zalpha and Chazelas, Bruno and Coelho, João and Cointepas, Marion and Conod, Uriel and Cristo, Eduardo and Silva, Ana Rita Costa and Darveau-Bernier, Antoine and Dauplaise, Laurie and Delisle, Jean-Baptiste and Gomes, Roseane de Lima and Forveille, Thierry and Frensch, Yolanda G. C. and Témich, Félix Gracia and Fontinele, Dasaev O. and Gagné, Jonathan and Genest, Frédéric and Genolet, Ludovic and Silva, João Gomes da and Grieves, Nolan and Hernandez, Olivier and Hobson, Melissa J. and Hoeijmakers, H. Jens and Hubin, Norbert and Jahandar, Farbod and Jayawardhana, Ray and Käufl, Hans-Ulrich and Kerley, Dan and Kolb, Johann and Krishnamurthy, Vigneshwaran and Kung, Benjamin and L'Heureux, Alexandrine and Larue, Pierre and Leath, Henry and Lim, Olivia and Curto, Gaspare Lo and Martins, Allan M. and Matthews, Jaymie and Mayer, Jean-Sébastien and Messias, Yuri S. and Metchev, Stan and Moranta, Leslie and Mounzer, Dany and Nari, Nicola and Nielsen, Louise D. and Osborn, Ares and Ouellet, Mathieu and Otegi, Jon and Parc, Léna and Pasquini, Luca and Passegger, Vera M. and Pelletier, Stefan and Peroux, Céline and Piaulet-Ghorayeb, Caroline and Plotnykov, Mykhaylo and Pompei, Emanuela and Poulin-Girard, Anne-Sophie and Rasilla, José Luis and Reshetov, Vladimir and Saint-Antoine, Jonathan and Sarajlic, Mirsad and Saviane, Ivo and Schnell, Robin and Segovia, Alex and Seidel, Julia and Silber, Armin and Sinclair, Peter and Sordet, Michael and Sosnowska, Danuta and Srivastava, Avidaan and Teixeira, Márcio A. and Thibault, Simon and Vallée, Philippe and Vaulato, Valentina and Wardenier, Joost P. and Wehbe, Bachar and Weisserman, Drew and Wevers, Ivan and Wildi, François and Yariv, Vincent and Zins, Gérard},
	month = aug,
	year = {2025},
	pages = {A11},
}

@article{pizzolato_stellar_2003,
	title = {The stellar activity-rotation relationship revisited: {Dependence} of saturated and non-saturated {X}-ray emission regimes on stellar mass for late-type dwarfs},
	volume = {397},
	copyright = {© ESO, 2003},
	issn = {0004-6361, 1432-0746},
	shorttitle = {The stellar activity-rotation relationship revisited},
	url = {https://www.aanda.org/articles/aa/abs/2003/01/aa2680/aa2680.html},
	doi = {10.1051/0004-6361:20021560},
	language = {en},
	number = {1},
	urldate = {2025-07-07},
	journal = {A\&A},
	author = {Pizzolato, N. and Maggio, A. and Micela, G. and Sciortino, S. and Ventura, P.},
	month = jan,
	year = {2003},
	pages = {147--157},
}

@article{maldonado_hades_2020,
	title = {{HADES} {RV} programme with {HARPS}-{N} at {TNG}. {XII}. {The} abundance signature of {M} dwarf stars with planets},
	volume = {644},
	issn = {0004-6361},
	url = {https://ui.adsabs.harvard.edu/abs/2020A%26A...644A..68M/abstract},
	doi = {10.1051/0004-6361/202039478},
	language = {en},
	number = {A48},
	urldate = {2025-06-10},
	journal = {A\&A},
	author = {Maldonado, J. and Micela, G. and Baratella, M. and D'Orazi, V. and Affer, L. and Biazzo, K. and Lanza, A. F. and Maggio, A. and González Hernández, J. I. and Perger, M. and Pinamonti, M. and Scandariato, G. and Sozzetti, A. and Locci, D. and Di Maio, C. and Bignamini, A. and Claudi, R. and Molinari, E. and Rebolo, R. and Ribas, I. and Toledo-Padrón, B. and Covino, E. and Desidera, S. and Herrero, E. and Morales, J. C. and Suárez-Mascareño, A. and Pagano, I. and Petralia, A. and Piotto, G. and Poretti, E.},
	month = dec,
	year = {2020},
	pages = {23},
}

@article{pillitteri_x-ray_2022,
	title = {X-ray flares of the young planet host {Ds} {Tucanae} {A}},
	volume = {666},
	copyright = {https://creativecommons.org/licenses/by/4.0},
	issn = {0004-6361, 1432-0746},
	url = {https://www.aanda.org/10.1051/0004-6361/202244268},
	doi = {10.1051/0004-6361/202244268},
	language = {en},
	urldate = {2024-08-05},
	journal = {A\&A},
	author = {Pillitteri, I. and Argiroffi, C. and Maggio, A. and Micela, G. and Benatti, S. and Reale, F. and Colombo, S. and Wolk, S. J.},
	month = oct,
	year = {2022},
	pages = {A198},
}

@article{maggio_xuv_2024,
	title = {{XUV} irradiation of young planetary atmospheres. {Results} from a joint {XMM}-{Newton} and {HST} observation of {HIP67522}},
	volume = {690},
	issn = {0004-6361},
	url = {https://ui.adsabs.harvard.edu/abs/2024A&A...690A.383M/abstract},
	doi = {10.1051/0004-6361/202451582},
	language = {en},
	urldate = {2025-07-25},
	journal = {A\&A},
	author = {Maggio, A. and Pillitteri, I. and Argiroffi, C. and Locci, D. and Benatti, S. and Micela, G.},
	month = oct,
	year = {2024},
	pages = {A383},
}

@article{loyd_muscles_2018,
	title = {The {MUSCLES} {Treasury} {Survey}. {V}. {FUV} {Flares} on {Active} and {Inactive} {M} {Dwarfs}},
	volume = {867},
	issn = {0004-637X},
	url = {https://ui.adsabs.harvard.edu/abs/2018ApJ...867...71L/abstract},
	doi = {10.3847/1538-4357/aae2bd},
	language = {en},
	number = {1},
	urldate = {2025-04-22},
	journal = {ApJ},
	author = {Loyd, R. O. Parke and France, Kevin and Youngblood, Allison and Schneider, Christian and Brown, Alexander and Hu, Renyu and Segura, Antígona and Linsky, Jeffrey and Redfield, Seth and Tian, Feng and Rugheimer, Sarah and Miguel, Yamila and Froning, Cynthia S.},
	month = nov,
	year = {2018},
	pages = {71},
}

@article{linsky_intrinsic_2014,
	title = {The {Intrinsic} {Extreme} {Ultraviolet} {Fluxes} of {F5} {V} {TO} {M5} {V} {Stars}},
	volume = {780},
	issn = {0004-637X},
	url = {https://ui.adsabs.harvard.edu/abs/2014ApJ...780...61L/abstract},
	doi = {10.1088/0004-637X/780/1/61},
	language = {en},
	number = {1},
	urldate = {2025-06-05},
	journal = {ApJ},
	author = {Linsky, Jeffrey L. and Fontenla, Juan and France, Kevin},
	month = jan,
	year = {2014},
	pages = {61},
}

@article{lee_effects_2018,
	title = {Effects of a {Solar} {Flare} on the {Martian} {Hot} {O} {Corona} and {Photochemical} {Escape}},
	volume = {45},
	copyright = {©2018. American Geophysical Union. All Rights Reserved.},
	issn = {1944-8007},
	url = {https://onlinelibrary.wiley.com/doi/abs/10.1029/2018GL077732},
	doi = {10.1029/2018GL077732},
	language = {en},
	number = {14},
	urldate = {2025-10-28},
	journal = {Geophys. Res. Lett.},
	author = {Lee, Yuni and Dong, Chuanfei and Pawlowski, Dave and Thiemann, Edward and Tenishev, Valeriy and Mahaffy, Paul and Benna, Mehdi and Combi, Michael and Bougher, Stephen and Eparvier, Frank},
	year = {2018},
	pages = {6814--6822},
}

@article{kuntz_epic-mos_2008,
	title = {The {EPIC}-{MOS} particle-induced background spectra},
	volume = {478},
	copyright = {© ESO, 2008},
	issn = {0004-6361, 1432-0746},
	url = {https://www.aanda.org/articles/aa/abs/2008/05/aa7912-07/aa7912-07.html},
	doi = {10.1051/0004-6361:20077912},
	language = {en},
	number = {2},
	urldate = {2025-01-09},
	journal = {A\&A},
	author = {Kuntz, K. D. and Snowden, S. L.},
	month = feb,
	year = {2008},
	pages = {575--596},
}

@article{kowalski_stellar_2024,
	title = {Stellar flares},
	volume = {21},
	issn = {1614-4961},
	url = {https://doi.org/10.1007/s41116-024-00039-4},
	doi = {10.1007/s41116-024-00039-4},
	language = {en},
	number = {1},
	urldate = {2025-06-05},
	journal = {Living rev. sol. phys.},
	author = {Kowalski, Adam F.},
	month = apr,
	year = {2024},
	pages = {1},
}

@article{kiraga_age-rotation-activity_2007,
	title = {Age-{Rotation}-{Activity} {Relations} for {M} {Dwarf} {Stars}},
	volume = {57},
	issn = {0001-5237},
	url = {https://ui.adsabs.harvard.edu/abs/2007AcA....57..149K/abstract},
	doi = {10.48550/arXiv.0707.2577},
	language = {en},
	urldate = {2025-06-10},
	journal = {Acta Astron.},
	author = {Kiraga, M. and Stepien, K.},
	month = jun,
	year = {2007},
	pages = {149--172},
}

@article{kipping_no_2017,
	title = {No {Conclusive} {Evidence} for {Transits} of {Proxima} b in {MOST} {Photometry}},
	volume = {153},
	issn = {0004-6256},
	url = {https://ui.adsabs.harvard.edu/abs/2017AJ....153...93K/abstract},
	doi = {10.3847/1538-3881/153/3/93},
	language = {en},
	number = {3},
	urldate = {2025-06-10},
	journal = {AJ},
	author = {Kipping, David M. and Cameron, Chris and Hartman, Joel D. and Davenport, James R. A. and Matthews, Jaymie M. and Sasselov, Dimitar and Rowe, Jason and Siverd, Robert J. and Chen, Jingjing and Sandford, Emily and Bakos, Gáspár Á and Jordán, Andrés and Bayliss, Daniel and Henning, Thomas and Mancini, Luigi and Penev, Kaloyan and Csubry, Zoltan and Bhatti, Waqas and Da Silva Bento, Joao and Guenther, David B. and Kuschnig, Rainer and Moffat, Anthony F. J. and Rucinski, Slavek M. and Weiss, Werner W.},
	month = mar,
	year = {2017},
	pages = {93},
}

@article{jethwa_when_2015,
	title = {When is pile-up important in the {XMM}-{Newton} {EPIC} cameras?},
	volume = {581},
	copyright = {© ESO, 2015},
	issn = {0004-6361, 1432-0746},
	url = {https://www.aanda.org/articles/aa/abs/2015/09/aa25579-14/aa25579-14.html},
	doi = {10.1051/0004-6361/201425579},
	language = {en},
	urldate = {2024-10-10},
	journal = {A\&A},
	author = {Jethwa, P. and Saxton, R. and Guainazzi, M. and Rodriguez-Pascual, P. and Stuhlinger, M.},
	month = sep,
	year = {2015},
	pages = {A104},
}

@article{howard_mouse_2022,
	title = {The {Mouse} that {Squeaked}: {A} small flare from {Proxima} {Cen} observed in the millimeter, optical, and soft {X}-ray with {Chandra} and {ALMA}},
	volume = {938},
	issn = {0004-637X, 1538-4357},
	shorttitle = {The {Mouse} that {Squeaked}},
	url = {http://arxiv.org/abs/2209.05490},
	doi = {10.3847/1538-4357/ac9134},
	language = {en},
	number = {2},
	urldate = {2024-08-07},
	journal = {ApJ},
	author = {Howard, Ward S. and MacGregor, Meredith A. and Osten, Rachel and Forbrich, Jan and Cranmer, Steven R. and Tristan, Isaiah and Weinberger, Alycia J. and Youngblood, Allison and Barclay, Thomas and Loyd, R. O. Parke and Shkolnik, Evgenya L. and Zic, Andrew and Wilner, David J.},
	month = oct,
	year = {2022},
	pages = {103},
}

@article{hazra_magnetic_2025,
	title = {Magnetic interaction of stellar coronal mass ejections with close-in exoplanets: implication on planetary mass-loss and {Ly} α transits},
	volume = {536},
	issn = {0035-8711},
	shorttitle = {Magnetic interaction of stellar coronal mass ejections with close-in exoplanets},
	url = {https://doi.org/10.1093/mnras/stae2559},
	doi = {10.1093/mnras/stae2559},
	number = {2},
	urldate = {2025-07-28},
	journal = {MNRAS},
	author = {Hazra, Gopal and Vidotto, Aline A and Carolan, Stephen and Villarreal D’Angelo, Carolina and Ó Fionnagáin, Dúalta},
	month = jan,
	year = {2025},
	pages = {1089--1103},
}

@article{haisch_einstein_1980,
	title = {Einstein {X}-ray observations of {Proxima} {Centauri} and the surrounding region},
	volume = {242},
	issn = {0004-637X},
	url = {https://ui.adsabs.harvard.edu/abs/1980ApJ...242L..99H/abstract},
	doi = {10.1086/183411},
	language = {en},
	urldate = {2025-07-08},
	journal = {ApJ},
	author = {Haisch, B. M. and Harnden, F. R. and Seward, F. D. and Vaiana, G. S. and Linsky, J. L. and Rosner, R.},
	month = dec,
	year = {1980},
	pages = {L99--L103},
}

@article{gudel_xmm-newton_2001,
	title = {The {XMM}-{Newton} view of stellar coronae: {X}-ray spectroscopy of the corona of {AB} {Doradus}},
	volume = {365},
	copyright = {© ESO, 2001},
	issn = {0004-6361, 1432-0746},
	shorttitle = {The {XMM}-{Newton} view of stellar coronae},
	url = {https://www.aanda.org/articles/aa/abs/2001/01/aaxmm25/aaxmm25.html},
	doi = {10.1051/0004-6361:20000220},
	language = {en},
	number = {1},
	urldate = {2025-11-14},
	journal = {A\&A},
	author = {Güdel, M. and Audard, M. and Briggs, K. and Haberl, F. and Magee, H. and Maggio, A. and Mewe, R. and Pallavicini, R. and Pye, J.},
	month = jan,
	year = {2001},
	pages = {L336--L343},
}

@article{johnstone_active_2021,
	title = {The active lives of stars: {A} complete description of the rotation and {XUV} evolution of {F}, {G}, {K}, and {M} dwarfs},
	volume = {649},
	issn = {0004-6361},
	shorttitle = {The active lives of stars},
	url = {https://ui.adsabs.harvard.edu/abs/2021A&A...649A..96J/abstract},
	doi = {10.1051/0004-6361/202038407},
	language = {en},
	urldate = {2025-05-23},
	journal = {A\&A},
	author = {Johnstone, C. P. and Bartel, M. and Güdel, M.},
	month = may,
	year = {2021},
	pages = {A96},
}

@article{gudel_x-ray_2009,
	title = {X-ray spectroscopy of stars},
	volume = {17},
	issn = {1432-0754},
	url = {https://doi.org/10.1007/s00159-009-0022-4},
	doi = {10.1007/s00159-009-0022-4},
	language = {en},
	number = {3},
	urldate = {2025-11-14},
	journal = {A\&AR},
	author = {Güdel, Manuel and Nazé, Yaël},
	month = sep,
	year = {2009},
	pages = {309--408},
}

@article{reale_modeling_2004,
	title = {Modeling an {X}-ray flare on {Proxima} {Centauri}: {Evidence} of two flaring loop components and of two heating mechanisms at work},
	volume = {416},
	issn = {0004-6361, 1432-0746},
	shorttitle = {Modeling an {X}-ray flare on {Proxima} {Centauri}},
	url = {http://www.aanda.org/10.1051/0004-6361:20034027},
	doi = {10.1051/0004-6361:20034027},
	language = {en},
	number = {2},
	urldate = {2024-08-07},
	journal = {A\&A},
	author = {Reale, F. and Güdel, M. and Peres, G. and Audard, M.},
	month = mar,
	year = {2004},
	pages = {733--747},
}

@article{gudel_x-ray_2004,
	title = {X-ray astronomy of stellar coronae},
	volume = {12},
	issn = {1432-0754},
	url = {https://doi.org/10.1007/s00159-004-0023-2},
	doi = {10.1007/s00159-004-0023-2},
	language = {en},
	number = {2},
	urldate = {2024-09-03},
	journal = {A\&AR},
	author = {Güdel, Manuel},
	month = sep,
	year = {2004},
	pages = {71--237},
}

@article{gudel_flares_2004,
	title = {Flares from small to large: {X}-ray spectroscopy of {Proxima} {Centauri} with {XMM}-{Newton}},
	volume = {416},
	issn = {0004-6361, 1432-0746},
	shorttitle = {Flares from small to large},
	url = {http://www.aanda.org/10.1051/0004-6361:20031471},
	doi = {10.1051/0004-6361:20031471},
	language = {en},
	number = {2},
	urldate = {2024-08-07},
	journal = {A\&A},
	author = {Güdel, M. and Audard, M. and Reale, F. and Skinner, S. L. and Linsky, J. L.},
	month = mar,
	year = {2004},
	pages = {713--732},
}

@article{gudel_x-ray_2002,
	title = {X-{Ray} {Evidence} for {Flare} {Density} {Variations} and {Continual} {Chromospheric} {Evaporation} in {Proxima} {Centauri}},
	volume = {580},
	issn = {0004637X, 15384357},
	url = {https://iopscience.iop.org/article/10.1086/345404},
	doi = {10.1086/345404},
	language = {en},
	number = {1},
	urldate = {2024-08-07},
	journal = {ApJ},
	author = {Güdel, Manuel and Audard, Marc and Skinner, Stephen L. and Horvath, Matthias I.},
	month = nov,
	year = {2002},
	pages = {L73--L76},
}

@article{garcia_munoz_physical_2007,
	title = {Physical and chemical aeronomy of {HD} 209458b},
	volume = {55},
	issn = {0032-0633},
	url = {https://www.sciencedirect.com/science/article/pii/S0032063307000852},
	doi = {10.1016/j.pss.2007.03.007},
	number = {10},
	urldate = {2024-08-06},
	journal = {P\&SS},
	author = {García Muñoz, A.},
	month = jul,
	year = {2007},
	pages = {1426--1455},
}

@article{garcia_munoz_modeling_2025,
	title = {Modeling helium in exoplanet atmospheres. {A} revised network with photoelectron-driven processes},
	volume = {698},
	issn = {0004-6361},
	url = {https://ui.adsabs.harvard.edu/abs/2025A&A...698A.199G/abstract},
	doi = {10.1051/0004-6361/202555145},
	language = {en},
	urldate = {2025-11-17},
	journal = {A\&A},
	author = {García Muñoz, A.},
	month = jun,
	year = {2025},
	pages = {A199},
}

@article{gaidos_trumpeting_2014,
	title = {Trumpeting {M} dwarfs with {CONCH}-{SHELL}: a catalogue of nearby cool host-stars for habitable exoplanets and life},
	volume = {443},
	issn = {0035-8711},
	shorttitle = {Trumpeting {M} dwarfs with {CONCH}-{SHELL}},
	url = {https://ui.adsabs.harvard.edu/abs/2014MNRAS.443.2561G/abstract},
	doi = {10.1093/mnras/stu1313},
	language = {en},
	number = {3},
	urldate = {2025-06-10},
	journal = {MNRAS},
	author = {Gaidos, E. and Mann, A. W. and Lépine, S. and Buccino, A. and James, D. and Ansdell, M. and Petrucci, R. and Mauas, P. and Hilton, E. J.},
	month = sep,
	year = {2014},
	pages = {2561--2578},
}

@article{gaia_collaboration_dr3_vizier_2020,
	title = {{VizieR} {Online} {Data} {Catalog}: {Gaia} {EDR3} ({Gaia} {Collaboration}, 2020)},
	volume = {1350},
	shorttitle = {{VizieR} {Online} {Data} {Catalog}},
	url = {https://ui.adsabs.harvard.edu/abs/2020yCat.1350....0G/abstract},
	doi = {10.26093/cds/vizier.1350},
	language = {en},
	urldate = {2025-06-10},
	journal = {VizieR},
	author = {{Gaia Collaboration (DR3)}},
	month = nov,
	year = {2020},
	pages = {I/350},
}

@article{fuhrmeister_multi-wavelength_2011,
	title = {Multi-wavelength observations of {Proxima} {Centauri}},
	volume = {534},
	copyright = {© ESO, 2011},
	issn = {0004-6361, 1432-0746},
	url = {https://www.aanda.org/articles/aa/abs/2011/10/aa17447-11/aa17447-11.html},
	doi = {10.1051/0004-6361/201117447},
	language = {en},
	urldate = {2025-01-09},
	journal = {A\&A},
	author = {Fuhrmeister, B. and Lalitha, S. and Poppenhaeger, K. and Rudolf, N. and Liefke, C. and Reiners, A. and Schmitt, J. H. M. M. and Ness, J.-U.},
	month = oct,
	year = {2011},
	pages = {A133},
}

@article{fuhrmeister_high_2022,
	title = {The high energy spectrum of {Proxima} {Centauri} simultaneously observed at {X}-ray and {FUV} wavelengths},
	volume = {663},
	copyright = {https://www.edpsciences.org/en/authors/copyright-and-licensing},
	issn = {0004-6361, 1432-0746},
	url = {https://www.aanda.org/10.1051/0004-6361/202243077},
	doi = {10.1051/0004-6361/202243077},
	language = {en},
	urldate = {2024-08-07},
	journal = {A\&A},
	author = {Fuhrmeister, B. and Zisik, A. and Schneider, P. C. and Robrade, J. and Schmitt, J. H. H. M. and Predehl, P. and Czesla, S. and France, K. and Muñoz, A. García},
	month = jul,
	year = {2022},
	pages = {A119},
}

@article{favata_spectroscopic_1999,
	title = {Spectroscopic analysis of a super-hot giant flare observed on {Algol} by {BeppoSAX} on 30 {August} 1997},
	volume = {350},
	issn = {0004-6361},
	url = {https://ui.adsabs.harvard.edu/abs/1999A&A...350..900F/abstract},
	doi = {10.48550/arXiv.astro-ph/9909041},
	language = {en},
	urldate = {2025-11-14},
	journal = {A\&A},
	author = {Favata, F. and Schmitt, J. H. M. M.},
	month = oct,
	year = {1999},
	pages = {900--916},
}

@article{favata_x-ray_2008,
	title = {The {X}-ray cycle in the solar-type star {HD} 81809},
	volume = {490},
	issn = {0004-6361, 1432-0746},
	url = {http://arxiv.org/abs/0806.2279},
	doi = {10.1051/0004-6361:200809694},
	number = {3},
	urldate = {2025-09-03},
	journal = {A\&A},
	author = {Favata, F. and Micela, G. and Orlando, S. and Schmitt, J. H. M. M. and Sciortino, S. and Hall, J.},
	month = nov,
	year = {2008},
	pages = {1121--1126},
}

@article{favata_stellar_2003,
	title = {Stellar {Coronal} {Astronomy}},
	volume = {108},
	issn = {0038-6308},
	url = {http://link.springer.com/10.1023/B:SPAC.0000007491.80144.21},
	doi = {10.1023/B:SPAC.0000007491.80144.21},
	language = {en},
	number = {4},
	urldate = {2024-08-31},
	journal = {Space Sci. Rev.},
	author = {Favata, Fabio and Micela, Giuseppina},
	year = {2003},
	pages = {577--708},
}

@misc{esa_xmm-newton_2023,
	title = {{XMM}-{Newton} {SAS} v.xmmsas\_20230412\_1735-21.0.0},
	url = {https://www.cosmos.esa.int/web/xmm-newton/sas},
	author = {ESA},
	year = {2023},
}

@article{dressing_occurrence_2015,
	title = {The {Occurrence} of {Potentially} {Habitable} {Planets} {Orbiting} {M} {Dwarfs} {Estimated} from the {Full} {Kepler} {Dataset} and an {Empirical} {Measurement} of the {Detection} {Sensitivity}},
	volume = {807},
	issn = {0004-637X},
	url = {https://ui.adsabs.harvard.edu/abs/2015ApJ...807...45D/abstract},
	doi = {10.1088/0004-637X/807/1/45},
	language = {en},
	number = {1},
	urldate = {2025-07-23},
	journal = {ApJ},
	author = {Dressing, Courtney D. and Charbonneau, David},
	month = jul,
	year = {2015},
	pages = {45},
}

@article{dressing_occurrence_2013,
	title = {The {Occurrence} {Rate} of {Small} {Planets} around {Small} {Stars}},
	volume = {767},
	issn = {0004-637X},
	url = {https://ui.adsabs.harvard.edu/abs/2013ApJ...767...95D/abstract},
	doi = {10.1088/0004-637X/767/1/95},
	language = {en},
	number = {1},
	urldate = {2025-07-23},
	journal = {ApJ},
	author = {Dressing, Courtney D. and Charbonneau, David},
	month = apr,
	year = {2013},
	pages = {95},
}

@article{czesla_h_2022,
	title = {Hα and {He} {I} absorption in {HAT}-{P}-32 b observed with {CARMENES}. {Detection} of {Roche} lobe overflow and mass loss},
	volume = {657},
	issn = {0004-6361},
	url = {https://ui.adsabs.harvard.edu/abs/2022A%26A...657A...6C/abstract},
	doi = {10.1051/0004-6361/202039919},
	language = {en},
	number = {A6},
	urldate = {2025-06-09},
	journal = {A\&A},
	author = {Czesla, S. and Lampón, M. and Sanz-Forcada, J. and García Muñoz, A. and López-Puertas, M. and Nortmann, L. and Yan, D. and Nagel, E. and Yan, F. and Schmitt, J. H. M. M. and Aceituno, J. and Amado, P. J. and Caballero, J. A. and Casasayas-Barris, N. and Henning, Th and Khalafinejad, S. and Molaverdikhani, K. and Montes, D. and Pallé, E. and Reiners, A. and Schneider, P. C. and Ribas, I. and Quirrenbach, A. and Zapatero Osorio, M. R. and Zechmeister, M.},
	month = jan,
	year = {2022},
	pages = {26},
}

@article{currie_direct_2023,
	title = {Direct {Imaging} and {Spectroscopy} of {Extrasolar} {Planets}},
	volume = {534},
	issn = {1050-3390},
	url = {https://ui.adsabs.harvard.edu/abs/2023ASPC..534..799C/abstract},
	doi = {10.48550/arXiv.2205.05696},
	language = {en},
	urldate = {2025-08-05},
	journal = {PP7},
	author = {Currie, T. and Biller, B. and Lagrange, A. and Marois, C. and Guyon, O. and Nielsen, E. L. and Bonnefoy, M. and De Rosa, R. J.},
	month = jul,
	year = {2023},
	pages = {799},
}

@incollection{colombo_stellar_2025,
	title = {Stellar {X}-{Ray}-{UV} {Coronal} {Activity} and {Its} {Impact} on {Planets}},
	isbn = {978-3-319-30648-3},
	url = {https://doi.org/10.1007/978-3-319-30648-3_19-2},
	language = {en},
	urldate = {2025-01-31},
	booktitle = {Handbook of {Exoplanets}},
	publisher = {Springer Nature Switzerland},
	author = {Colombo, Salvatore and Locci, Daniele and Spinelli, Riccardo and Petralia, Antonino and Cecchi-Pestellini, Cesare and Micela, Giuseppina},
	editor = {Deeg, Hans J. and Belmonte, Juan Antonio},
	year = {2025},
	pages = {1--19},
}

@article{cohen_dynamics_2011,
	title = {The {Dynamics} of {Stellar} {Coronae} {Harboring} {Hot} {Jupiters}. {II}. {A} {Space} {Weather} {Event} on a {Hot} {Jupiter}},
	volume = {738},
	issn = {0004-637X},
	url = {https://ui.adsabs.harvard.edu/abs/2011ApJ...738..166C/abstract},
	doi = {10.1088/0004-637X/738/2/166},
	language = {en},
	number = {2},
	urldate = {2025-07-28},
	journal = {AJ},
	author = {Cohen, O. and Kashyap, V. L. and Drake, J. J. and Sokolov, I. V. and Gombosi, T. I.},
	month = sep,
	year = {2011},
	pages = {166},
}

@article{ciaravella_loop_1997,
	title = {Loop modeling of coronal {X}-ray spectra. {III}. {Fitting} loop spectra with one- and two-component thermal models.},
	volume = {320},
	issn = {0004-6361},
	url = {https://ui.adsabs.harvard.edu/abs/1997A&A...320..945C/abstract},
	language = {en},
	urldate = {2025-07-03},
	journal = {A\&A},
	author = {Ciaravella, A. and Maggio, A. and Peres, G.},
	month = apr,
	year = {1997},
	pages = {945--956},
}

@article{chen_effects_2025,
	title = {Effects of {Transient} {Stellar} {Emissions} on {Planetary} {Climates} of {Tidally} {Locked} {Exo}-{Earths}},
	volume = {170},
	issn = {1538-3881},
	url = {https://doi.org/10.3847/1538-3881/add33e},
	doi = {10.3847/1538-3881/add33e},
	language = {en},
	number = {1},
	urldate = {2025-10-28},
	journal = {AJ},
	author = {Chen, Howard and De Luca, Paolo and Hochman, Assaf and Komacek, Thaddeus D.},
	month = jun,
	year = {2025},
	pages = {40},
}

@book{burnham_model_1998,
	title = {Model {Selection} and {Inference}},
	copyright = {http://www.springer.com/tdm},
	isbn = {978-1-4757-2919-1 978-1-4757-2917-7},
	url = {http://link.springer.com/10.1007/978-1-4757-2917-7},
	language = {en},
	urldate = {2025-06-20},
	publisher = {Springer},
	author = {Burnham, Kenneth P. and Anderson, David R.},
	year = {1998},
}

@article{boyajian_stellar_2012,
	title = {Stellar {Diameters} and {Temperatures}. {II}. {Main}-sequence {K}- and {M}-stars},
	volume = {757},
	issn = {0004-637X},
	url = {https://ui.adsabs.harvard.edu/abs/2012ApJ...757..112B/abstract},
	doi = {10.1088/0004-637X/757/2/112},
	language = {en},
	number = {2},
	urldate = {2025-06-10},
	journal = {ApJ},
	author = {Boyajian, Tabetha S. and von Braun, Kaspar and van Belle, Gerard and McAlister, Harold A. and ten Brummelaar, Theo A. and Kane, Stephen R. and Muirhead, Philip S. and Jones, Jeremy and White, Russel and Schaefer, Gail and Ciardi, David and Henry, Todd and López-Morales, Mercedes and Ridgway, Stephen and Gies, Douglas and Jao, Wei-Chun and Rojas-Ayala, Bárbara and Parks, J. Robert and Sturmann, Laszlo and Sturmann, Judit and Turner, Nils H. and Farrington, Chris and Goldfinger, P. J. and Berger, David H.},
	month = oct,
	year = {2012},
	pages = {112},
}

@article{binder_x-ray_2024,
	title = {X-{Ray} {Emission} of {Nearby} {Low}-mass and {Sunlike} {Stars} with {Directly} {Imageable} {Habitable} {Zones}},
	volume = {275},
	issn = {0067-0049},
	url = {https://ui.adsabs.harvard.edu/abs/2024ApJS..275....1B},
	doi = {10.3847/1538-4365/ad71d6},
	urldate = {2025-02-12},
	journal = {ApJS},
	author = {Binder, Breanna A. and Peacock, Sarah and Schwieterman, Edward W. and Turnbull, Margaret C. and Virgen, Azariel Y. and Kane, Stephen R. and Farrish, Alison and Garcia-Sage, Katherine},
	month = nov,
	year = {2024},
	pages = {1},
}

@article{bessell_late_1991,
	title = {The {Late} {M} {Dwarfs}},
	volume = {101},
	issn = {0004-6256},
	url = {https://ui.adsabs.harvard.edu/abs/1991AJ....101..662B/abstract},
	doi = {10.1086/115714},
	language = {en},
	urldate = {2025-06-10},
	journal = {AJ},
	author = {Bessell, M. S.},
	month = feb,
	year = {1991},
	pages = {662},
}

@article{bazot_uncertain_2016,
	title = {On the uncertain nature of the core of α {Cen} {A}},
	volume = {460},
	issn = {0035-8711, 1365-2966},
	url = {https://academic.oup.com/mnras/article-lookup/doi/10.1093/mnras/stw921},
	doi = {10.1093/mnras/stw921},
	language = {en},
	number = {2},
	urldate = {2025-07-22},
	journal = {MNRAS},
	author = {Bazot, M. and Christensen-Dalsgaard, J. and Gizon, L. and Benomar, O.},
	month = aug,
	year = {2016},
	pages = {1254--1269},
}

@article{ballet_pile-up_1999,
	title = {Pile-up on {X}-ray {CCD} instruments},
	volume = {135},
	copyright = {© European Southern Observatory (ESO), 1999},
	issn = {0365-0138, 1286-4846},
	url = {https://aas.aanda.org/articles/aas/abs/1999/05/h0886/h0886.html},
	doi = {10.1051/aas:1999179},
	language = {en},
	number = {2},
	urldate = {2024-10-10},
	journal = {A\&AS},
	author = {Ballet, J.},
	month = mar,
	year = {1999},
	pages = {371--381},
}

@article{aschwanden_irradiance_1994,
	title = {Irradiance observations of the 1 8 Å solar soft {X}-ray flux from goes},
	volume = {152},
	issn = {0038-0938},
	url = {https://ui.adsabs.harvard.edu/abs/1994SoPh..152...53A/abstract},
	doi = {10.1007/BF01473183},
	language = {en},
	number = {1},
	urldate = {2025-09-03},
	journal = {Sol. Phys.},
	author = {Aschwanden, Markus J.},
	month = jun,
	year = {1994},
	pages = {53--59},
}

@article{ayres_landscape_2025,
	title = {Landscape of {Coronal} {X}-{Ray} {Variability} and {Cycles}},
	volume = {169},
	issn = {0004-6256},
	url = {https://ui.adsabs.harvard.edu/abs/2025AJ....169..281A/abstract},
	doi = {10.3847/1538-3881/adc570},
	language = {en},
	number = {5},
	urldate = {2025-10-27},
	journal = {AJ},
	author = {Ayres, Thomas},
	month = may,
	year = {2025},
	pages = {281},
}

@article{arnaud_xspec_1996,
	title = {{XSPEC}: {The} {First} {Ten} {Years}},
	volume = {101},
	issn = {1050-3390},
	shorttitle = {{XSPEC}},
	url = {https://ui.adsabs.harvard.edu/abs/1996ASPC..101...17A/abstract},
	language = {en},
	urldate = {2025-04-16},
	journal = {ADASS},
	author = {Arnaud, K. A.},
	year = {1996},
	pages = {17},
}

@article{amaral_impact_2025,
	title = {The {Impact} of {Stellar} {Flares} on the {Atmospheric} {Escape} of {Exoplanets} {Orbiting} {M} {Stars}. {I}. {Insights} from the {AU} {Mic} {System}},
	volume = {985},
	issn = {0004-637X},
	url = {https://ui.adsabs.harvard.edu/abs/2025ApJ...985..100A/abstract},
	doi = {10.3847/1538-4357/adc932},
	language = {en},
	number = {1},
	urldate = {2025-11-17},
	journal = {ApJ},
	author = {Amaral, Laura N. R. do and Shkolnik, Evgenya L. and Loyd, R. O. Parke and Peacock, Sarah},
	month = may,
	year = {2025},
	pages = {100},
}

@article{amaral_contribution_2022,
	title = {The {Contribution} of {M}-dwarf {Flares} to the {Thermal} {Escape} of {Potentially} {Habitable} {Planet} {Atmospheres}},
	volume = {928},
	issn = {0004-637X},
	url = {https://doi.org/10.3847/1538-4357/ac53af},
	doi = {10.3847/1538-4357/ac53af},
	language = {en},
	number = {1},
	urldate = {2025-10-28},
	journal = {ApJ},
	author = {Amaral, Laura N. R. and Barnes, Rory and Segura, Antígona and Luger, Rodrigo},
	month = mar,
	year = {2022},
	pages = {12},
}

@article{vidotto_star-planet_2025,
	title = {Star-{Planet} {Interactions}: {A} {Computational} {View}},
	issn = {0066-4146, 1545-4282},
	shorttitle = {Star-{Planet} {Interactions}},
	url = {http://arxiv.org/abs/2506.00470},
	doi = {10.1146/annurev-astro-021225-030604},
	urldate = {2025-06-03},
	journal = {ARA\&A},
	author = {Vidotto, A. A.},
	month = may,
	year = {2025},
}

@article{segura_effect_2010,
	title = {The effect of a strong stellar flare on the atmospheric chemistry of an earth-like planet orbiting an {M} dwarf},
	volume = {10},
	issn = {1557-8070},
	doi = {10.1089/ast.2009.0376},
	language = {eng},
	number = {7},
	journal = {Astrobiology},
	author = {Segura, Antígona and Walkowicz, Lucianne M. and Meadows, Victoria and Kasting, James and Hawley, Suzanne},
	month = sep,
	year = {2010},
	pmid = {20879863},
	pmcid = {PMC3103837},
	pages = {751--771},
}

@article{yan_extended_2018,
	title = {An extended hydrogen envelope of the extremely hot giant exoplanet {KELT}-9b},
	volume = {2},
	issn = {2397-3366},
	url = {http://arxiv.org/abs/1807.00869},
	doi = {10.1038/s41550-018-0503-3},
	number = {9},
	urldate = {2025-06-09},
	journal = {Nature Astronomy},
	author = {Yan, Fei and Henning, Thomas},
	month = jul,
	year = {2018},
	pages = {714--718},
}

@article{molendi_assessing_2003,
	title = {Assessing the {EPIC} spe tral alibration in the hard band with a {3C} 273 observation},
	language = {en},
	journal = {XMM-SOC-CAL-TN-0036},
	author = {Molendi, Silvano and Bassini, Via and Sembay, Steve},
	year = {2003},
}

@article{haisch_solar-like_1995,
	title = {Solar-{Like} {M}-{Class} {X}-ray {Flares} on {Proxima} {Centauri} {Observed} by the {ASCA} {Satellite}},
	volume = {268},
	url = {https://www.science.org/doi/10.1126/science.268.5215.1327},
	doi = {10.1126/science.268.5215.1327},
	number = {5215},
	urldate = {2025-07-08},
	journal = {Science},
	author = {Haisch, Bernhard and Antunes, A. and Schmitt, J. H. M. M.},
	month = jun,
	year = {1995},
	pages = {1327--1329},
}

@misc{esa_xmm-newton_soc_users_2023,
	title = {Users {Guide} to the {XMM}-{Newton} {Science} {Analysis} {System}, issue 18.0},
	url = {https://xmm-tools.cosmos.esa.int/external/xmm_user_support/documentation/sas_usg/USG/},
	author = {ESA: XMM-Newton SOC},
	year = {2023},
}

@article{anglada-escude_terrestrial_2016,
	title = {A terrestrial planet candidate in a temperate orbit around {Proxima} {Centauri}},
	volume = {536},
	issn = {0028-0836},
	url = {https://ui.adsabs.harvard.edu/abs/2016Natur.536..437A/abstract},
	doi = {10.1038/nature19106},
	language = {en},
	number = {7617},
	urldate = {2025-06-10},
	journal = {Nature},
	author = {Anglada-Escudé, Guillem and Amado, Pedro J. and Barnes, John and Berdiñas, Zaira M. and Butler, R. Paul and Coleman, Gavin A. L. and de La Cueva, Ignacio and Dreizler, Stefan and Endl, Michael and Giesers, Benjamin and Jeffers, Sandra V. and Jenkins, James S. and Jones, Hugh R. A. and Kiraga, Marcin and Kürster, Martin and López-González, María J. and Marvin, Christopher J. and Morales, Nicolás and Morin, Julien and Nelson, Richard P. and Ortiz, José L. and Ofir, Aviv and Paardekooper, Sijme-Jan and Reiners, Ansgar and Rodríguez, Eloy and Rodríguez-López, Cristina and Sarmiento, Luis F. and Strachan, John P. and Tsapras, Yiannis and Tuomi, Mikko and Zechmeister, Mathias},
	month = aug,
	year = {2016},
	pages = {437--440},
}

@article{kasting_habitable_1993,
	title = {Habitable {Zones} around {Main} {Sequence} {Stars}},
	volume = {101},
	issn = {0019-1035},
	url = {https://www.sciencedirect.com/science/article/pii/S0019103583710109},
	doi = {10.1006/icar.1993.1010},
	number = {1},
	urldate = {2025-07-24},
	journal = {Icarus},
	author = {Kasting, James F. and Whitmire, Daniel P. and Reynolds, Ray T.},
	month = jan,
	year = {1993},
	pages = {108--128},
}

@article{yelle_aeronomy_2004,
	title = {Aeronomy of extra-solar giant planets at small orbital distances},
	volume = {170},
	issn = {0019-1035},
	url = {https://ui.adsabs.harvard.edu/abs/2004Icar..170..167Y/abstract},
	doi = {10.1016/j.icarus.2004.02.008},
	language = {en},
	number = {1},
	urldate = {2025-06-09},
	journal = {Icarus},
	author = {Yelle, Roger V.},
	month = jul,
	year = {2004},
	pages = {167--179},
}

@misc{ballet_xmm-ccf-rel-0406_2024,
	title = {{XMM}-{CCF}-{REL}-0406},
	url = {https://xmmweb.esac.esa.int/docs/documents/CAL-SRN-0406-1-1.pdf},
	author = {Ballet, Jean and Motch, Christian and Batalha, Rebeca},
	month = feb,
	year = {2024},
}
    
    \appendix

    \section{Dealing with pile-up}\label{appendix:pile-up}
    Pile-up is an instrumental effect in CCD photon-counting detectors that reduces the measured counts and hardens the spectrum. It distorts spectral shapes and can compromise model fitting.
    
    When a photon hits a CCD photon-counting detector, it generates an electronic cloud within a pixel. This cloud often spreads in adjacent pixels, and the resulting pixel configuration is called “pattern” in XMM-Newton nomenclature (or also “grade” in other instruments, e.g. Chandra). Higher-energy photons tends to produce broader charge distributions, resulting in larger pattern values. Spurious events, such as those from cosmic rays or charged particles, usually yield the highest pattern values and are filtered automatically by the onboard software. Valid patterns (e.g., 0–12 for pn, 0-4 for MOS) correspond to events most likely caused by X-rays. These are often grouped into singles (pattern $=$ 0) and doubles or greater (0 $\le$ pattern $\leq$ 12 or 4).
    
    When two or more photons hit the same or neighboring pixels within a single readout frame, they are recorded as a single, more energetic event, a phenomenon known as pile-up. This effect lowers the total count rate, distorts the spectral energy distribution, and shifts the pattern distribution toward higher values. It may also produce non-standard patterns (e.g. diagonal ones). For point sources, most photons concentrate in the core of the (PSF), making the pile-up rate dependent not only on source brightness but also on radial distance from the PSF center. Many high-pattern and high-energy events are flagged as spurious and excluded by onboard filtering. This leads to additional count loss beyond that strictly caused by pile-up itself, often visible as a depletion in the core of the PSF. Both PSF and pattern distribution are strongly energy-dependent, and complex corrections are applied to take this into account. Therefore, pile-up distortion of both the spectral and pattern distributions introduces complex and non-linear effects. Because of these complications, there is currently no universal method to correct for pile-up. Different correction strategies are described in the literature.
    
    \begin{enumerate}
        \item PSF core excising: when the pile-up fraction is high, the most effective strategy is to excise the core of the PSF. This works because the count rate per frame per pixel is higher at the core, whereas the wings of the PSF are usually unaffected. After excluding the core, the total flux can be reconstructed using knowledge of the PSF shape using the wings, where the pile-up fraction is negligible. The method can be applied even when core depletion is not visually apparent, although its accuracy depends on the quality of the PSF calibration and the number of events available in the wings. Two main challenges arise: (a) determining the appropriate excision radius, and (b) reconstructing the excluded counts to recover the total flux. There is no universal criterion for selecting the excision radius, and during periods of high variability it may be necessary to vary it over time.
        \item Single-pattern filtering: Single-pixel events are much less affected by pile-up for geometric and probabilistic reasons. A common strategy is therefore to retain only single-pixel patterns\footnote{Using single-pixel events also helps reject soft proton contamination that may have been missed by onboard filtering, particularly in pn small window mode (as in some of our observations), since such invalid events rarely appear as singles but may appear as doubles \citep{schellenberger_using_2025}.}. In general, the probability that single-pixel events are affected by pile-up is negligible \citep{ballet_pile-up_1999, jethwa_when_2015}. As with the PSF excision method, calibration files are needed to correctly generate a response matrix that takes into account the single-pixel event filtering. However, this correction is automatically applied by SAS without accounting for the pattern migration caused by pile-up, which reduces the fraction of single-pixel events \citep{ballet_pile-up_1999}. Because the standard pattern calibration assumes a fixed singles-to-doubles ratio, it does not correct for this shift, leading to an underestimation of the flux in already pile-up-affected data. Consequently, this method corrects spectral distortions but does not recover the true flux. This limitation is not always explicitly stated. Even without pile-up, the single-pattern filtering method inherently biases against high-energy photons, which are more likely to produce broader charge distributions and higher pattern values. Although this effect is accounted for in the calibration, the filtering still reduces the photon statistics at high energies.
        \item SAS response matrix correction: XMM-Newton guidelines recommend a correction (implemented in SAS) that incorporates pile-up effects into the response matrix. However, this method is implemented only for the pn instrument. Moreover, when performing time-resolved analyses, the response matrix should be recomputed for each time interval individually, that would be extremely time-consuming and, we suspect, also affected by the low statistics. For this reasons, we do not consider this option.
    \end{enumerate}
    
    We tested method 1), applying a variable excision radius determined from the count rate and using the {\it epatplot} output as a reference. However, we found that this correction introduced uncertainties larger than those caused by pile-up itself, likely because of the low number of counts in the PSF wings during most of the affected intervals. Consequently, we discarded this approach.
    
    We adopted method 2), developing a new method to correct the effects of pattern migration. Our correction builds on the relations presented in \cite{jethwa_when_2015}, linking count rate, flux loss, and spectral distortion for XMM-Newton (see their Figure 5). This represents an original adaptation of their framework. The flux loss, $F$, is defined as
    \begin{equation}
    F = 1-\frac{M}{\Lambda}
    \end{equation}
    where $\Lambda$ is the expected number of events and $M$ is the measured number of events. We can invert the relation to obtain
    \begin{equation}
    \Lambda = \frac{M}{1-F}.
    \end{equation}
    Then the number of events counts lost due to pile-up will be $P$:
    \begin{equation}\label{equation:losttopileup}
    P = \frac{MF}{1-F}.
    \end{equation}
    We also know the expected ratio between different pattern events as a function of their energies. For example, from the calibration files, we can extract the fraction of expected single-pattern events as a function of energy, $\alpha(E)$. Assuming that the number of piled-up singles is negligible, the expected number of single-pixel events can be written as:
    \begin{equation}
    \Lambda_s(E) = \alpha(E) \Lambda(E)
    \end{equation}
    We do not know $\Lambda(E)$ but we know the spectrum of the measured singles $M_s(E)$, and because we assume that singles are not pile-up, when the number of events is reasonably high, we can assume 
    \begin{equation}
    \Lambda_s(E) = constant \cdot M_s(E).
    \end{equation}
    If we do so, we can write
    \begin{equation}
    \int \Lambda(E)dE = \Lambda = \int constant\cdot\alpha^{-1}(E)M_s(E)dE
    \end{equation}
    and solve for the constant. This is also suggested in a different way in \cite{ballet_pile-up_1999}.
    Given the complexity of this approach, before testing it, we deem more reasonable to test the extreme limit in which we assume
    \begin{equation}
    \Lambda_s - M_s = \Lambda - M = P
    \end{equation}
    id est, we consider the extreme case in which all the lost counts are considered as single counts. We don't know the impact of this correction on the results and it could be negligible compared to our error, only introducing more unpredictable errors, so we will evaluate the worst case scenario.
    
    We apply this correction directly on the raw data in the following way:
    \begin{itemize}
        \item we divide the observation into 30 seconds time intervals, then we proceed for each interval;
        \item we use data from Figure 5 of \cite{jethwa_when_2015} and equation \ref{equation:losttopileup} to estimate the number of missing events due to pile-up;
        \item we construct a cumulative distribution from $M_s(E)$, the measured singles, and we extract $P$ events energy randomly;
        \item we assign a random time within the interval to the events energies and we get a set of events;
        \item we add those events to the original event list to obtain a half-synthetic event list corrected for single pattern events pile-up loss.
    \end{itemize}
    
    A limitation of this correction, beyond it representing an upper limit, is that single-pattern filtering typically excludes the most energetic photons, which are generally doubles or greater patterns. Although this effect should be accounted for by SAS in the calibration steps, it nonetheless reduces the number of high-energy counts and, therefore, their statistical significance and our ability to constraint the plasma higher-temperature components. Moreover, since pile-up occurs primarily during flares, which produce harder spectra, this effect may particularly increase uncertainties at the high-energy end of the reconstructed spectra during such events when filtering for single-pattern events.
    
    We end up with three different source event lists for each observation and each instrument:
    \begin{enumerate}
       \item NOCORR: all valid-pattern events are included; no correction is applied. This list suffers from pile-up during bright intervals and serves as a lower bound as well as the uncorrected reference.
       \item STDCORR: only single-pattern events are used. This is the simplest correction, designed to correct spectral distortion only, though it is sometimes used as a general correction. Results from this list help estimate flux loss due to pattern migration.
       \item NEWCORR: our new method, which applies single-pattern filtering combined with generated events. Results from this list represent an upper limit because we are assuming that all the events affected by pile-up would have resulted in single pattern events.
    \end{enumerate}
    
    In Figure \ref{fig:lc_pu_comparison}, we show the estimated count-loss percentage for the different instruments during the two major events in our sample (observations 0049350101 and 0551120401). Pile-up count loss reaches up to 22{\%} for the pn. This occurs during the less intense of the two events because of different frame mode used by XMM-Newton during observation 0049350101. It directly reflects the correction dependence on the frame time coming from the relations of \cite{jethwa_when_2015}. In contrast, MOS1 and MOS2 were operated with the same setup throughout all observations, and can therefore serve as a reference. For these instruments, pile-up reach its maximum (13{\%}) during the greatest flare (observation 0049350101). 
                
    \begin{figure}[htbp]
        \centering
        \includegraphics[width=0.9\linewidth]{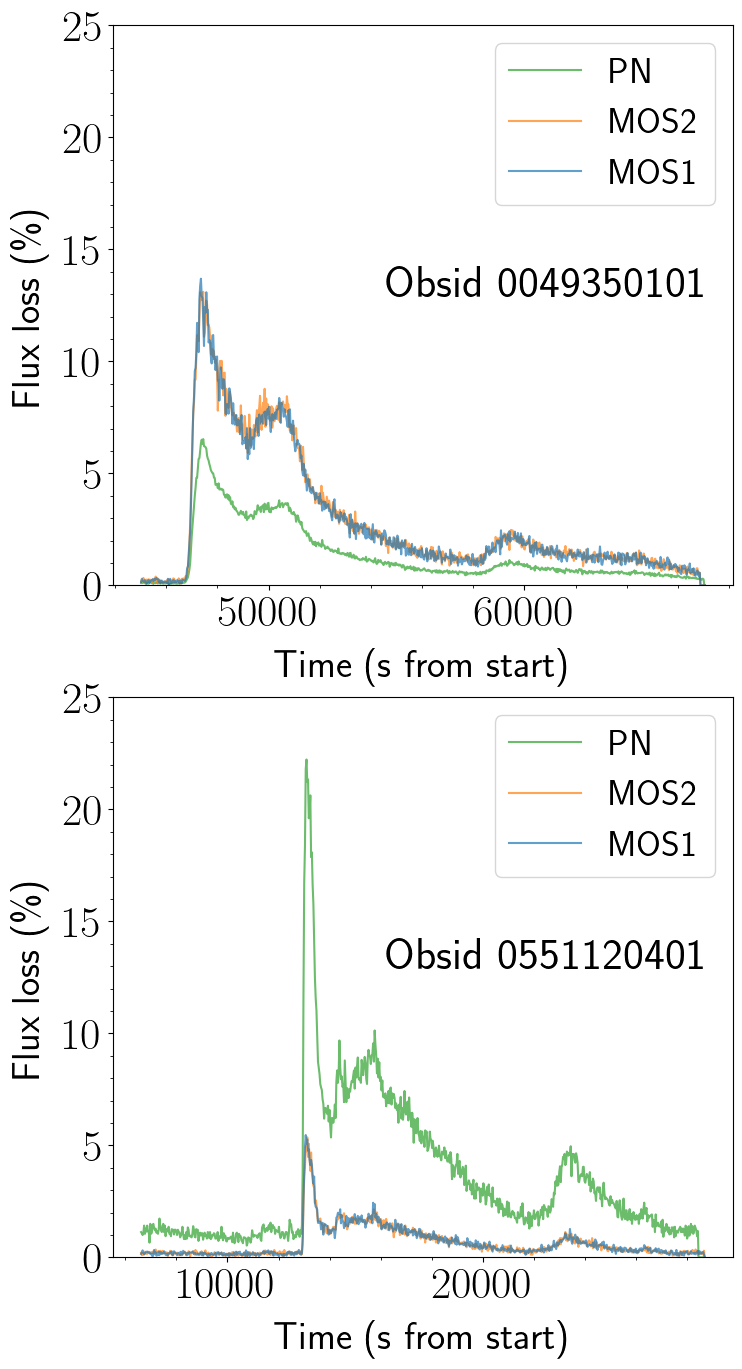}
        \caption{A comparison between the pile-up flux-loss inferred for the different instruments during observations 0049350101 and 0551120401. The flux loss is defined as in \cite{jethwa_when_2015}.}
        \label{fig:lc_pu_comparison}
    \end{figure}
    
    We now examine how the subsequent steps in the reduction process affect the total flux, defining the effects of pile-up at the end of the analysis chain as a function of time and flux. We adopt the 3-temperature model as a reference. We use uppercase abbreviations to refer to the total fluxes derived from each correction method. As a metric for pile-up-induced flux loss, we define the quantity $1 - \frac{\text{NOCORR}}{\text{XXXCORR}}$, where $\text{XXX} = \text{STD}$ or $\text{NEW}$. In Figures \ref{fig:pileup_all_single} and \ref{fig:pileup_all_singlecorr}, we plot this quantity for the STDCORR and NEWCORR cases, respectively.
    
    \begin{figure}[htbp]
        \centering    
        \includegraphics[width=0.99\linewidth]{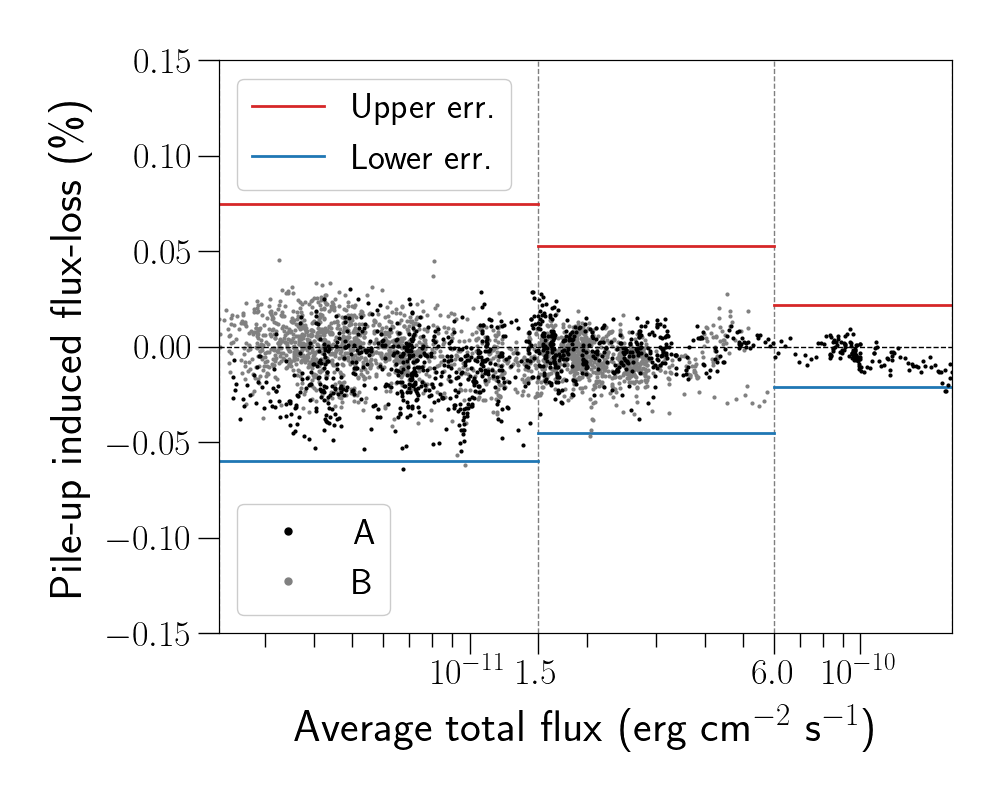}
        \caption{Comparison between fluxes obtained without correcting for pile-up and correcting by selecting only single patterns. Pile-up induced flux-loss defined as (1 - NOCORR/STDCORR) is plotted with black/gray dots over the total flux. Black dots represent intervals of the observation with id 0049350101, that was observing with pn in small frame mode. The gray dots are used for all the other observations, that we observing with the standard frame-time. Relative upper/lower total flux 1-$\sigma$ errors from the reduction procedure are depicted as red/blue segments for the same flux ranges as in Figure \ref{fig:total_flux_errors}.}
        \label{fig:pileup_all_single}   
    \end{figure}
    
    \begin{figure}[htbp]
        \centering
        \includegraphics[width=0.99\linewidth]{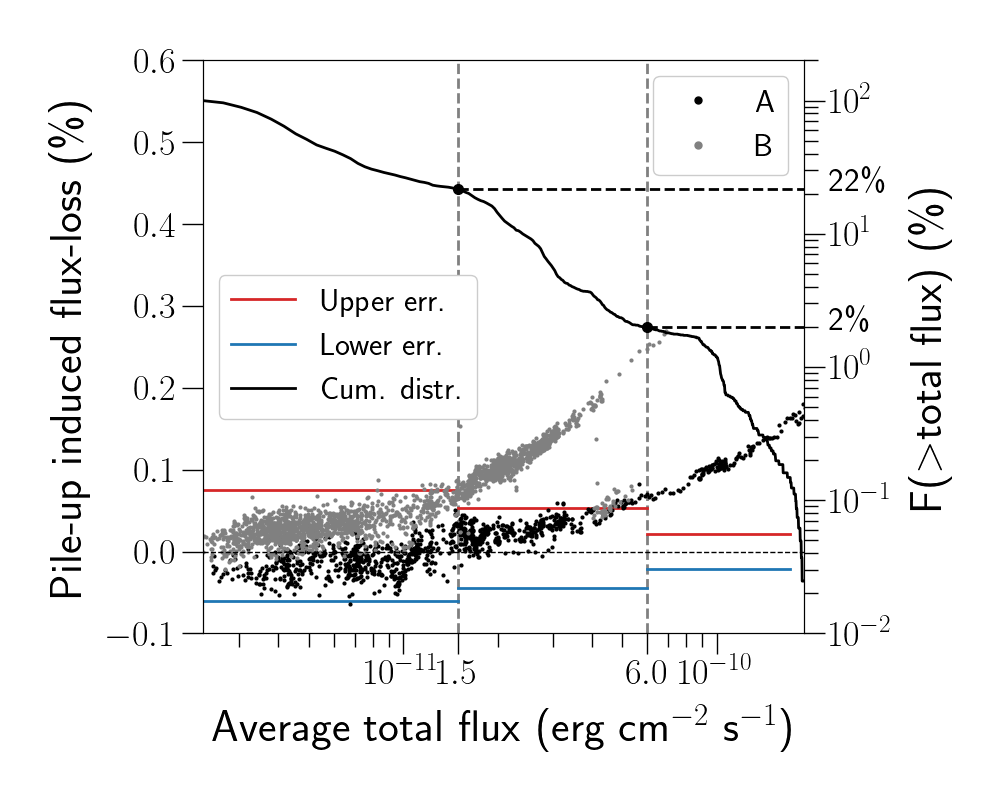}
        \caption{Same of Figure \ref{fig:pileup_all_single} but here the flux loss is defined as (1 - NOCORR/NEWCORR), using the fluxes obtained with the new correction method. The right y-axis shows the inverse cumulative distribution of the total fluxes.}
        \label{fig:pileup_all_singlecorr}
    \end{figure}
    
    Figure \ref{fig:pileup_all_single} shows that, when comparing fluxes computed using all valid patterns (NOCORR) with those using only single-pixel events (STDCORR), the differences induced by pile-up remain well within the flux uncertainties. This indicates that STDCORR does not effectively correct for flux loss: the retrieved closely matches the NOCORR one, implying that applying the correction has negligible impact on the total flux.
    
    Figure \ref{fig:pileup_all_singlecorr} clearly shows flux loss when using the NEWCORR correction. Two distinct populations of time intervals, labeled (A) and (B) in black and gray, correspond to different frame modes of the pn detector. This reflects both the frame-mode-dependent nature of the correction model and the dominant contribution of pn relative to MOS1 and MOS2 in the total signal.
    
    When the pile-up fraction is low, the correction may introduce uncertainties larger than the flux loss itself. Moreover, at low flux levels, pile-up-induced losses are generally smaller than the statistical uncertainties. For this reason, we identify the time intervals for which the correction is meaningful and omit it when unnecessary.
    
    Our strategy is as follows: we apply the pile-up correction only when the estimated flux loss exceeds the total flux error. This allows us to preserve the higher signal-to-noise ratio of the full-pattern dataset in low-count intervals. We find that the pile-up effect can be neglected in the low-flux, eye-selected range, where it is consistently smaller than the average instrumental error and approximately symmetric around zero. For intervals in this range, we use the NOCORR reduction. For total fluxes above $1.5 \times 10^{-11}$ erg s$^{-1}$ cm$^{-2}$, corresponding to roughly 22{\%} of the observing time, we apply the correction.
    
    In principle, separate thresholds could be defined for each detector configuration (i.e., combinations of pn, MOS1, and MOS2 modes). However, based on our tests, a single threshold across all configurations provides sufficiently accurate correction for our purposes.
    
    \section{Further comments on model selection}\label{appendix:further_comments_on_model_selection}
    In Figure \ref{fig:spectra_model_comparisons}, we compare results obtained by applying a single model to all time intervals versus selecting the best model for each time bin. To show the wide range covered and the frequency of occurrence, we divide the spectra according to total flux percentiles. For each percentile interval we plot the average spectrum. For wavelengths shorter than $\sim 5$ {\AA}, different models yield spectra that differ by many orders of magnitude. This is due to the low signal we have at those frequencies and to their sensitivity to the high temperatures, which changes in different models. Fluxes at these energies are better constrained only during flares.
        
    \begin{figure}
        \centering
        \includegraphics[width=0.99\linewidth]{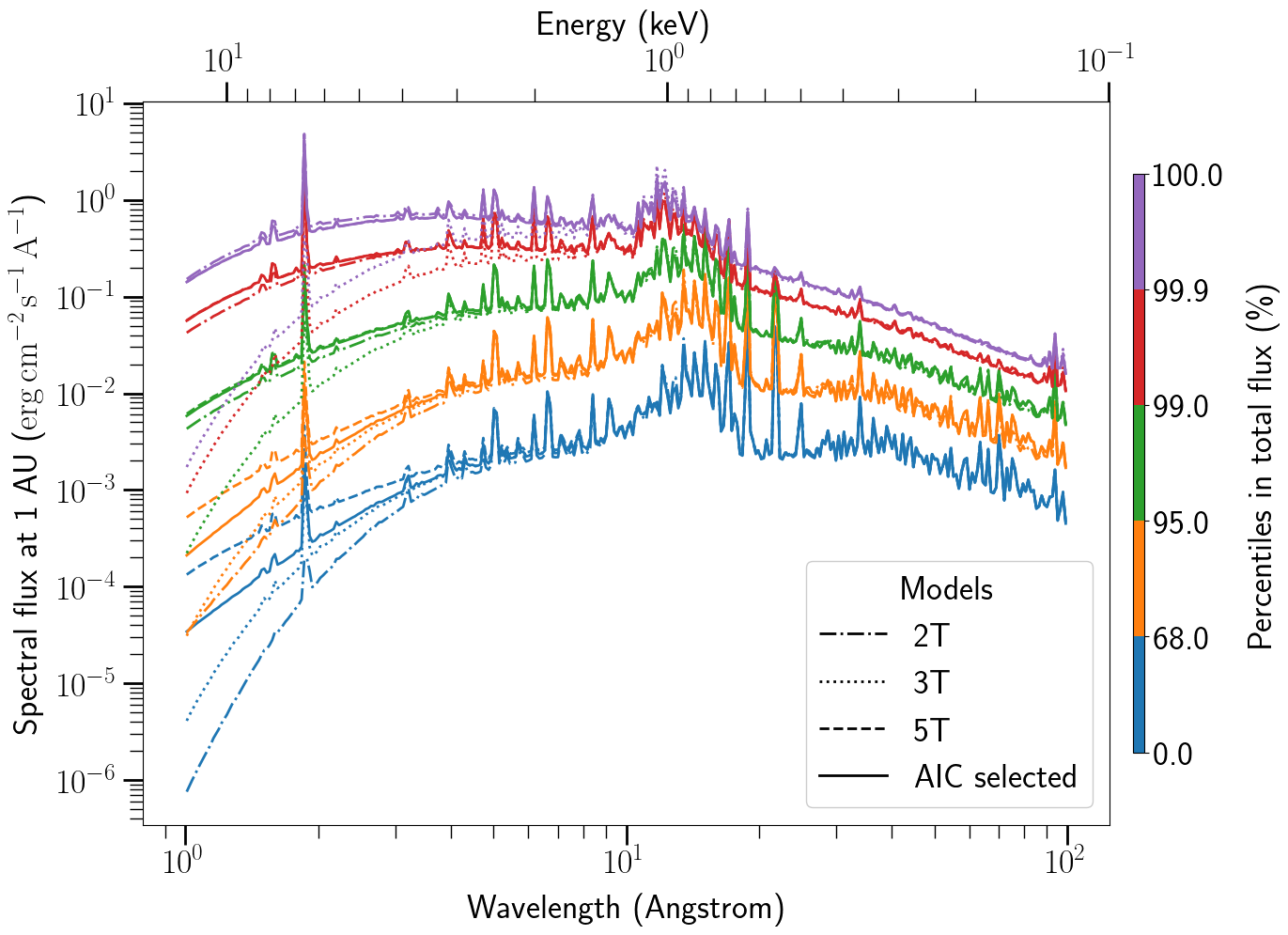}
        \caption{Different models and AIC-selected flux density averages over various total-flux-percentiles are compared. The color bar represents the percentile interval. }
        \label{fig:spectra_model_comparisons}
    \end{figure}
    
    Figure \ref{fig:spectra_limits} shows the AIC-selected spectra only. Averages are plotted with continuous lines. Shaded areas, delimited by dotted lines, represent the extremes reached in each energy for the specific percentile interval. To avoid confusion, we only plot three of the five percentile intervals, but similar behaviors are found also for the others. We note that the red set reaches higher flux density values than the purple one in the short-wavelength bins. This highlights the wide variety of spectral shapes, and shows that a higher total flux does not necessarily correspond to a higher high-energy partial flux. We warn against interpreting dotted lines as spectra. They are not, but they represent extreme values reached independently in each wavelength bin.
    
    \begin{figure}
        \centering
        \includegraphics[width=0.99\linewidth]{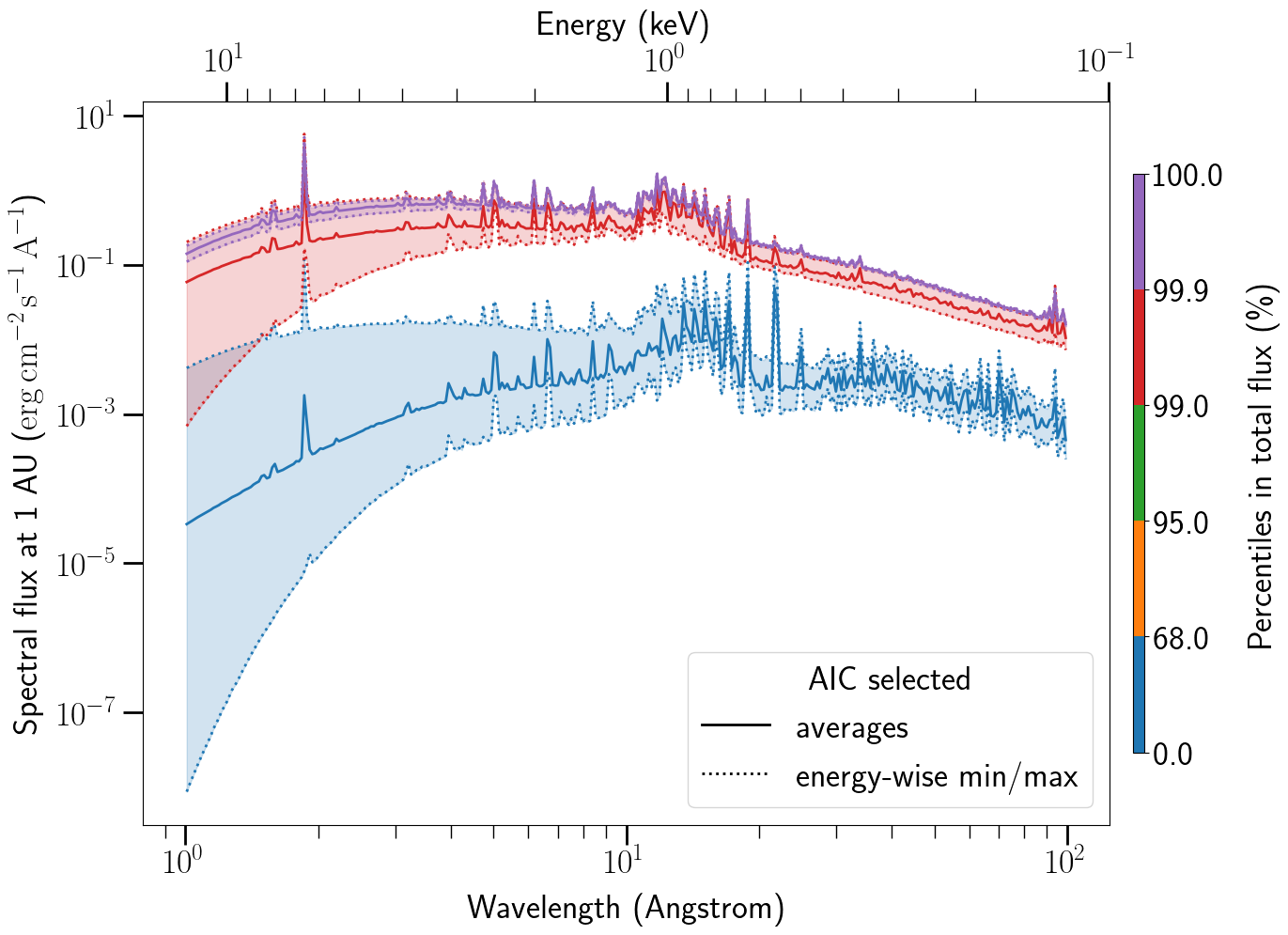}
        \caption{AIC-selected flux density averages over various total-flux-percentiles are plotted with a continuous line. The shaded area delimited by dotted lines represents maximum and minimum values reached in that percentile interval at that wavelength. For clarity's sake only three percentile intervals are reported.}
        \label{fig:spectra_limits}
    \end{figure}
    
\end{document}